\documentclass{aa}  

\usepackage{graphicx}
\usepackage{txfonts}
\usepackage[colorlinks=true, linkcolor=blue, citecolor=blue, urlcolor=blue]{hyperref}
\usepackage{placeins}
\usepackage{subcaption}
\usepackage{natbib}
\bibpunct{(}{)}{;}{a}{}{,}
\begin{document}

	   \title{Measuring the expansion velocities of broad-line Ic supernovae}
	   
	   \subtitle{An investigation of neglected sources of error in two popular methods}

       \author{
           G. Finneran\,\href{https://orcid.org/0000-0001-7590-2920}{\includegraphics[height=10pt]{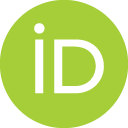}}\inst{1 \thanks{gabriel.finneran@ucdconnect.ie}} \and 
           A. Martin-Carrillo\,\href{https://orcid.org/0000-0001-5108-0627}{\includegraphics[height=10pt]{ORCIDLogomark128x128.png}}\inst{1}
        }
       \institute{School of Physics and Centre for Space Research, University College Dublin, D04 V1W8 Dublin, Ireland}
  
  		\date{}
 
  \abstract
   {The velocities of Ic-BL supernovae (measured from the Fe II feature near 5000\,\AA), are commonly determined using two techniques, which can sometimes result in different velocities for the same event. The template fitting method blueshifts and broadens a template Ic spectrum to match the Ic-BL spectrum under investigation, while the spline fitting method uses spline interpolation to pinpoint the minimum wavelength of a given feature. Both methods have been widely applied, but little attention has been devoted to determining their merits and quantifying their uncertainties.}
   {This work directly compares and contrasts both methods qualitatively and quantitatively to determine the best practices for their application. It identifies sources of error which are not accounted for by most authors and quantifies their impact on the final velocity measurement, highlighting the strengths and weaknesses of each method. Finally, it identifies the cause of velocity discrepancies for events measured using both methods.} 
   {We quantified the impact of smoothing the spectra prior to the application of each method using a range of smoothing parameters applied to two well-sampled cases. Additional sources of error were identified from the assumptions required by each method. To identify the source of velocity discrepancies, the velocities of two cases were measured and directly compared.}
   {Additional sources of error for template fitting arise due to the choice of phase of the template spectrum ($\sim$1000 km/s) and smoothing of the input spectrum ($\sim$500 km/s). These errors are relatively small compared to typical errors of 3000 km/s on this method. The impact of phase shifts is minimised at peak time. The spline fitting method tends to underestimate uncertainties by around 1000 km/s. This method can also be impacted by fine tuning of the smoothing parameters ($\sim$500-1000 km/s), but careful selection of these parameters can mitigate this. Optimum smoothing parameters for different cases are presented along with suggestions for best practice. Direct comparison of both methods showed that velocity discrepancies are not always present, debunking the claim that the template fitting method always handles blending better than spline fitting. Spline fitting seems to struggle to handle blending only in cases where the Fe II features are superimposed on a red continuum, which biases the minimum of this feature towards the bluest line of the triplet, creating an artificially higher velocity. This situation may be relatively rare among Ic-BLs, based on typical temperature evolution.}
   {Both methods tend to underestimate the uncertainty on velocity, however the spline fitting method has a lower error overall. Both methods can be applied under certain circumstances with similar results. Further research into the conditions under which spline fitting may produce larger velocities than the template fitting method is required, including comparisons of the results obtained with both methods to simulations. The morphology of the velocity evolution of an SN appears to be the same regardless of the method used.}
  		
   		\maketitle
\nolinenumbers
\section{Introduction}
	The optical spectra of supernovae tell us about their composition, the composition of their environment and their geometry. By measuring the observed wavelengths of prominent lines or features in these spectra, it is possible to infer the velocity of certain elements in the ejecta. The spectra of broad-line Ic supernovae (Ic-BLs) lack hydrogen and helium and are dominated by broad features of Fe, Si and Ca \citep[see e.g.][]{Ho.2022}. Their broad lines are the result of high expansion velocities, which are significantly larger than those of Ic supernovae \citep[e.g.][]{Modjaz.2016}. Ic-BL supernovae are of particular interest as they are the only supernova type linked with gamma-ray bursts (GRBs) \citep[e.g][]{Cano.201723}. It has been proposed that the influence of the GRB jet may lead to differences in velocity or velocity evolution between Ic-BLs with and without GRBs (e.g. \citet{Finneran.2024B}; \citet{Modjaz.2016}).
	
	The velocities of Ic-BL features are most commonly measured using either the template fitting method \citep[e.g.][]{Srinivasaragavan.2024, Taddia.2019, Modjaz.2016} or by spline fitting methods \citep[e.g.][]{Finneran.2024B,DElia.2015, Schulze.2014,Patat.2001q8t}. The template fitting method uses a Ic template spectrum which is broadened and blue-shifted until it matches the input Ic-BL spectrum \citep{Modjaz.2016}. Methods such as spline fitting interpolate the shape of the feature in question to determine the wavelength of the minimum flux, applying the doppler formula to determine the velocity \citep[e.g.][]{Silverman.2012, Patat.2001q8t}. Although both methods are widely used to measure the velocities of Ic-BLs, they are often implemented without a discussion of their uncertainties or their compatibility. 
	
	The rapid expansion of the supernova ejecta means that the features of Ic-BL spectra are heavily blended. In regions where a feature is composed of multiple lines, it can be difficult to ascribe the minimum flux of a large feature to one line; for example the Fe II feature near 5000 \AA. This is likely to impact the velocities and velocity evolution produced by spline fitting methods, as shown by \cite{Prentice.2018}. The template fitting method was developed in order to handle exactly this scenario \citep{Modjaz.2016}. However, it tends to produce larger errors than spline fitting, and often produces visually poor fits. The question remains whether these errors represent the true uncertainty on the velocity, or are over-estimates. Although this method relies on several assumptions, the impact of these, in particular their potential contribution to the overall error, has not yet been studied in detail.
	
	Although it may produce inaccurate results in regions with heavy blending, the spline fitting method offers some advantages. The errors produced by this method are generally smaller than those of template fitting, and the fits are generally of higher quality. However, smaller errors may be an indiction of under-estimation, which could produce false trends in velocity evolution. Another advantage is that the spline fitting method makes no assumptions about the underlying spectrum, and can be applied to all features in the Ic-BL spectrum. In contrast, the template fitting method cannot be applied to the silicon or calcium features of Ic-BLs. This is due to the differences in the spectra of Ic and Ic-BL supernovae in this region \citep{Modjaz.2016}. This makes the spline fitting method much more useful for the analysis of multiple features for a large population of Ic-BLs \citep[e.g.][]{Finneran.2024B}. 
	
Previous studies have looked for evidence of differences in velocity between Ic-BLs with and without GRBs. \cite{Modjaz.2016} found that are slight differences in overall velocities, however \cite{Finneran.2024B} found these differences to be marginal to non-existent. At present these studies are hampered by small sample sizes, but it is expected that in the coming years the number of Ic-BLs, particularly those with a GRB detection, will grow. As a result it will be important to determine the optimum strategy for measuring the velocities of these events. \cite{Modjaz.2016} noted that these two methods produce statistically different velocities for the same supernova, and this was also observed for some SNe in the analysis presented by \cite{Finneran.2024B}. Specifically, for some SNe, the spline-fitting method seems to produce velocities that are larger than those obtained by the template-fitting method, however this is not the case in all instances. This discrepancy raises the issue of which method should be trusted for Ic-BL supernova velocities. Indeed, this discussion is also relevant to supernova velocity determination more generally, as methods similar to the spline fitting method are often applied to other supernova subtypes \citep[e.g.][]{Silverman.2012}. 
		
		Given the widespread usage of these two techniques, this paper focusses on a review of potential sources of error \textbf{that} are not currently accounted for in either method. In Sect. \ref{sec:smoothingerrors} the impact of smoothing the spectra in both methods is determined, and some heuristics for determining the optimum smoothing parameters are developed. Additional sources of error are identified and quantified for the template fitting method in Sect. \ref{sec:templatefitting}, and for the spline fitting method in Sect. \ref{sec:splinesystematics}. In Sect. \ref{sec:directcompare} a direct comparison of both techniques for two test supernovae is used to identify potential factors leading to the velocity discrepancies found by previous studies. Section \ref{sec:discussion} suggests possible use cases for both methods, as well as mitigation techniques to reduce the impact of these novel sources of error. Finally the conclusions are presented in Sect. \ref{sec:conclusion}.

\section{Impact of spectrum smoothing on velocities}\label{sec:smoothingerrors}

\begin{figure*}[ht!]
	\centering
	\includegraphics[width=0.49\linewidth]{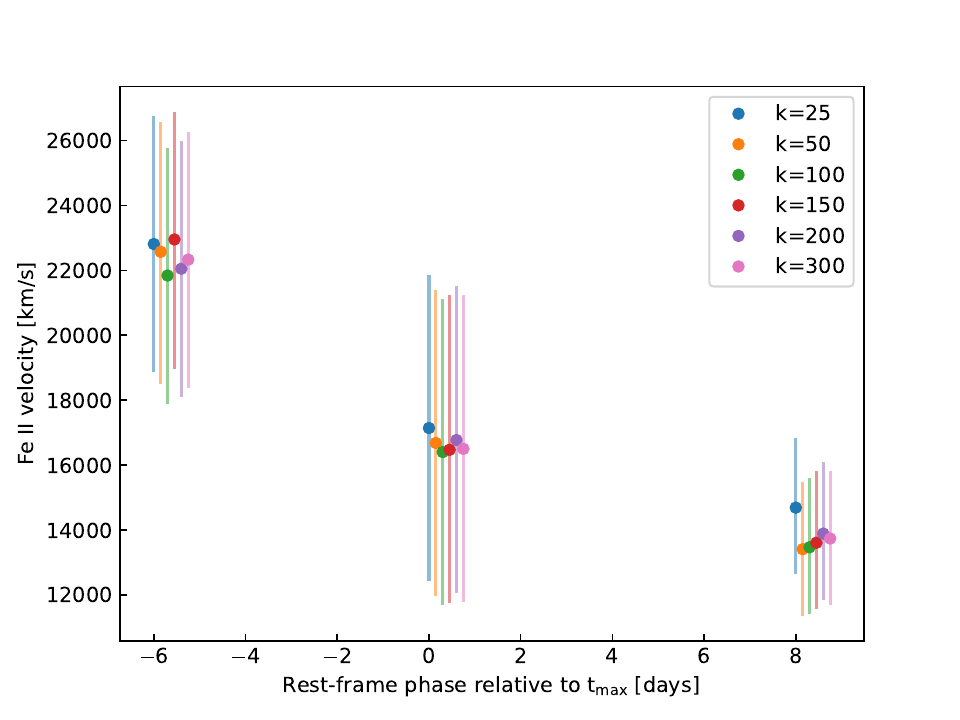}
	\includegraphics[width=0.49\linewidth]{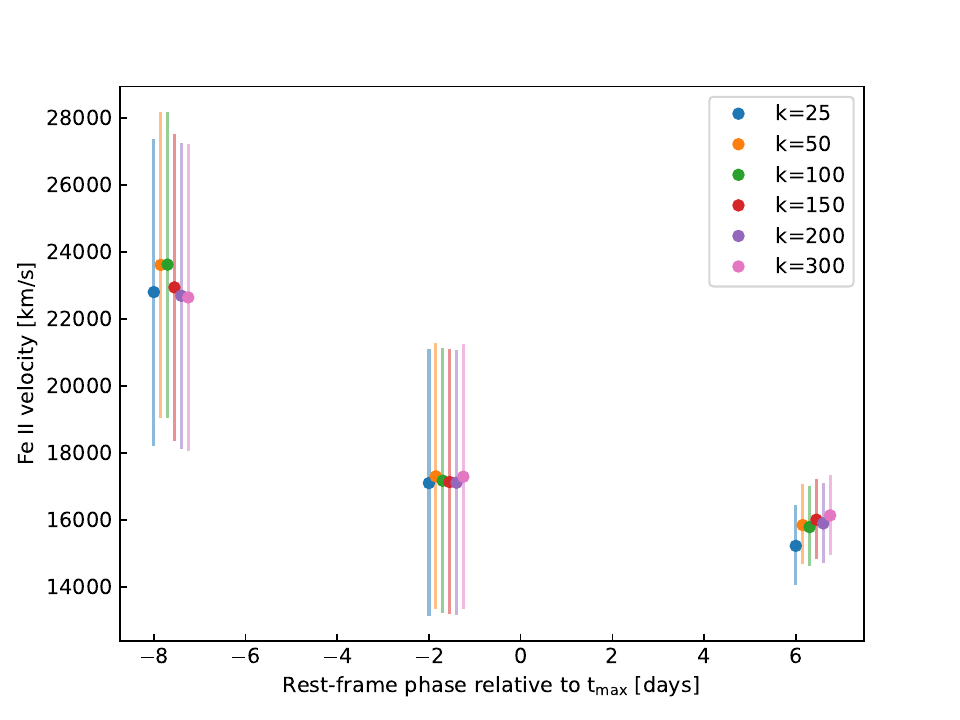}
	\caption[Effects of smoothing on the template fitting method 1]{Influence of the smoothing parameter, $k$, on velocity measured by the template fitting method for GRB980425-SN1998bw (left panel) and GRB130702A-SN2013dx (right panel). A smaller value of $k$ implies more severe smoothing.}
	\label{fig:keffects1}
\end{figure*}

\begin{figure*}[h!]
	\centering
	\includegraphics[width=0.49\linewidth]{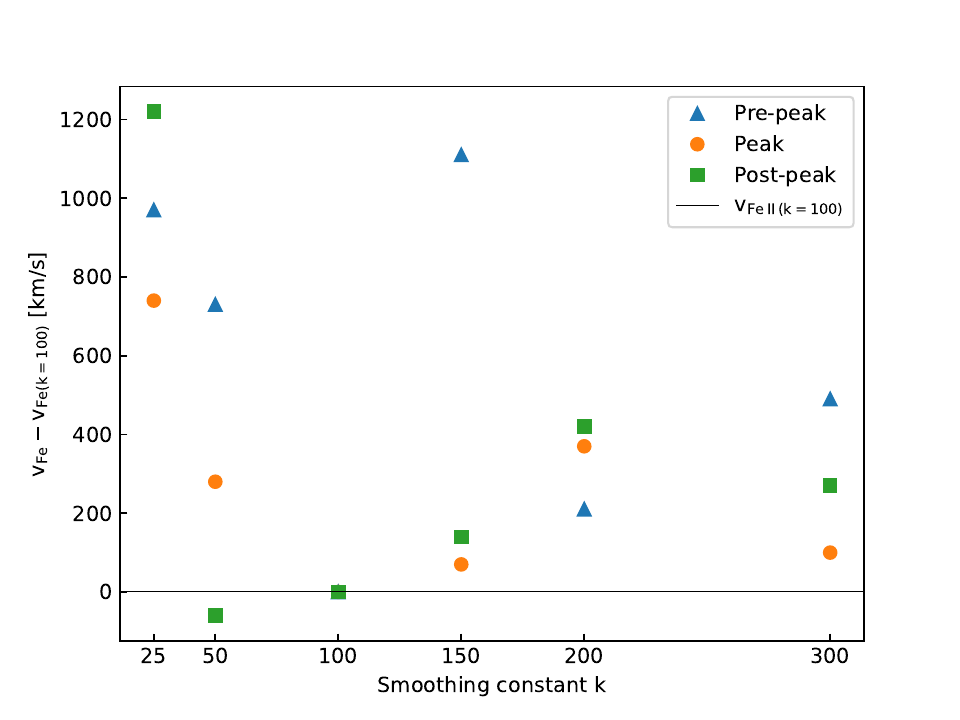}
	\includegraphics[width=0.49\linewidth]{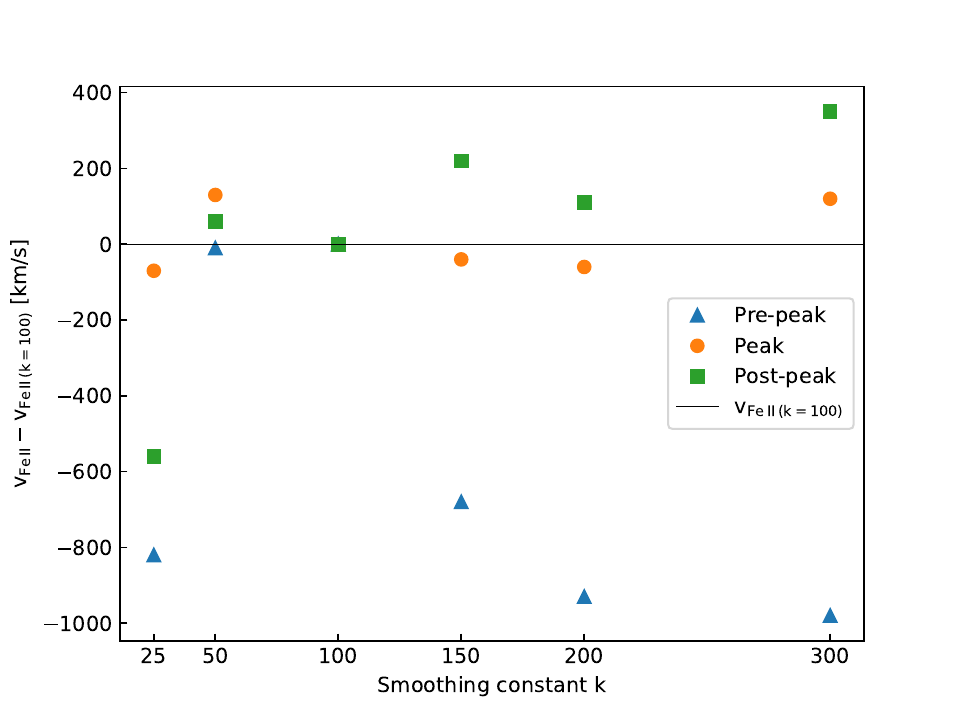}
	\caption[Effects of smoothing on the template fitting method 2]{Similar to Fig. \ref{fig:keffects1}, with the velocities scaled by the velocity for $k$=100. GRB980425-SN1998bw (left panel) and GRB130702A-SN2013dx (right panel)}
	\label{fig:keffects2}
\end{figure*}

To reliably quantify the velocity of a given feature in a spectrum, some sort of smoothing algorithm must normally be applied to the data. The goal of all smoothing algorithms is to remove as much noise as possible from the raw data without introducing significant new features into the spectrum, or modifying existing features substantially. Spectral noise can complicate the process of determining the edges and minimum wavelengths of spectral features; for instance by creating local maxima or minima that do not truly represent the overall feature. Smoothing processes can leave visible differences in the processed spectrum, such as reduced peak heights and broadened features. This is particularly true for algorithms such as the Savitzky-Golay filter (SG, \cite{SG.1964}), which preserves the peak area but not the height. \cite{Finneran.2024B} measured the velocities of a large sample of Ic-BL supernovae with and without GRBs using a spline-fitting method, which relies on the SG filter for smoothing. The template-fitting method, developed by \cite{Modjaz.2016}, uses the Fourier smoothing method described in \cite{Liu.2016}. We proceed to describe the effects of smoothing on the velocities inferred from these two methods.

\subsection{Smoothing in the template-fitting method}

A Fourier smoothing method similar to that used by \cite{Liu.2016} was developed for this analysis\footnote{An analysis was carried out to confirm that the results of the implementation used here match those of the method presented in the literature. This code is freely available at \url{https://github.com/GabrielF98/fouriersmooth}. This is the first publicly available Python version of the original IDL code \citep{Liu.2016} that implements this smoothing algorithm for supernova spectra.}. The parameter $k$ can be varied to control the smoothing, with larger $k$ values implying weaker smoothing. This smoothing method was applied to GRB980425-SN1998bw and GRB130702-SN2013dx for this study; with three spectra for each GRB-SN: one pre-peak, one at peak and one post-peak. Their rest-frame phases were computed based on maximum light times given in \cite{Galama.1998} and \cite{DElia.2015} respectively. The redshifts and spectra are sourced from the GRBSN webtool\footnote{\url{grbsn.watchertelescope.ie}} \citep{Finneran.2024A} and WISeREP\footnote{\url{https://www.wiserep.org/}} \citep{WISEREP}. 

Each spectrum was smoothed using a range of $k$ parameters, before their iron velocities were estimated using the template-fitting method. \cite{Modjaz.2016} used a value of $k$=100 for their smoothing; this corresponds to a velocity of 3000 km/s, which serves as the lower limit on supernova feature velocity in the smoothing (any features above this velocity are classed as noise and are removed during smoothing). \cite{Liu.2016} use a more conservative $k$=300 (1000 km/s) value for their smoothing, since they were studying slower Ib and Ic supernovae. In this analysis the following $k$ values were used: 25, 50, 100, 150, 200 and 300. These were chosen to include the values used in the literature, and to sample across a wider range. 

Figures \ref{fig:keffects1} and \ref{fig:keffects2} show the impact of changing the smoothing parameter ($k$) on the Fe II velocity measured by this method. As can be seen, velocities measured by the template-fitting method generally do not change significantly when the value of $k$ is changed. Perhaps one exception to this is $k$=25 which seems to have a large impact for both SNe at late times. This could be because $k$=25 represents a velocity of 12000 km/s, which is close to the iron velocity at late times in these supernovae \citep[e.g.][]{Finneran.2024B}. This could result in real features being removed by smoothing at this level. In other cases, as shown in Fig. \ref{fig:keffects2}, the scatter introduced by changing $k$ contributes $\sim$500-1000 km/s to the uncertainty in velocity. The uncertainties produced by the template fitting method will dominate over this effect. This appears to be the case both before, at and after the SN peak time. In conclusion, the choice of $k$ is not overly relevant in determining the velocity of a supernova using this method, although its contribution to the final errors cannot be fully neglected.

\subsection{Smoothing in the spline fitting method}

\begin{table*}[h!]
 \caption{Listing of events and their spectra used in the analysis of spline density and filter width effects.}
 \label{tab:specrefs}
 \centering
 \begin{tabular}{llll}
	\hline\hline
	Event & Date & Instrument & Source\\
	\hline
	GRB980425-SN1998bw &  1998-05-16 & DFOSC & \cite{Patat.2001q8t}\\ 
	SN2003jd & 2003-10-28 & FLWO & \cite{Modjaz.2014}\\
	SN2016P & 2016-01-21 & SPRAT & \cite{Prentice.2019}\\
	SN2020bvc & 2020-02-21 & SEDM & \cite{Ho.2020}\\
    \hline
 \end{tabular}
\end{table*}
There are two tuneable smoothing parameters in the spline fitting method: the number of knots used for the spline, known as ``spline density''; and the width of the SG filter used to smooth the spectrum, called ``filter width''. We aim to find heuristics for choosing the optimum filter width and spline density. In real applications these parameters should be determined on a case-by-case basis, using the original spectrum as a reference point. This section also looks at the typical errors introduced by the smoothing of the input spectrum. 

\begin{figure}[h!]
	\centering
	\includegraphics[width=\linewidth]{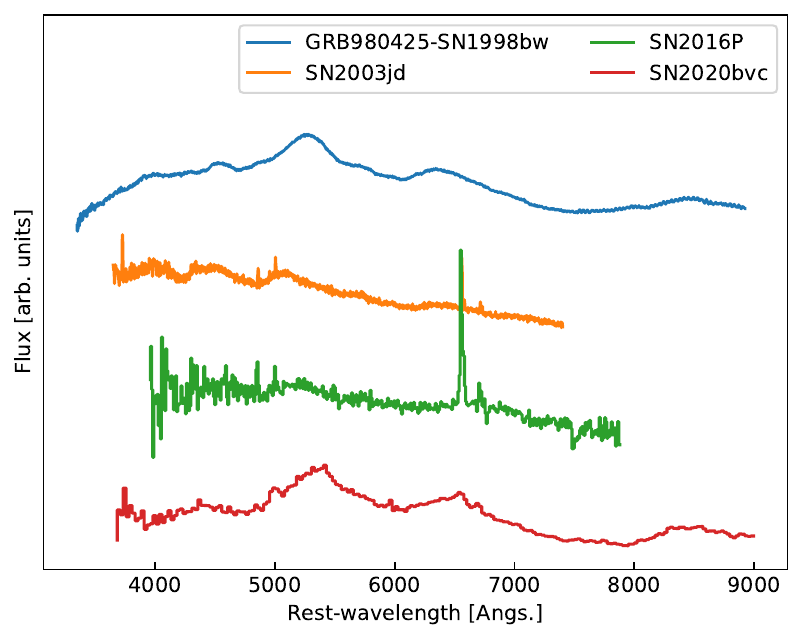}
	\caption{Sample spectra used to determine the optimum parameters for the spline fitting method. Data credits: GRB980425-SN1998bw \citep{Patat.2001q8t}; SN2003jd \citep{Modjaz.2014}; SN2016P \citep{Prentice.2016}; SN2020bvc \citep{Ho.2020}.}
	\label{fig:samplespectra}
\end{figure}

\begin{figure*}[h!]\includegraphics[width=\linewidth]{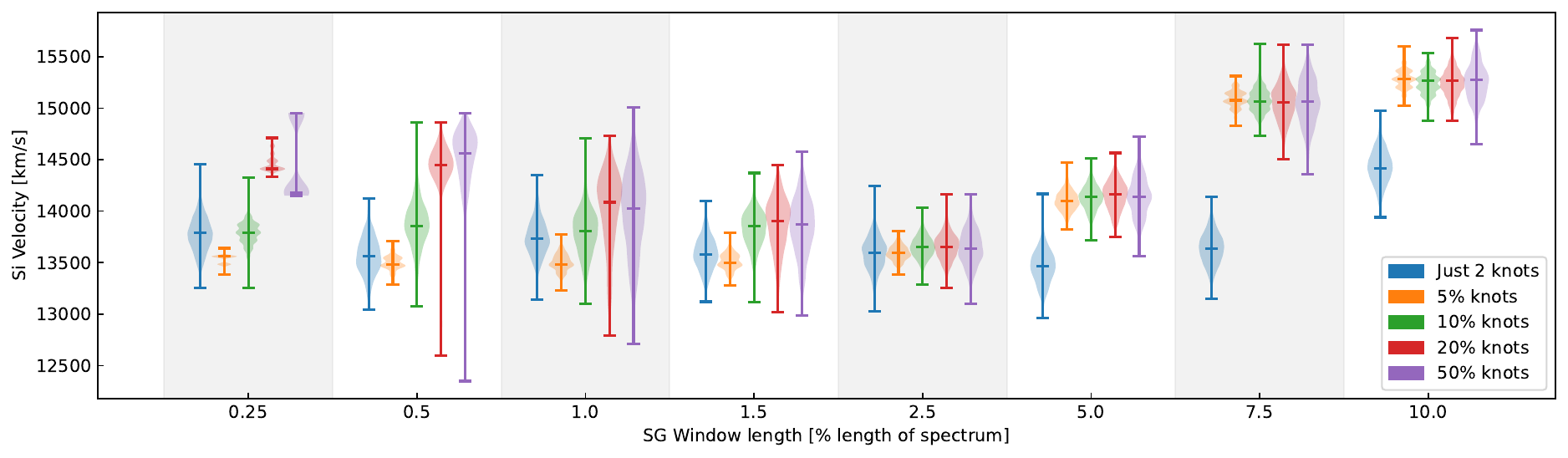}
	\caption[Variation of Si II velocity with filter width and spline density for a sample spectrum of GRB980425-SN1998bw]{Variation of Si II velocity with filter width and spline density for a sample spectrum of GRB980425-SN1998bw; this spectrum has high resolution and high S/N. Fits with a small filter width occasionally produce multi-modal velocity distributions, but a large filter width eventually changes the velocity; optimum filter width is around 1.5-2.5\%. Increasing the number of knots has little impact on velocity, except in the base case of 2 knots which should not be used. Within the space of optimal parameters, the variation of mean velocity is $<$1000 km/s.}
	\label{fig:knotsmoothSi1998bw}
\end{figure*}

The violin plots presented in Fig. \ref{fig:knotsmoothSi1998bw} show the effect of changing the smoothing filter width or the spline density on the distribution of Si II velocities for a high-resolution, high S/N spectrum of GRB980425-SN1998bw. Figure \ref{fig:knotsmoothSi1998bw} shows that non-normal distributions are common at filter widths $<$1\% of the spectrum length, indicating that this source of error is a legitimate concern when determining optimum parameters. 

Figure \ref{fig:knotsmoothSi1998bw} also shows that uncertainties are largest when filter widths are $<$2.5\%. At higher filter widths there is nothing gained in terms of precision, so the optimum filter width is around 2.5-5\%. Within this range the variation in the mean velocity is around 500 km/s. For a given filter width, there is little difference between 5-50\% spline density. In this range there are less than 500 km/s shifts in the mean velocity. In the case of using two knots, as expected, there is not enough flexibility in the fit to determine the minimum value accurately, so the velocity produced in the 2 knot case tends to differ more from other spline densities.

	\begin{figure*}[h!]

	\centering
	\begin{subfigure}{0.33\textwidth}
	\includegraphics[width=\linewidth, trim=1.2cm 0cm 1.2cm 0.4cm, clip]{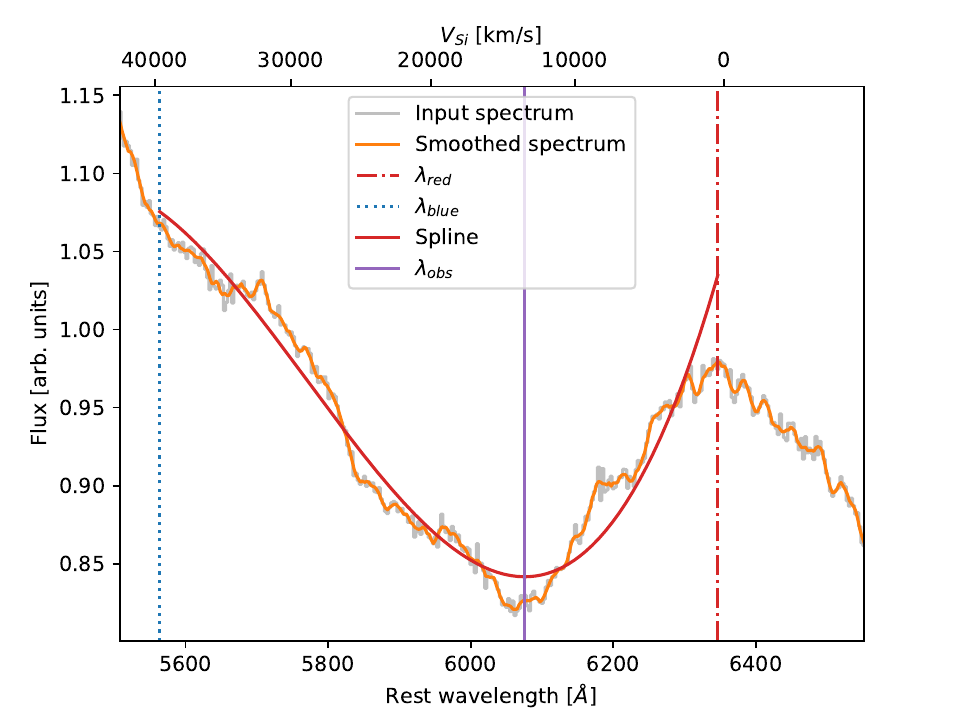}
	\caption{Knots: 2; SG: 0.5\%}
	\end{subfigure}
	\begin{subfigure}{0.33\textwidth}
	\includegraphics[width=\linewidth, trim=1.2cm 0cm 1.2cm 0.4cm, clip]{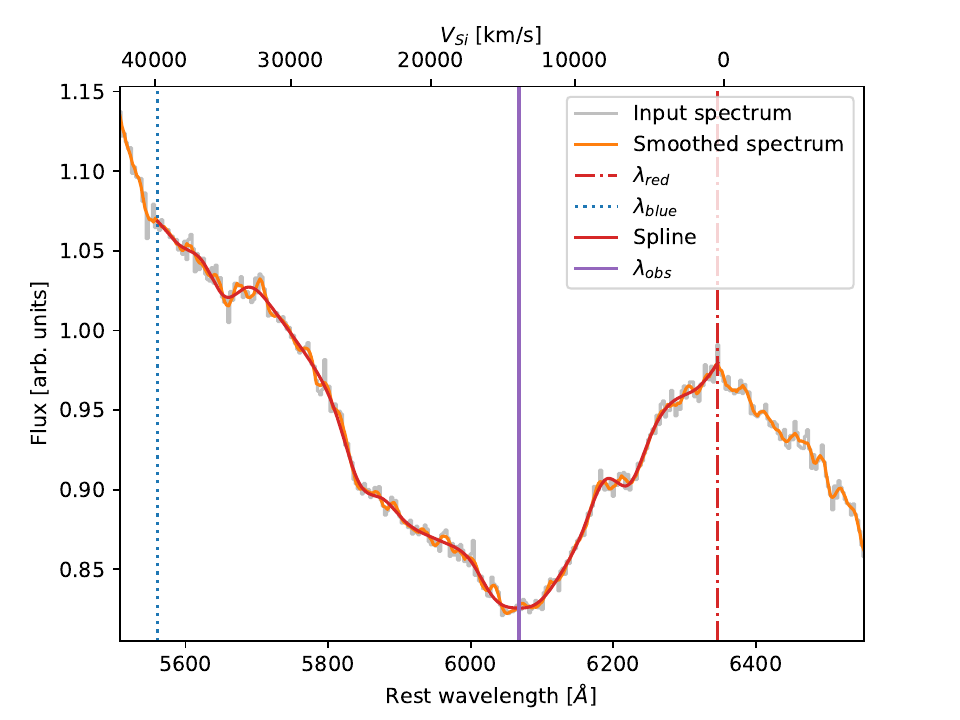}
	\caption{Knots: 10\%; SG: 0.5\%}
	\end{subfigure}
	\begin{subfigure}{0.33\textwidth}
	\includegraphics[width=\linewidth, trim=1.2cm 0cm 1.2cm 0.4cm, clip]{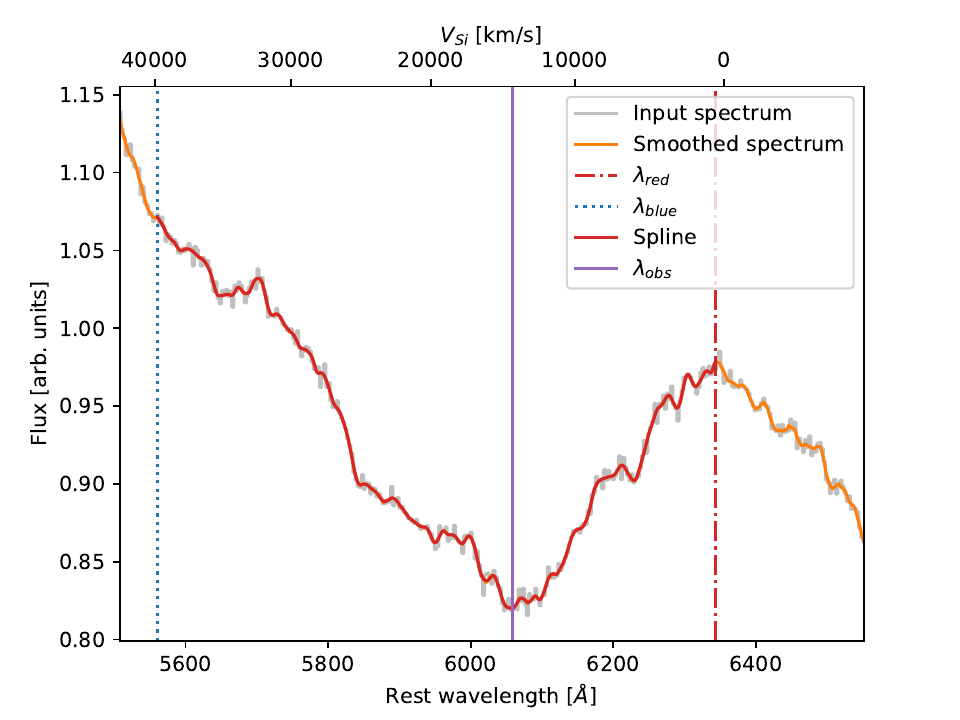}
	\caption{Knots: 50\%; SG: 0.5\%}
	\end{subfigure}
	
	\vspace{0.5cm}
	\begin{subfigure}{0.33\textwidth}
	\includegraphics[width=\linewidth, trim=1.2cm 0cm 1.2cm 0.4cm, clip]{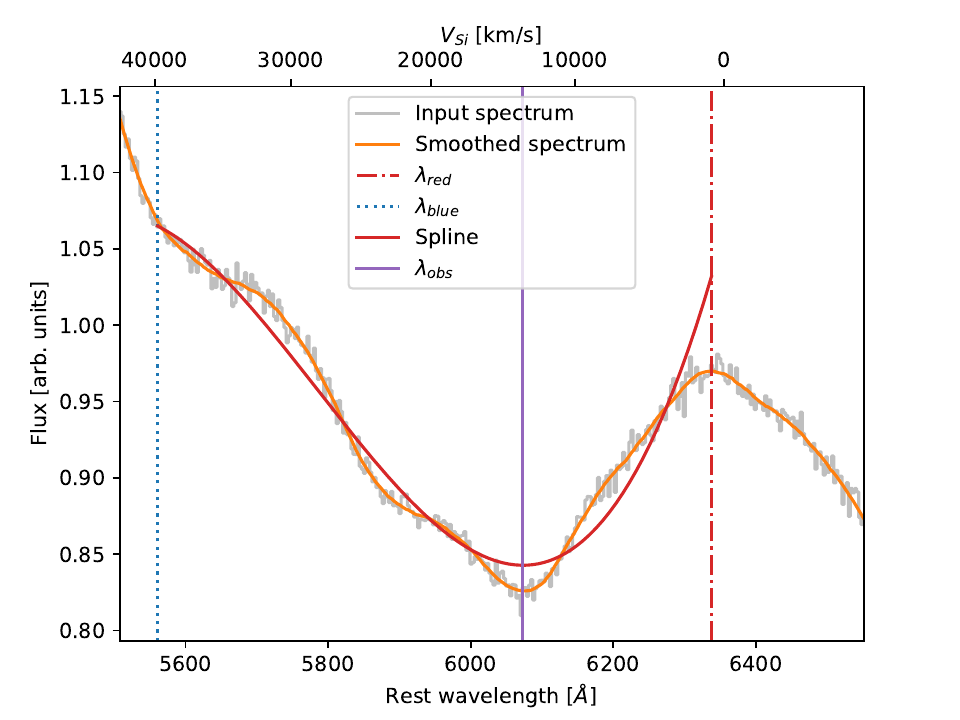}
	\caption{Knots: 2; SG: 2.5\%}
	\end{subfigure}
	\begin{subfigure}{0.33\textwidth}
	\includegraphics[width=\linewidth, trim=1.2cm 0cm 1.2cm 0.4cm, clip]{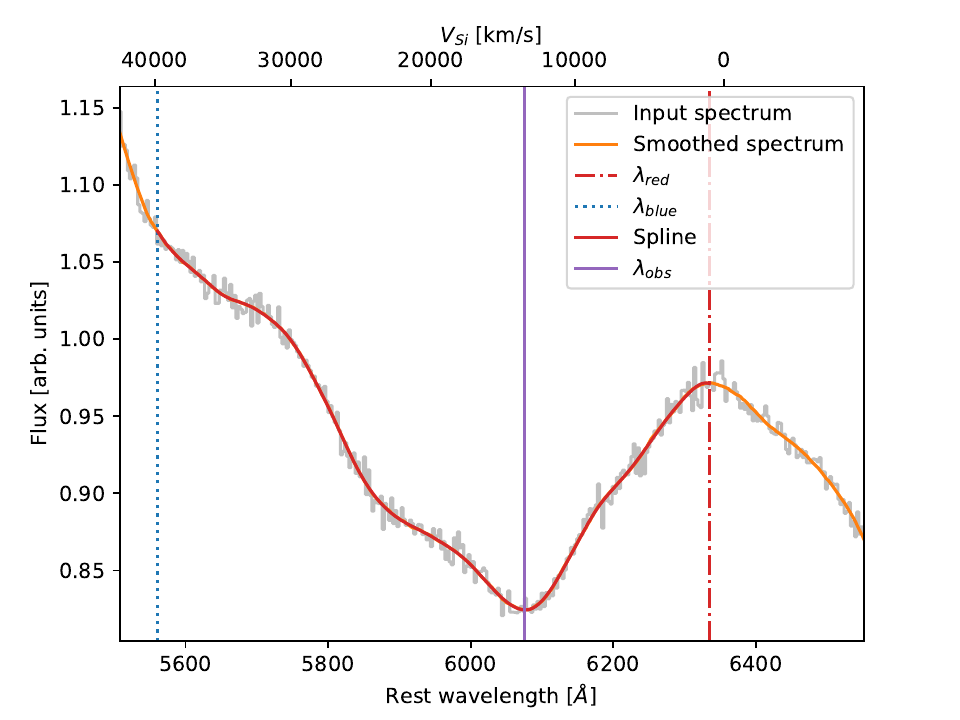}
	\caption{Knots: 10\%; SG: 2.5\%}
	\end{subfigure}
	\begin{subfigure}{0.33\textwidth}
	\includegraphics[width=\linewidth, trim=1.2cm 0cm 1.2cm 0.4cm, clip]{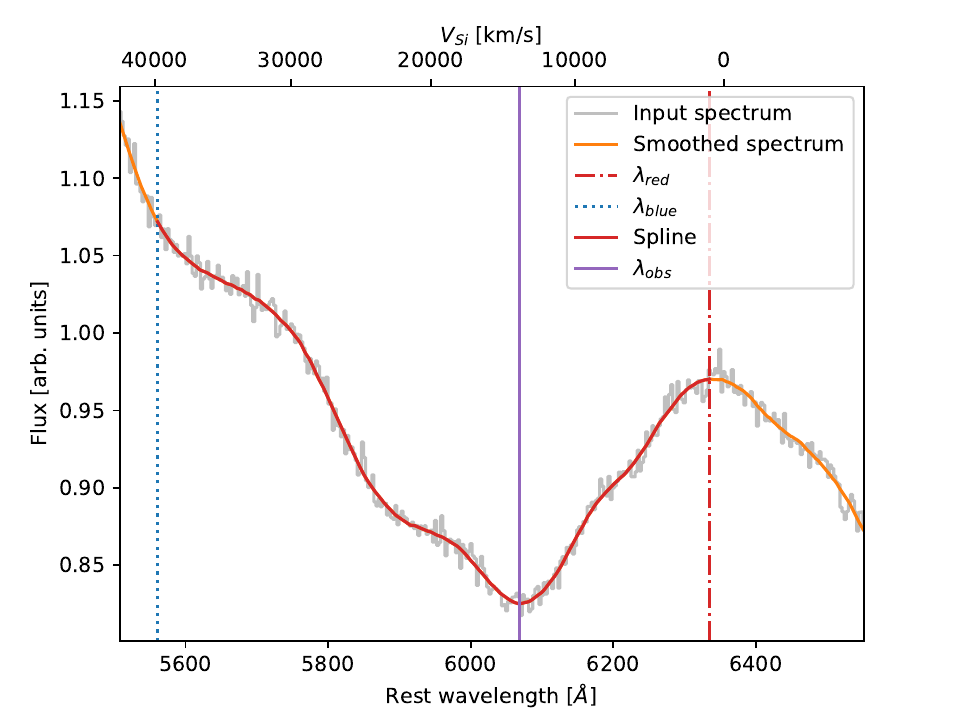}
	\caption{Knots: 50\%; SG: 2.5\%}
	\end{subfigure}
	
	\vspace{0.5cm}
	\begin{subfigure}{0.33\textwidth}
	\includegraphics[width=\linewidth, trim=1.2cm 0cm 1.2cm 0.4cm, clip]{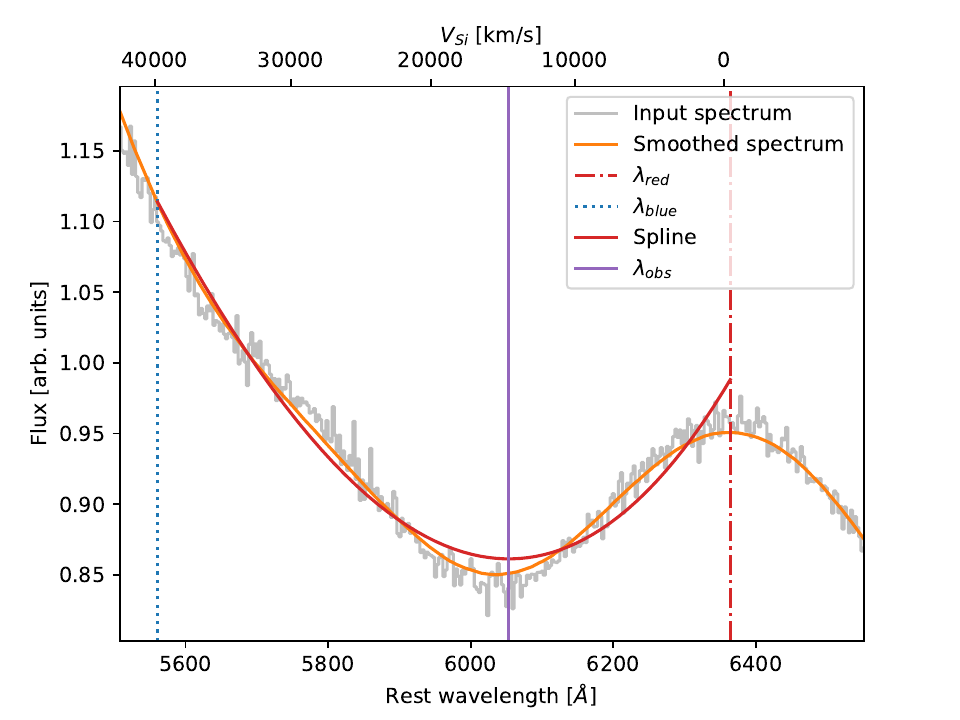}
	\caption{Knots: 2; SG: 10\%}
	\end{subfigure}
	\begin{subfigure}{0.33\textwidth}
	\includegraphics[width=\linewidth, trim=1.2cm 0cm 1.2cm 0.4cm, clip]{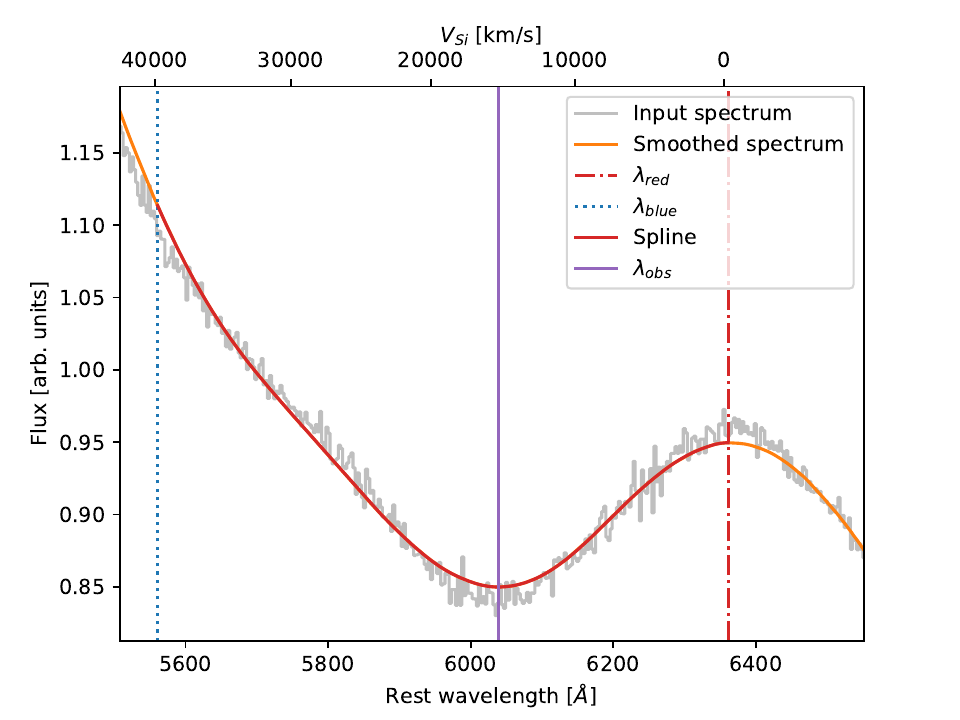}
	\caption{Knots: 10\%; SG: 10\%}
	\end{subfigure}
	\begin{subfigure}{0.33\textwidth}
	\includegraphics[width=\linewidth, trim=1.2cm 0cm 1.2cm 0.4cm, clip]{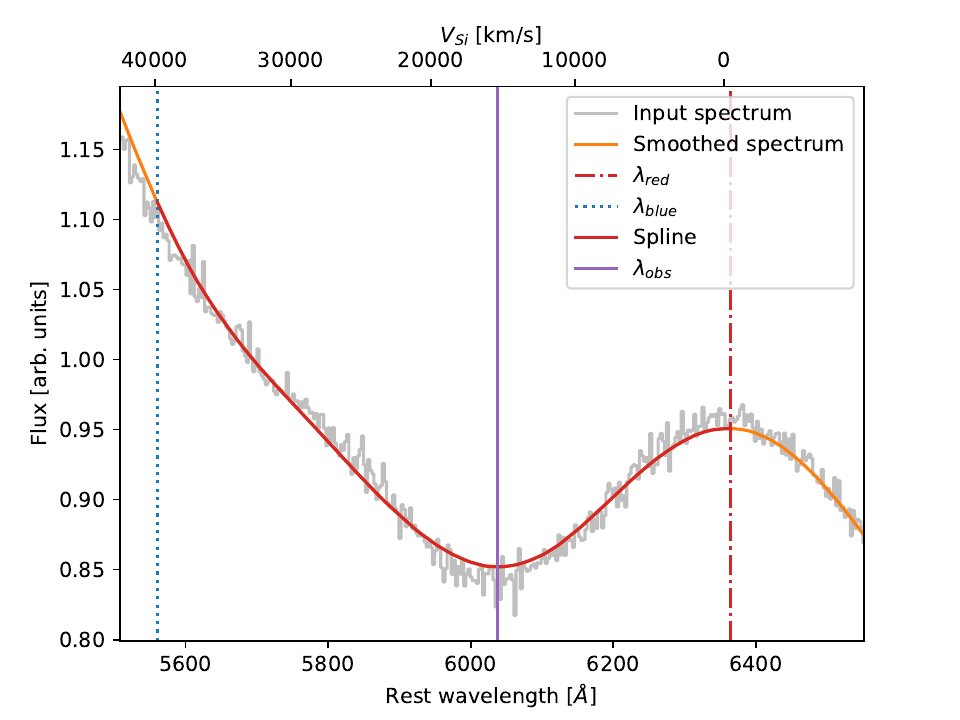}
	\caption{Knots: 50\%; SG: 10\%}
	\end{subfigure}

\caption[Impact of filter width and spline density on the shape and fidelity of the spline fitting method - GRB980425-SN1998bw]{Impact of changing the filter width and number of spline knots on the shape and fidelity of the spline fitting method. These plots show the Si II region for GRB980425-SN1998bw; this is a typical example of a high resolution spectrum with excellent S/N. The number of knots used in the spline increases from the left to the right of the grid; likewise the filter width increases from top to bottom. Increases to filter width create a smoother spectrum; eventually this leads to the smearing out of real features. The fidelity of the spline to the smoothed spectrum is improved by increasing the number of knots used for the spline, but eventually meets diminishing returns. This analysis suggests that an optimum spline density of 10\% should be paired with a filter width of $\sim$2.5\%. For a full description of the method and figures, refer to \cite{Finneran.2024B}.}
\label{fig:1998bwknotswidth}
\end{figure*}

Figure \ref{fig:1998bwknotswidth} offers an insight into the fits that underlie the results for the violin plots of GRB980425-SN1998bw, confirming the optimum parameters. In these plots the filter width increases from top to bottom; the spline density increases going from left to right. In cases with low spline density, the fidelity of the spline is poor; which explains the difference in velocity between the case of two knots and many knots. In the case of 10\% spline density, there are enough knots in the spline to track the spectral shape very well. The case of 50\% spline density also produces good fits, supporting the observation from the violin plot showing similar velocity distributions from 5-50\% spline density. However, there is no justification for choosing a high spline density, as it will not improve the fit, and may in fact over-fit the spectrum in some cases. 

Fig. \ref{fig:1998bwknotswidth} also shows the impact of too little or too much smoothing. When smoothing is too weak, there are multiple local minima at which the fit may become trapped, possibly contributing to the multi-modal distributions. When too much smoothing is applied, as in the 10\% filter width case, all of the features in the spectrum have been removed, and in this instance this has resulted in a shift of the minimum velocity to the blue end of the spectrum. This explains the observation from the violin plot shown in Figure \ref{fig:knotsmoothSi1998bw}, which shows a difference in velocity at high filter widths compared to low and medium filter widths. This  supports a choice of 10\% spline density paired with 2.5\% filter width, as it is reasonably similar to the input spectrum and the fit follows this profile very accurately. Additionally, any change to the minimum wavelength compared with the original spectrum is minimal. 

A similar violin plot for the Fe II feature of SN2016P is shown in Fig. \ref{fig:knotsmoothFe2016P}; which is a low-resolution, low S/N spectrum. It is clear that the spline fitting method struggles in this case, probably because the low S/N and low-resolution may lead to the removal of real signal when smoothing (smoothing algorithms tend to preserve area, rather than preserving peaks, which can smear out a weak signal). In this case there appears to be no optimal smoothing level. Once again there is no preference between 5-50\% spline density; however, there is also no bias against using the case of 2 knots. This is probably because smoothing is producing a very simple spectral shape in this region, which a single spline segment can approximate very well. It may also be because there is so much noise that most fits will be reasonable fits. The typical shift in mean velocity introduced when changing the smoothing parameters is around 2000 km/s, which is comparable to the one-sigma errors of this method. 

Support for these results can be found in Fig. \ref{fig:2016Pknotswidth}, which shows the grid of fits for a range of fitting parameters. There are significant changes to the minimum wavelength depending on the parameters. In this instance, it is probably reasonable to use a larger smoothing filter width in an attempt to recover the trend. In cases such as this, concerns around smoothing are probably relevant, since distinguishing between two features which are adjacent is nearly impossible. Although it is still possible to measure a velocity in such cases, choosing optimum smoothing parameters will be difficult and the uncertainty will be larger. 

This analysis was also performed for the Fe II features of SN2020bvc, GRB030329-SN2003dh and GRB980425-SN1998bw, and for Si features of GRB030329-SN2003dh and GRB980425-SN1998bw. Typically, variation of the smoothing parameters contributes an additional error of $\sim$1000 km/s in high resolution, high S/N spectra. However, this may rise to 3000-4000 km/s in low resolution or low S/N spectra. These errors must be added in quadrature with the one sigma errors produced by this method. The dependence on the spectral quality is of particular concern, because many spectra will be of medium S/N and medium resolution. A typical additional source of error may therefore be in the 2000 km/s range. 

Based on the four test spectra studied here, it is difficult to recommend general optimum parameters, and thus it is advisable to determine them on a case by case basis. For all but the lowest S/N spectra, the optimum parameters are likely in the range of 0.5-5\% for smoothing, and around 5-20\% for spline density. These appear to be reasonable ranges for both the Fe II and Si II features. Given that the Ca II feature is much broader and often more heavily blended than either of these features, the optimum parameters may be slightly different. Additional violin plots and smoothing grid plots for other SNe considered in this analysis can be found in Appendix \ref{sec:appendix}. 

\section{Additional sources of error in template fitting}\label{sec:templatefitting}
\subsection{Choosing a reference epoch}
Accurate knowledge of the age of a supernova spectrum is required for comparisons between supernovae, classification, and studying the evolution of the spectral properties over time. The age of a spectrum or photometric datapoint is defined relative to a reference epoch. Supernova reference epochs are typically determined from the lightcurve\footnote{Though it is possible to estimate the age from the spectrum, for example \cite{Mazzali.2000}.}. As there are several epochs to choose from, astronomers must balance the ease of determining the date of a particular epoch against its accuracy.

\begin{figure}[h!]
	\centering
	\includegraphics[width=\linewidth]{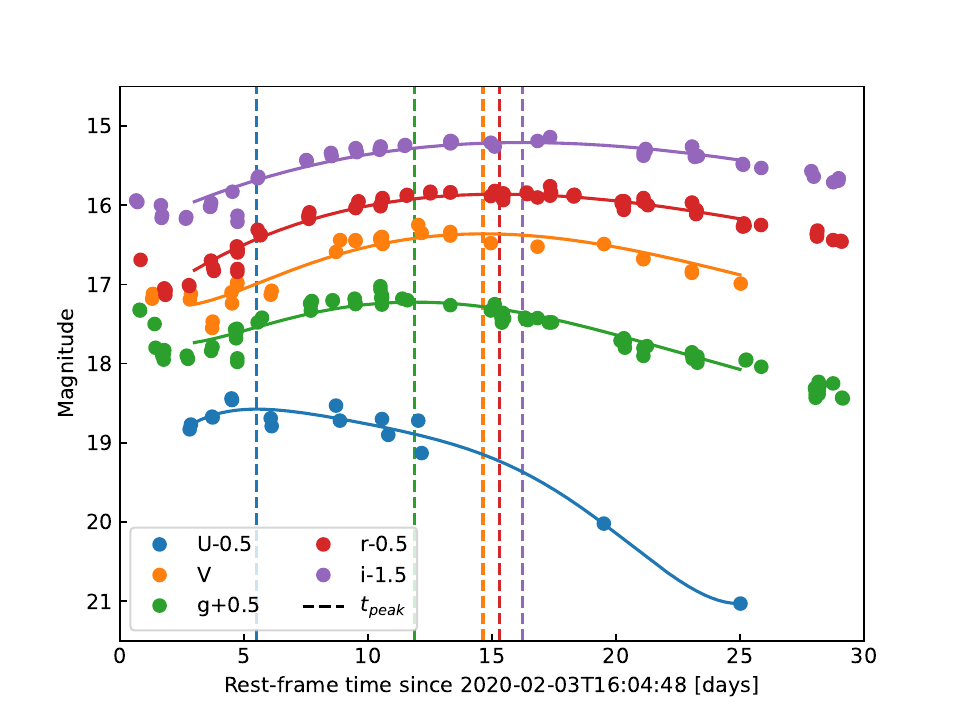}	
	\caption[SN2020bvc photometry]{U, V, g, r and i band photometry for the Ic-BL SN2020bvc, showing the energy-dependence of the peak time. The peak time in the i-band is almost 11 days later than the peak in U. Data credit: \citet{Rho.2021, Ho.2020, Izzo.2020}.}
	\label{fig:maxepoch}
\end{figure}

One of the most common supernova epochs is the date of maximum light, also called the peak time. Peak times can be determined using an analytical formula such as the Bazin model \citep[e.g.][]{Taddia.2015, Bazin.2011}, by quadratic fitting \citep[e.g.][]{Bianco.2014}, or by interpolation. The benefit of using peak time is that it is relatively straightforward to measure, since photometry near the peak is normally well-sampled with low scatter. However, the epoch of maximum light is energy dependent; with the date differing between observing filters. This is shown in Table \ref{tab:peaktimes} and is illustrated in Fig. \ref{fig:maxepoch}, where the peak time for several filters has been determined using spline interpolation for SN2020bvc. 

\begin{table}[h!]
\centering
\caption{Peak times for SN2020bvc in several filters.}\label{tab:peaktimes}
\begin{tabular}{p{2cm}ccccc}
\hline\hline
 Band & U & g & V & r & i \\
\hline
 $\mathrm{t_{peak}}$ [days] & 5.5 & 11.9 & 14.6 & 15.3 & 16.2\\
\hline
 \end{tabular}
\tablefoot{Times are in the rest-frame of SN2020bvc.}
\end{table}

The energy dependence of the peak time is a result of the integration of the flux from the underlying spectral energy distribution over the wavelength range of each filter. The spectral energy distribution can be approximated by a blackbody distribution, with a modification at UV wavelengths, where line blanketing by iron-group elements is significant. This explains why the U-band peak time differs significantly from the other filters in Table \ref{tab:peaktimes}. This pseudo-blackbody fit is defined by the `effective' temperature. As the effective temperature decreases, the peak of the emission shifts to redder wavelengths, and so in general redder bands reach their peak emission at later times than bluer bands. 

Peak time can typically be determined to within two days, though the degree of Nickel mixing can broaden the peak of the lightcurve \citep[e.g.][]{Piro.2013} and make it more difficult to precisely determine the peak time. \cite{Modjaz.2016} use the peak time in the V-band as the reference epoch for their sample. This is motivated by \cite{Bianco.2014}, who suggested that V is the best-sampled lightcurve for the majority of supernovae. Where V is not available, the conversions in \cite{Bianco.2014} can be used to determine the peak time in V using other bands. However, these conversions were computed from a sample containing just 3 Ic-BL supernovae, and have typical standard deviations of one to two days.

An alternative to maximum light is the explosion date, the beginning of the supernova. In the case of GRB-SNe, gamma-ray satellites provide a strong constraint on this time\footnote{Although recently SN2020bvc \citep{Izzo.2020, Ho.2020} and SN2023lcr \citep{2023GCN.34370....1M} lacked associated gamma-ray emission, and so their epochs are not known so precisely.}. Common methods used to determine the explosion time include: the mid-time between the first detection of the source and the closest prior constraining non-detection observation; fitting the early lightcurve using a simple function \citep[e.g.]{Taddia.2015}; fitting the SN lightcurve using a template SN lightcurve \citep[e.g.][]{Taddia.2019}; or modelling of the nuclear processes responsible for the lightcurve (for example with the model proposed by \citet{Arnett.1982}). When using the last non-detection method, the uncertainty on the explosion time can be large, unless the field has been visited recently. When fitting of the early lightcurve, the result of the fit is dependent on how well the lightcurve is sampled, the better the sampling the better the fit. Unfortunately, the early part of the lightcurve is often the most poorly sampled, since the supernovae tend to be quite dim at this stage, and many telescopes will not have followed them up yet. Non detections are also common in the early phase of the lightcurve and may not provide a good constraint for the polynomial fit. Estimates obtained from modelling tend to be the most precise, in the absence of an accurate GRB time, however these can also be impacted by assumptions about nickel mixing \citep{Piro.2013}. The major benefit of explosion times is that they do not depend on the observing band. In this respect, direct comparisons are easier to make between supernovae, as the explosion time represents a physical change in the source. Although the peak time may be more accurately determined than the explosion epoch, it may be less useful for comparisons of large datasets, since it depends on intrinsic properties of the source. 

\subsection{Impact of epoch accuracy on velocity}\label{sec:phase}

\begin{figure*}[h!]
    \centering
    \includegraphics[width=0.49\textwidth]{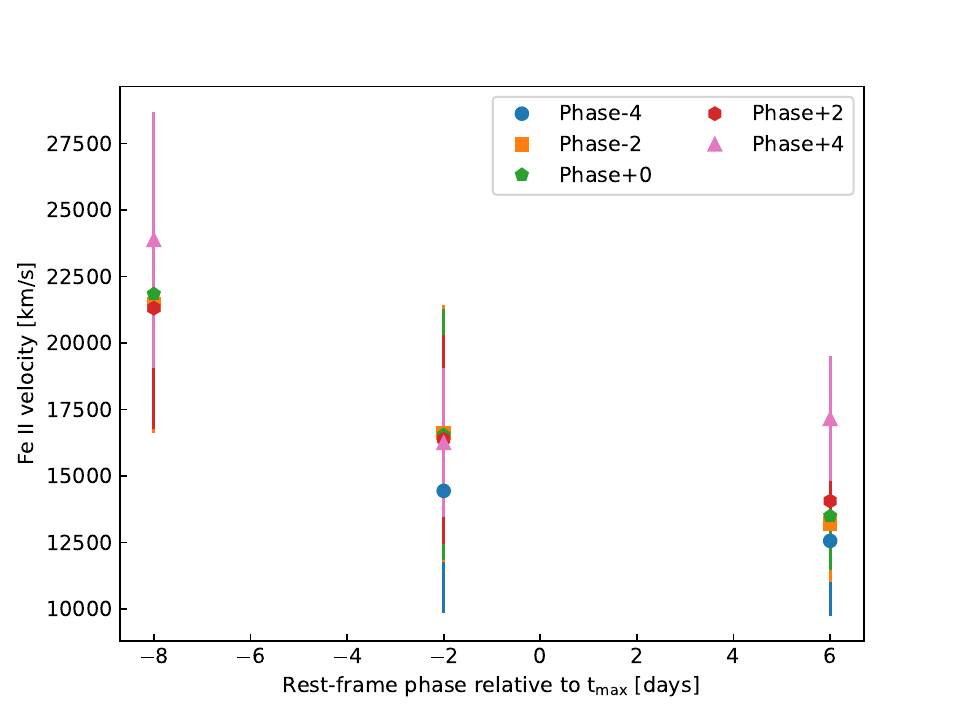}
    \includegraphics[width=0.49\textwidth]{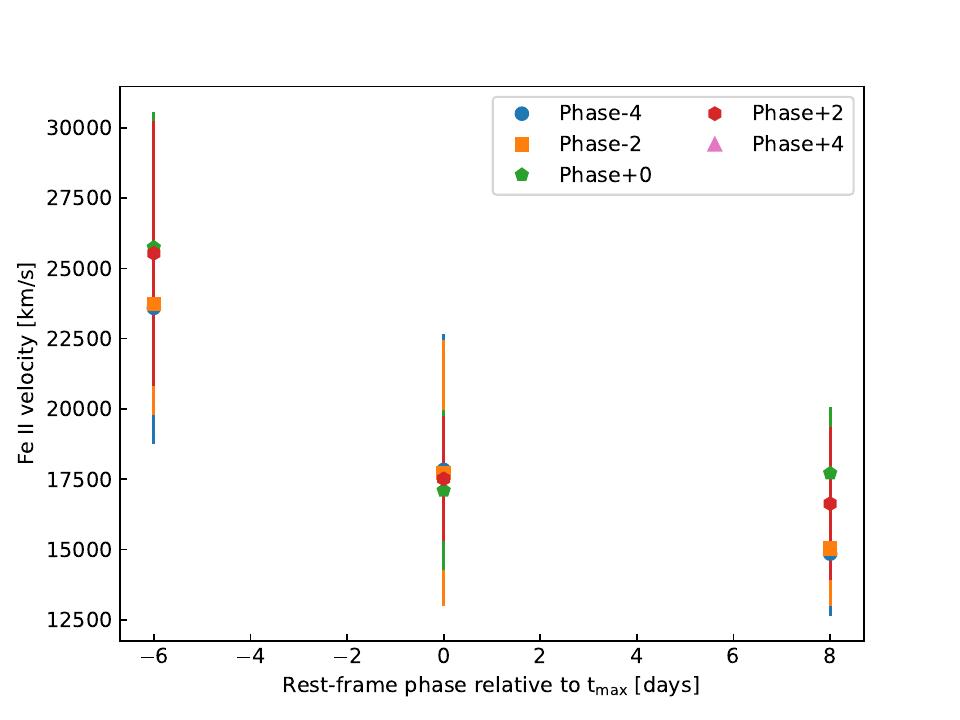}
    
    \caption[Influence of phase shifts on velocities measured by the template fitting method 1]{Influence of small shifts in phase on the Fe II velocity for GRB980424-SN1998bw (left panel) and GRB130702A-SN2013dx (right panel).}
    \label{fig:phaseeffects1}
\end{figure*}

\begin{figure*}[h!]
    \centering
    \includegraphics[width=0.49\textwidth]{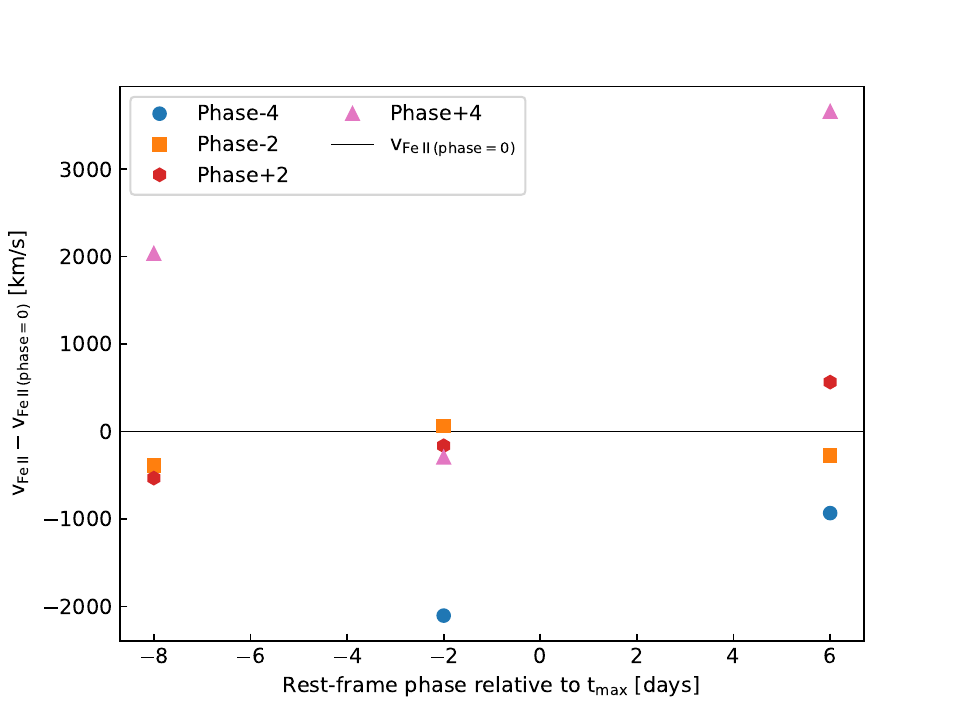}
    \includegraphics[width=0.49\textwidth]{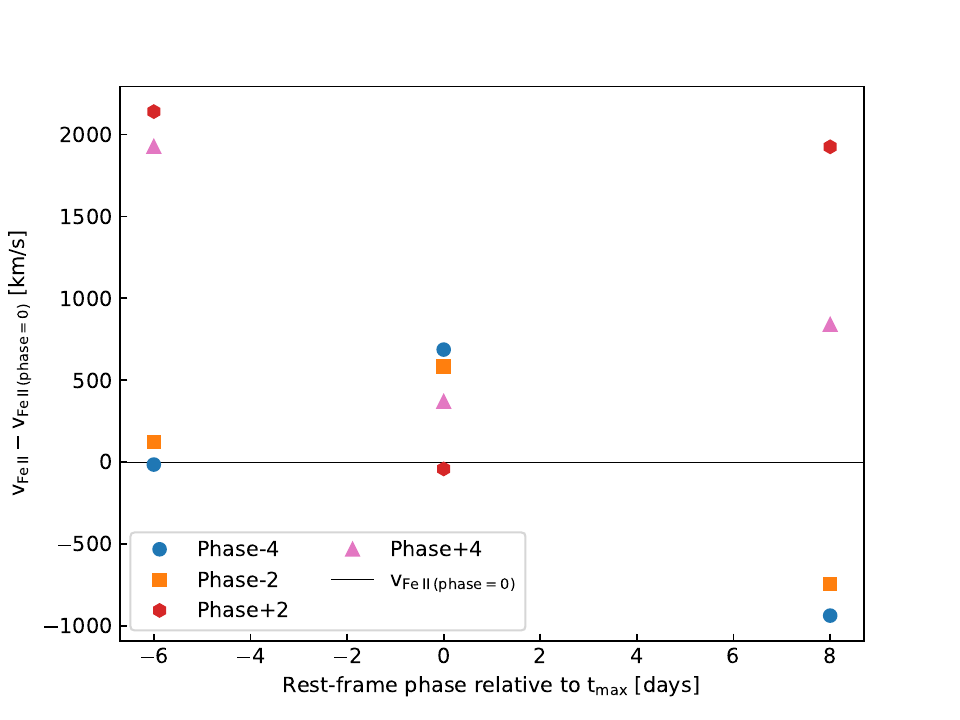}
    
    \caption[Influence of phase shifts on velocities measured by the template fitting method 2]{Similar to Fig. \ref{fig:phaseeffects1} but with the velocity shift shown relative to the phase=0 velocity. GRB980424-SN1998bw (left panel) and GRB130702A-SN2013dx (right panel).}
    \label{fig:phaseeffects2}
\end{figure*}

\cite{Modjaz.2016} created a set of template Ic spectra for use with their velocity measurement code. For every 2-day increment between -10 and 72 days of V-band maximum light, an average spectrum is created using a Gaussian weighted average of spectra within a 5-day window. The template fitting method measures the blue-shift and broadens the features of these templates to match the Ic-BL spectrum being fit. As a result, when using this method the user needs to specify the age of the input spectrum with respect to maximum light in the V-filter. This is called the spectral phase and is used to determine which template is fit to the input spectrum. The velocity of the iron lines in each template spectrum is added to the blue-shift determined by the method to compute the final Ic-BL iron velocity. Because of this, choosing the correct phase is crucial to determining the correct velocity with this method. 

Using the peak time as a reference epoch can lead to situations where the phase of the input spectrum is shifted by a couple of days relative to the true phase. Compounding this is the rounding of the phase, which is introduced because the template spectra are only available in 2 day increments. Figures \ref{fig:phaseeffects1} and \ref{fig:phaseeffects2} illustrate the impact of a shifted phase on the measured velocity. The velocity of each spectrum was measured five times, applying an artificial phase-shift of -4, -2, 0, 2 or 4 days. Not only does this change the template velocity being added to the blue-shift velocity, but the template will be a different shape at different phases, which could further change the results. This was performed for spectra around one week prior to the SN peak time in the V filter, at the peak time and one week after the peak time. 

This analysis was performed for two GRB-SNe: GRB980425-SN1998bw and GRB130702A-SN2013dx. For both events, and at every epoch, phase shifts introduce a spread in the measured velocities. The magnitude of the change is small enough that it could conceivably be masked by the measurement uncertainties, as shown in Fig. \ref{fig:phaseeffects1}. To highlight the effect, Fig. \ref{fig:phaseeffects2} shows the scatter in velocities relative to the velocity measurement for a phase shift of 0 days. Phase shifts of around 2 days alter the velocity by 500-1000 km/s, while this increases to over 2000 km/s for a 4 day phase shift. The influence of phase seems to be weakest near the peak time, which could be because of the use of the peak time as the epoch. The impact of phase one week before and one week after the peak seems to be roughly the same. This is interesting because the velocity is changing more rapidly prior to peak than after the peak, and one would expect shifts in time here to contribute a greater offset in velocity space. This additional source of error is not accounted for by the template fitting method, and should be added in quadrature to the errors produced by the code.

\section{Additional sources of error in spline fitting}\label{sec:splinesystematics}
\begin{figure*}[h!]
	\centering
	\begin{subfigure}{0.49\linewidth}
		\includegraphics[width=\linewidth]{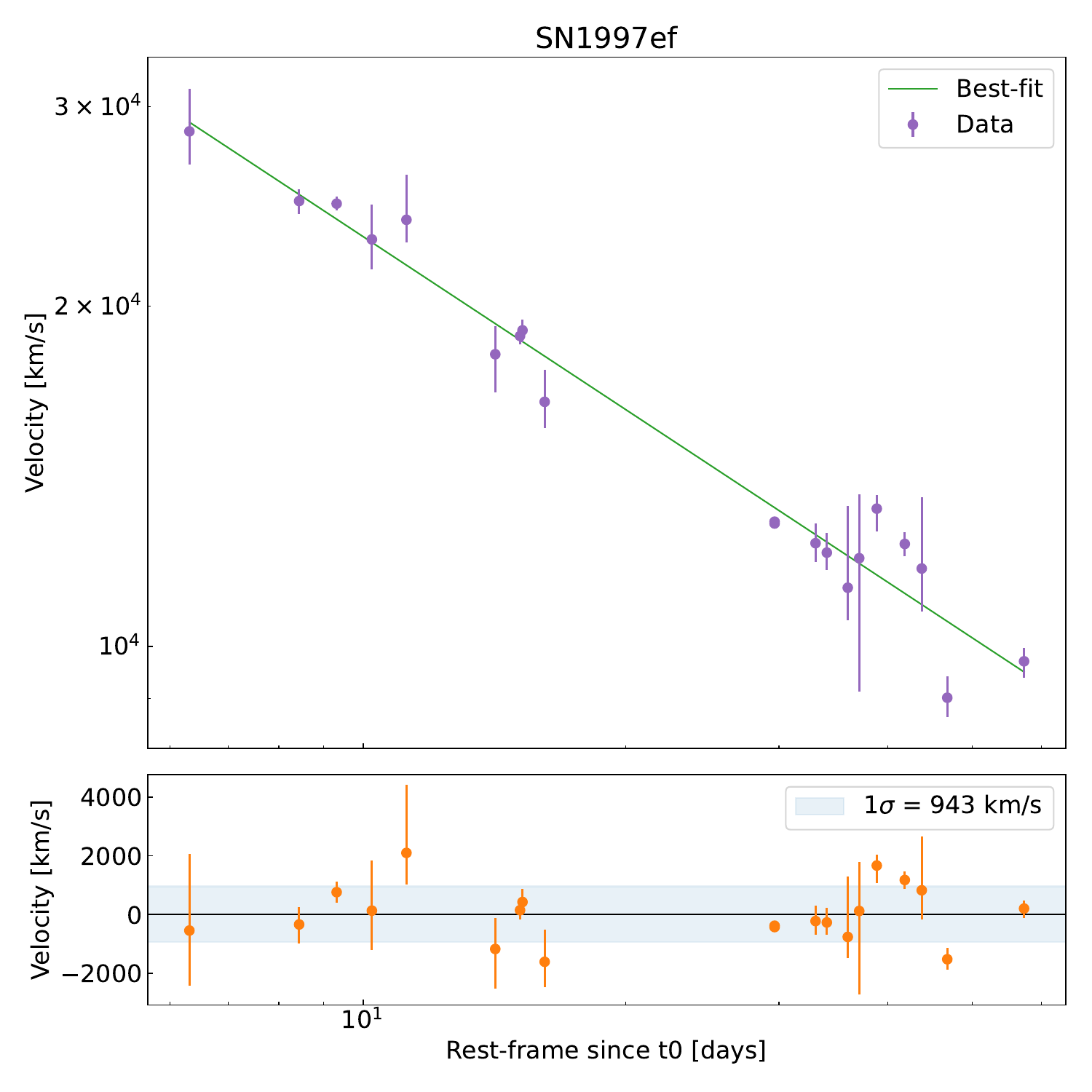}
	\end{subfigure}
	\begin{subfigure}{0.49\linewidth}
		\includegraphics[width=\linewidth]{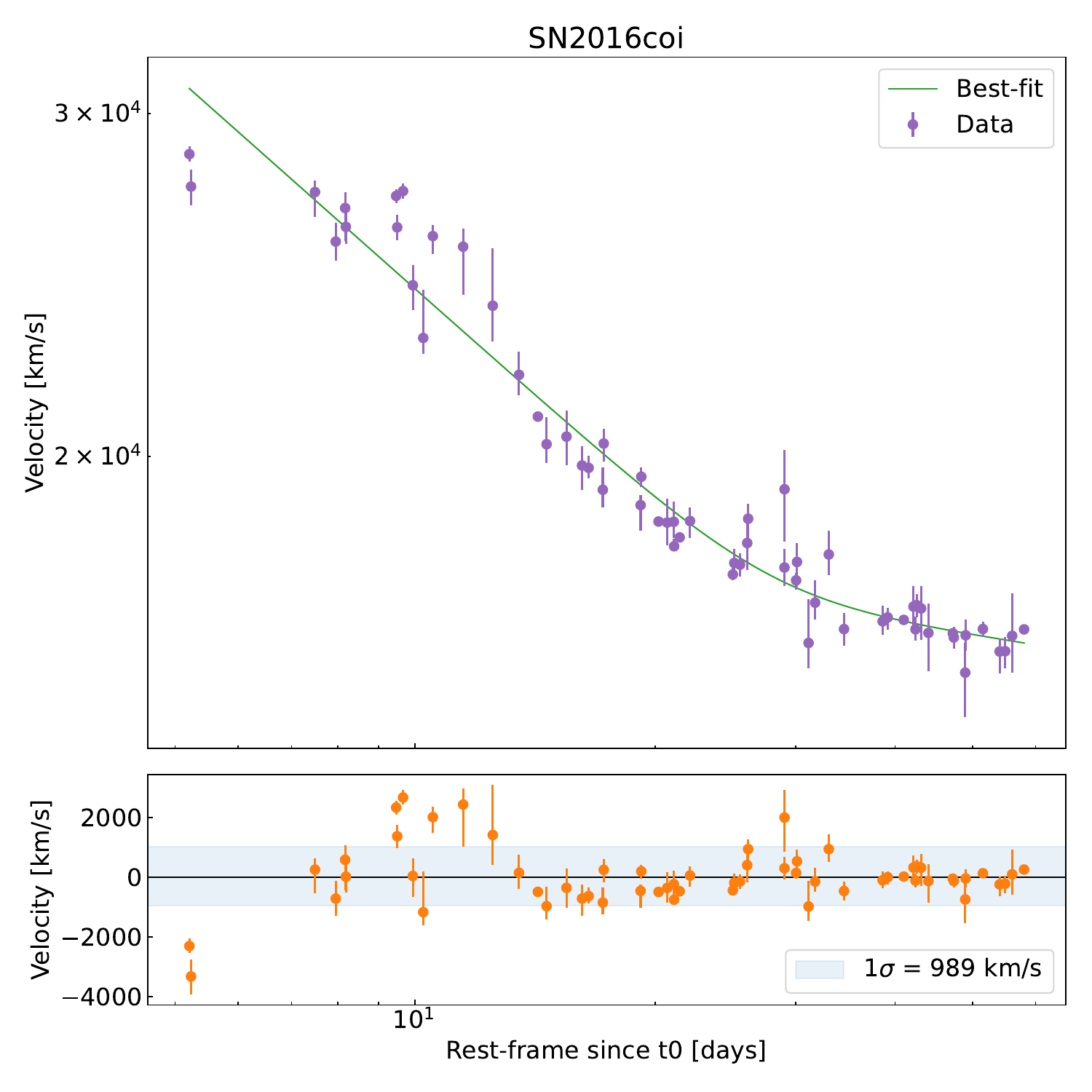}
	\end{subfigure}
	\begin{subfigure}{0.49\linewidth}
		\includegraphics[width=\linewidth]{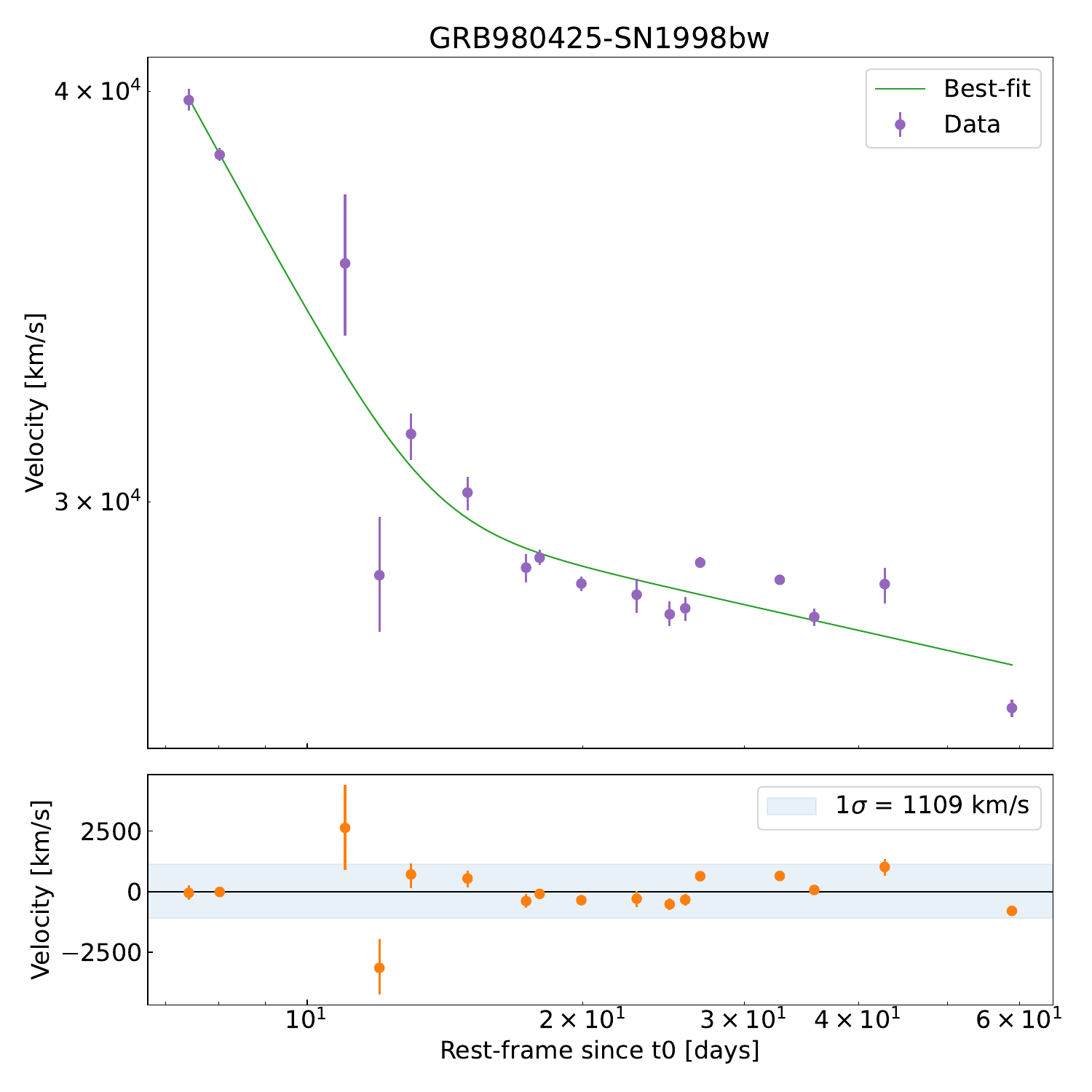}
	\end{subfigure}
	\begin{subfigure}{0.49\linewidth}
		\includegraphics[width=\linewidth]{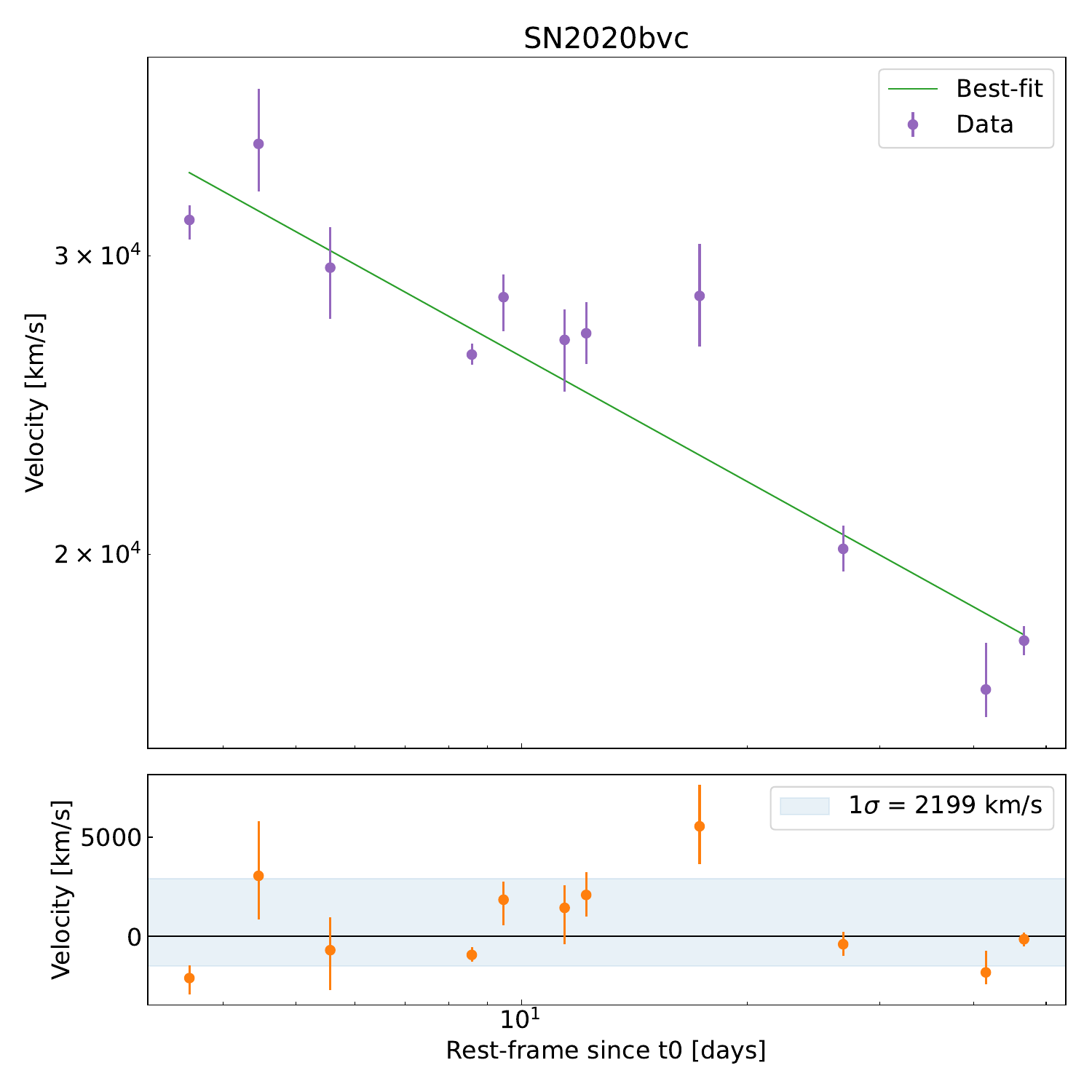}
	\end{subfigure}
	\caption[Best fits and residual plots for a selection of Ic-BL supernovae with and without GRBs]{Best fits and residual plots for a selection of Ic-BL supernovae with and without GRBs. The velocities are typically scattered around the best fit in a 1000 km/s range, though it can rise as high as 2000 km/s. Power-law and broken power-law fits display similar levels of scatter, as do Ic-BLs with and without GRBs, as indicated by the 1$\sigma$ error range.}
	\label{fig:scatteranalysis}
\end{figure*}

The spline fitting method tends to produce smaller errors than the template fitting method. However, it is possible that there are additional sources of error that the spline fitting method does not explicitly account for. Although it is not possible to account for all sources of error, it is possible to estimate their impact by looking at the residuals of (broken) power-law fits to the velocity evolution. Figure \ref{fig:scatteranalysis} demonstrates this for a sample of four Ic-BLs. The sample was selected to show examples of power-law and broken power-law fits for well-sampled Ic-BLs with and without GRBs. Figure \ref{fig:scatteranalysis} shows that the residuals for well sampled events typically exhibit a mean scatter of 1000 km/s, with SN2020bvc showing a higher scatter of $\sim$2000 km/s. This may have been because of the low resolution of many of the SN2020bvc spectra, which were taken with the Spectral Energy Distribution (SEDM) instrument \citep{SEDM}. There is no evidence for larger scatter at early vs. late times. Both power-law and broken power-law fits exhibit similar levels of scatter among their residuals. This additional source of error should be added in quadrature with the error produced by the spline fitting method. Comparing this error source with errors produced by the spline, they are equivalent in magnitude. This indicates that errors produced by pure spline fitting are significantly under-estimating the total error. This is surprising given the Monte-Carlo method which was utilised to perform resampling of the spectra. 

\section{Direct comparison}\label{sec:directcompare}
\begin{figure*}[h!]
	\centering
	\includegraphics[width=0.49\linewidth]{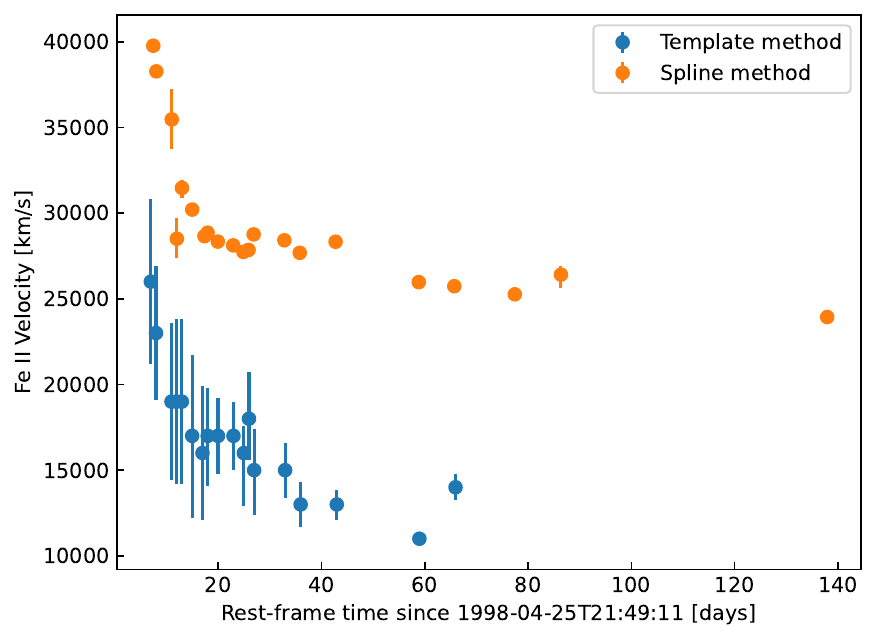}
	\includegraphics[width=0.49\linewidth]{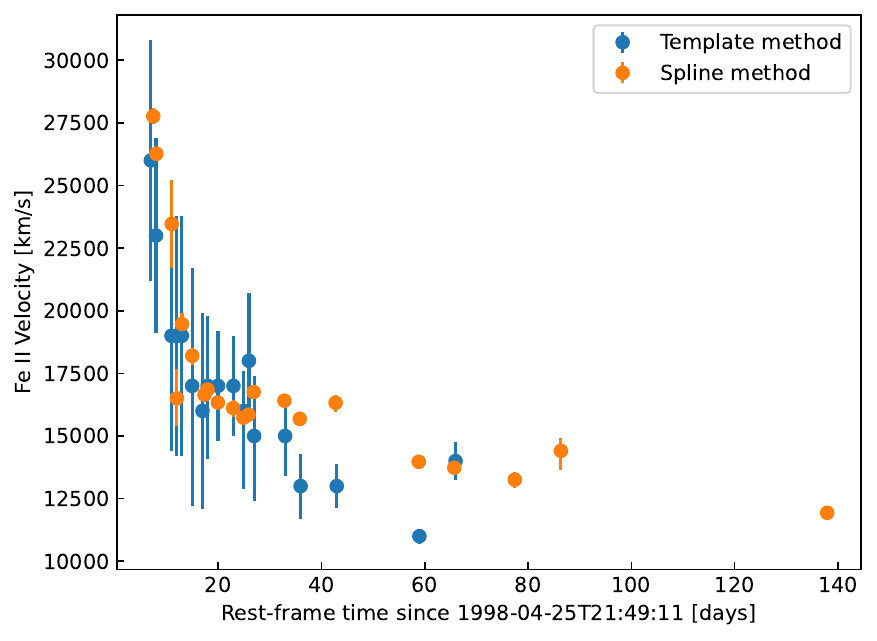}
	\caption[Template fitting vs. spline fitting velocities GRB980425-SN1998bw]{Comparison of results produced by the spline-fitting and template-fitting procedures; template fitting velocities from \cite{Modjaz.2016}. \textit{Left:} Velocity evolution for GRB980425-SN1998bw. The velocities produced by the spline fit are significantly higher than the template fitting method; this discrepancy persists at all times. \textit{Right:} Same as left, but having the spline fitting velocities reduced by 12000 km/s to aid comparison. The evolutionary trends are consistent within errors, particularly when accounting for additional sources of error discussed elsewhere in the text.}
	\label{fig:modjazsplinecompare1998bw}
\end{figure*}

The promise of the template fitting method is that it will handle cases of severe line blending; cases which, in theory, would be handled poorly by the spline fitting method, as this method will fit to the overall feature rather than the contribution of one line. To date, a detailed comparison has not been performed to prove that this is indeed the case. Before comparing the spline and template fitting methods to assess this, it should be noted that these methods do not measure the same thing. The spline fitting method determines the wavelength of the minimum flux of a feature, which can then be converted to a velocity by assuming a rest-wavelength for that feature, normally 5169 \AA\,in the case of Fe II. In contrast, the template fitting method blue-shifts and broadens the three iron lines near 5000 \AA\,in an average Ic spectrum, before combining the blue-shift and template velocities to compute the feature velocity. A direct comparison between these methods is presented here for GRB980425-SN1998bw and GRB130702A-SN2013dx, which provide examples of the possible scenarios encountered during velocity measurement. 

Figure \ref{fig:modjazsplinecompare1998bw} shows the velocities measured by the spline fitting method and the template fitting method for the iron feature for GRB980425-SN1998bw. The left panel clearly demonstrates that the velocities produced by the spline fitting method are systematically larger than those produced by template fitting. This is the expected result for cases of severe blending; the 5169\AA\,line is the reddest of the three lines in the iron feature, so the wavelength found by the spline method likely will be bluer than it should be, resulting in a higher measured velocity. In this instance, the spline fitting velocities are 1.5-2 times larger than the template fitting velocities at all times, including during the plateau. In the right panel of Figure \ref{fig:modjazsplinecompare1998bw}, the spline method velocities have been shifted by 12000 km/s to aid comparison. The decay rate for both is similar, however at late times (>35 days) the decay may be shallower for the spline fitting method. Bearing in mind the sources of error discussed in the preceding sections, it seems likely that the velocity evolution is the same for both methods, supporting the idea that the two methods are probing different tracers of the same velocity component. 

\begin{figure*}[h!]
	\centering
	\includegraphics[width=0.49\linewidth]{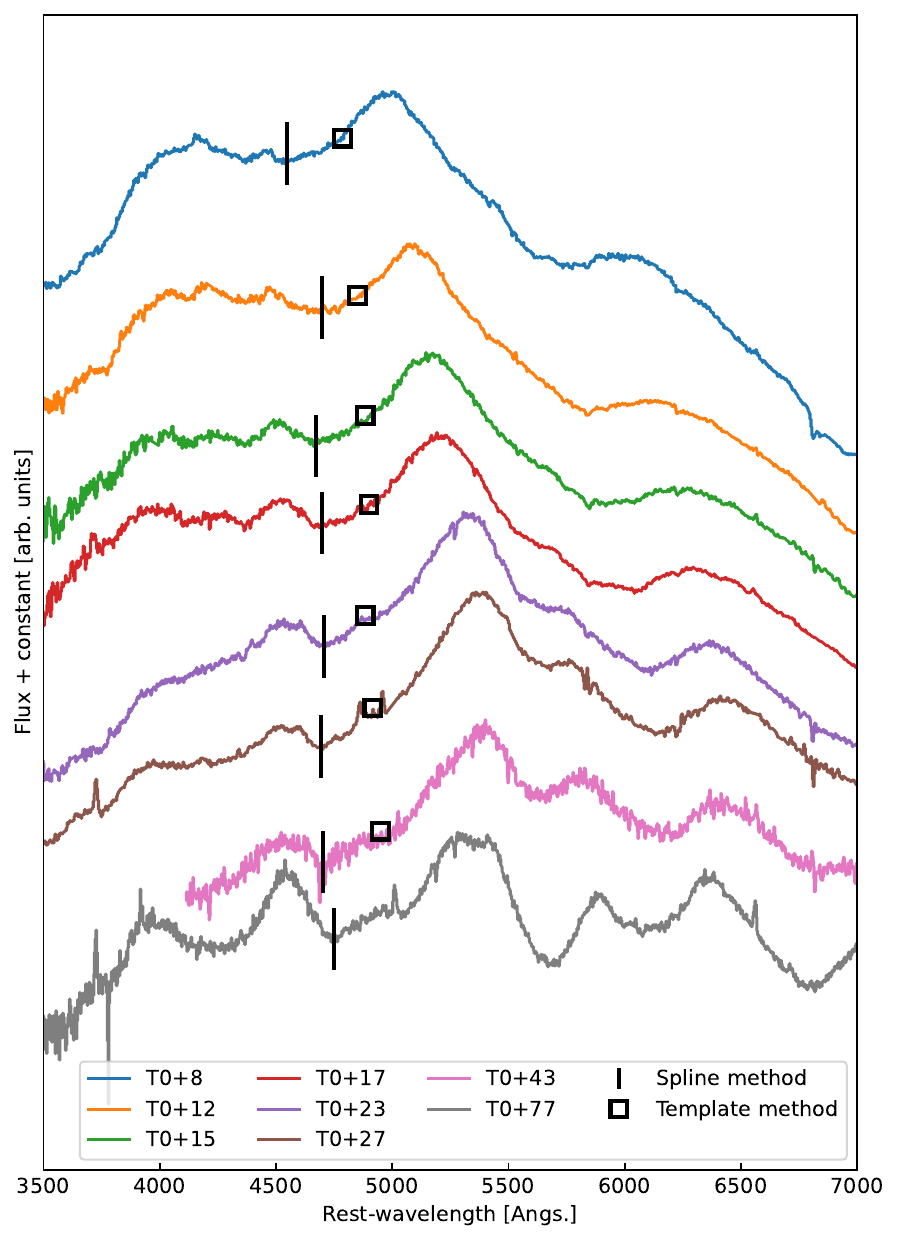}
	\includegraphics[width=0.49\linewidth]{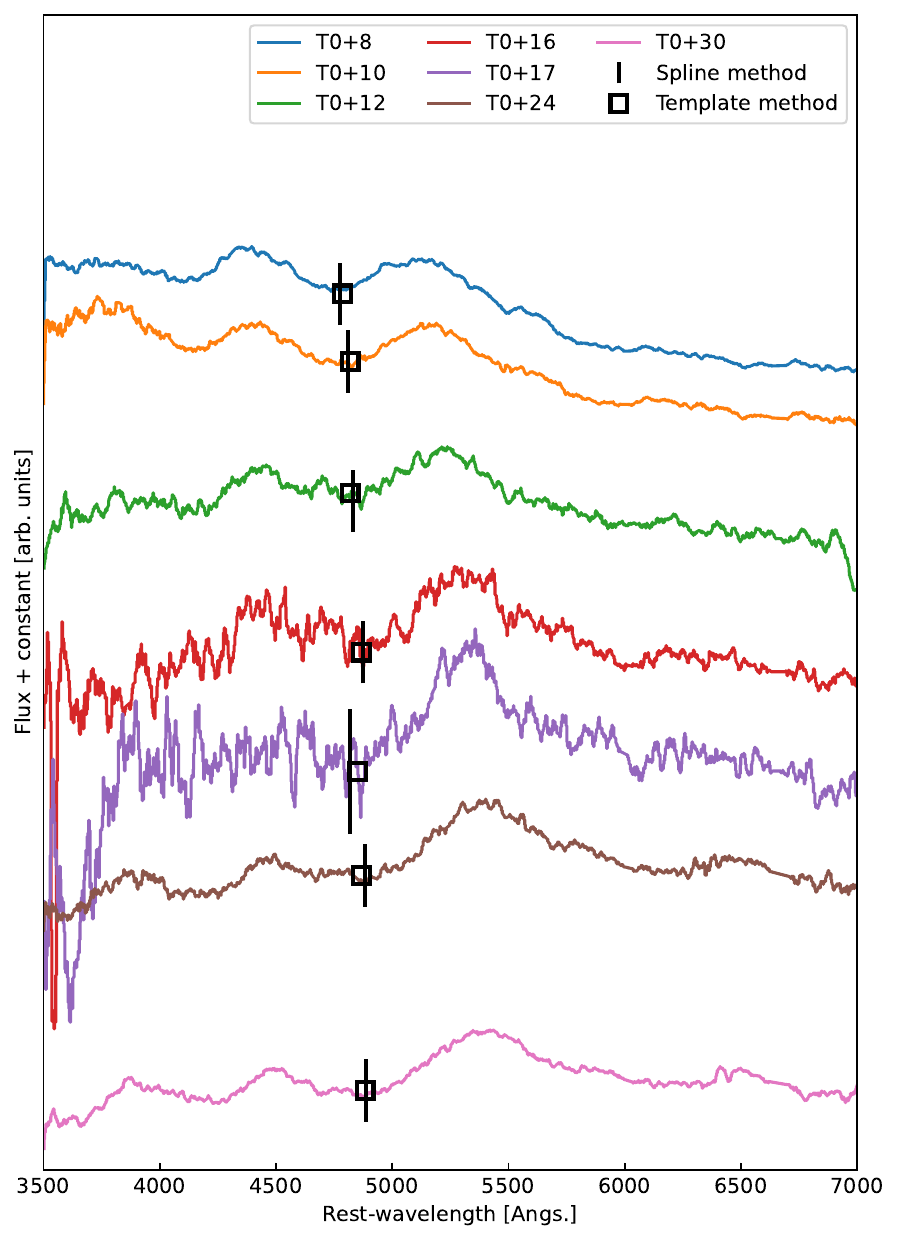}
	\caption[Direct comparison of spline and template fitting velocities]{Direct comparison of spline and template fitting velocities. Times of spectra are relative to T0 in the rest-frame of the source. Template fitting velocities from \cite{Modjaz.2016}. \textit{Left:} Spectral sequence of GRB980425-SN1998bw \citep{Patat.2001q8t}; the template fitting method produces a lower velocity than spline fitting. The template fitting method is tracking a different feature to that tracked by spline fitting. \textit{Right:} Spectral sequence of GRB130702A-SN2013dx \citep{DElia.2015}, binned at 20\AA \, resolution; both methods are tracking the same feature here, resulting in the same velocity. }
	\label{fig:1998bwtracingvels}
\end{figure*}

Evidence for this theory can be found in Fig. \ref{fig:1998bwtracingvels}, whose left panel shows the spectral evolution of GRB980425-SN1998bw, with the feature wavelengths from both methods over-plotted. At early times (T0+8 to T0+17), the spline fitting method targets the minimum of the large trough near 4500 \AA. This is the rightmost of two features visible in this region at this epoch, and is the most likely of the two to be associated with iron. The spline fitting method produces a wavelength that does not correspond to a significant feature at this stage, but which is located on the red edge of the major feature, explaining the lower velocity. Later, this feature becomes more distinct, and from T0+23 there is a sub-feature on the red side of the iron feature, which corresponds to the wavelength of the template fitting result. 

\begin{figure*}[h!]
	\centering
	\includegraphics[trim={2cm 2cm 1cm 4.4cm}, clip, width=0.49\linewidth]{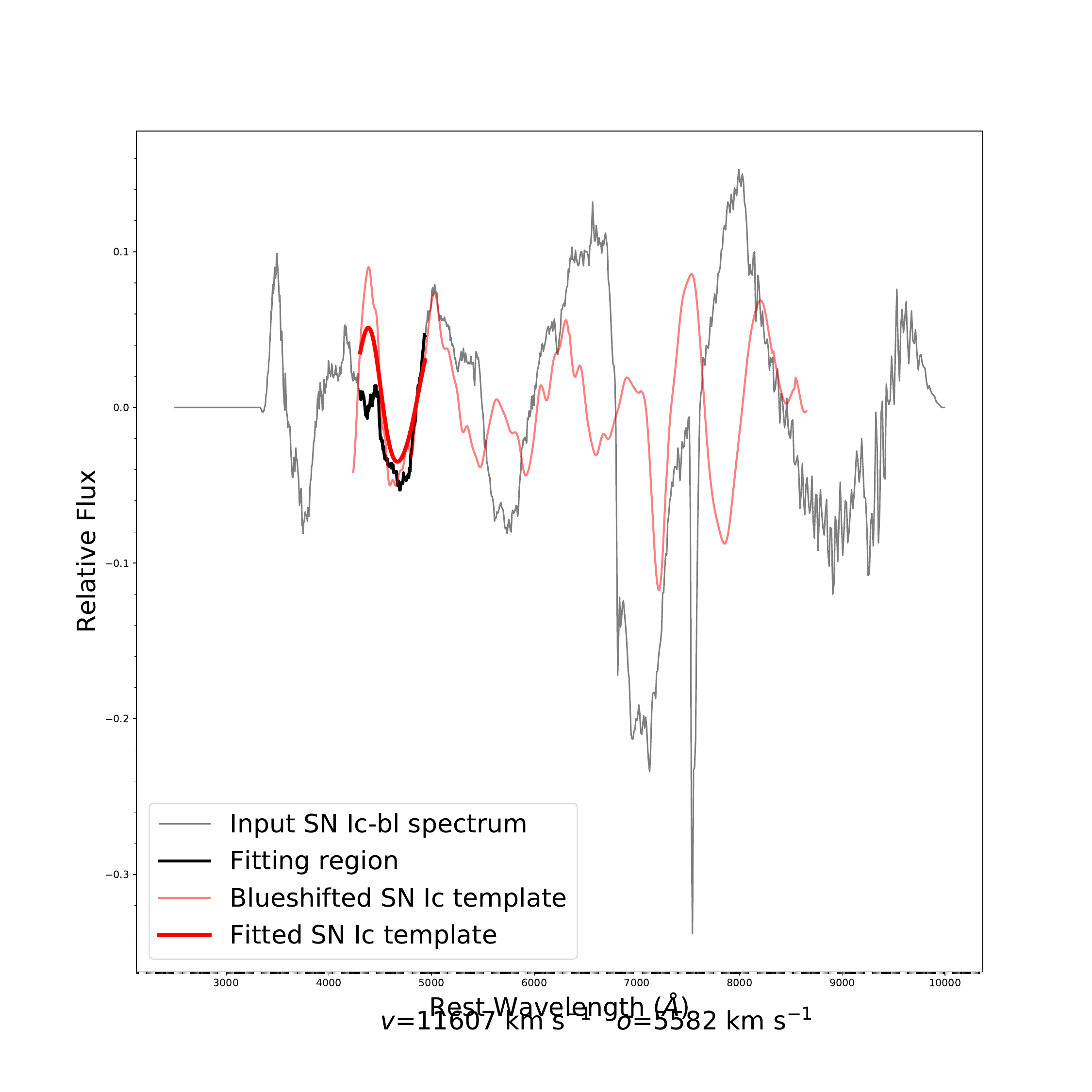}
	\includegraphics[trim={0 -2cm 0 0}, clip, width=0.49\linewidth]{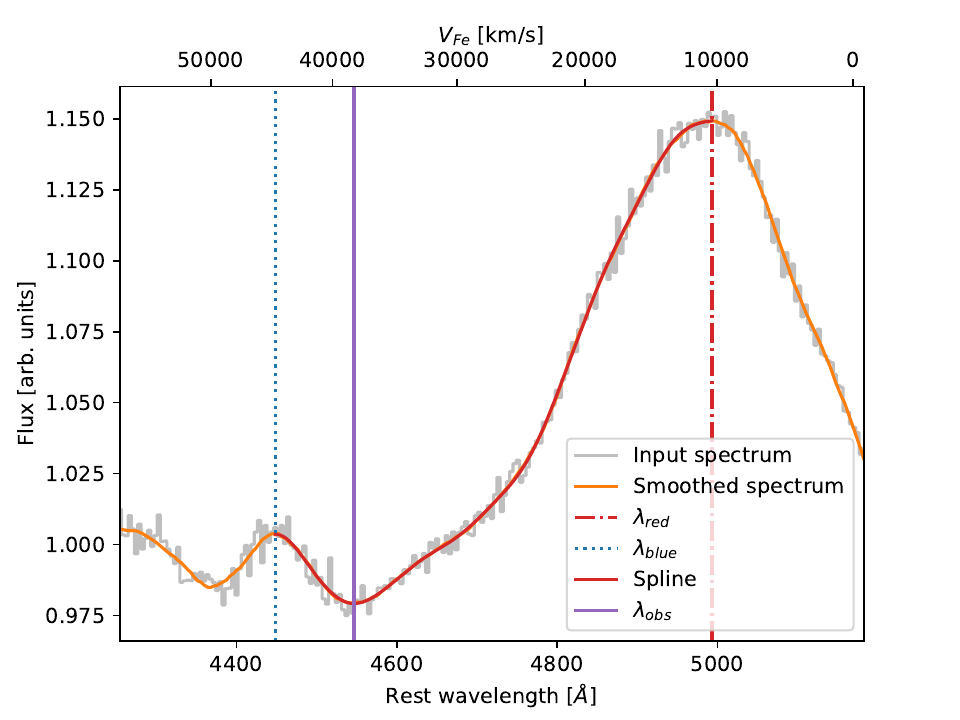}
	\caption[Fits for the T0+8 spectrum of GRB980425-SN1998bw]{Fits produced by the template fitting method (left) and spline fitting method (right) for the spectrum of GRB980425-SN1998bw at T0+8 days \citep{Patat.2001q8t}. No deblending of the iron lines is visible here.}
	\label{fig:comparefits1}
\end{figure*}

\begin{figure*}[h!]
\centering
\includegraphics[trim={2cm 2cm 1cm 4.4cm}, clip, width=0.49\linewidth]{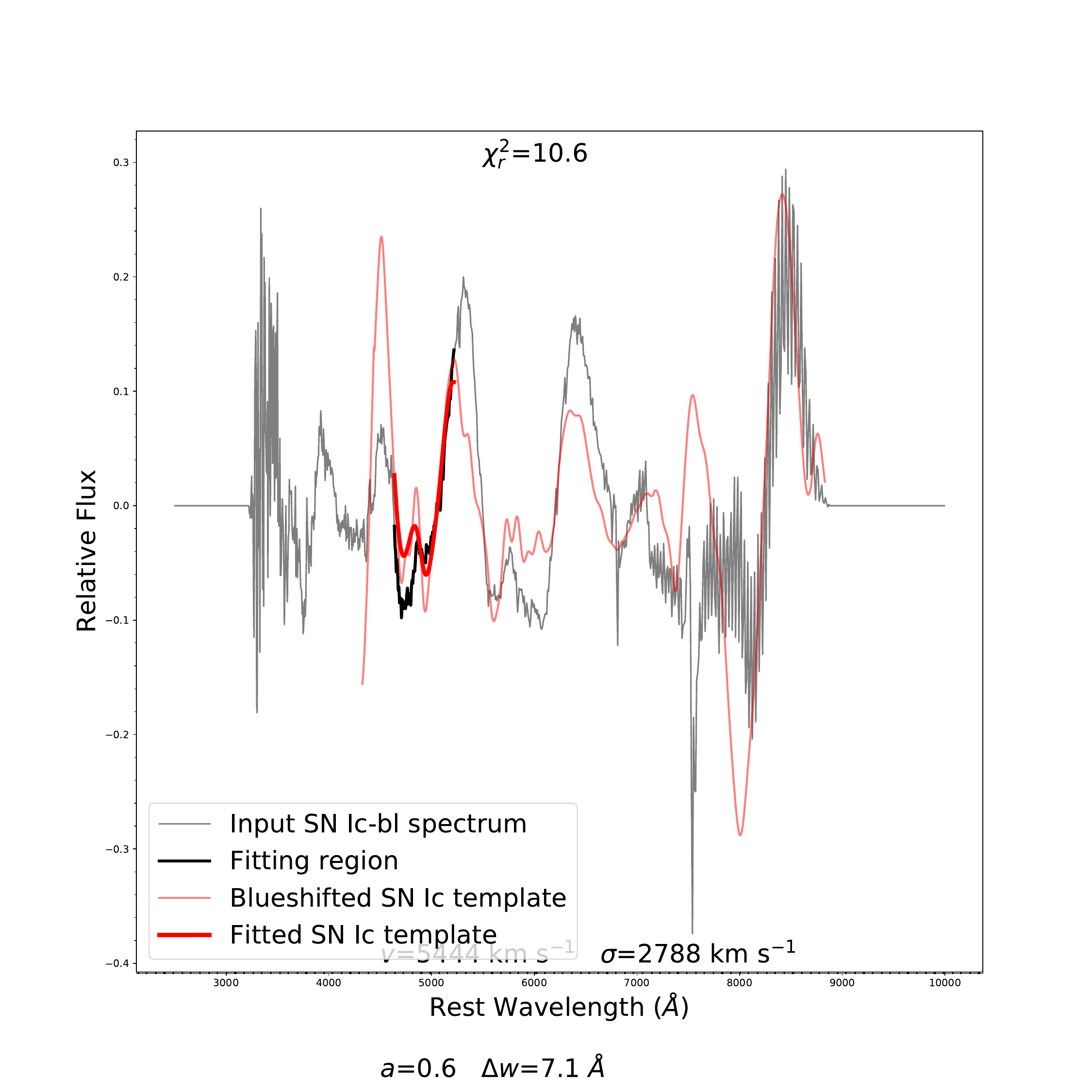}
\includegraphics[trim={0 -2cm 0 0}, clip, width=0.49\linewidth]{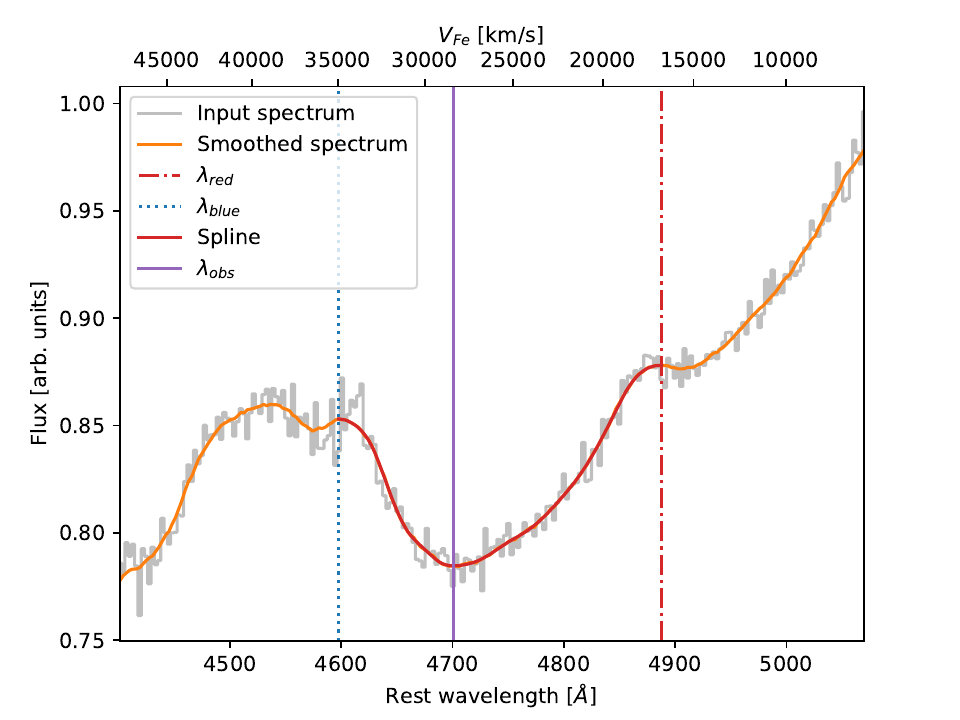}
	\caption[Fits for the T0+23 spectrum of GRB980425-SN1998bw]{Fits produced by the template fitting method (left) and spline fitting method (right) for the spectrum of GRB980425-SN1998bw at T0+23 days \citep{Patat.2001q8t}. The iron lines have begun deblending here. The template fitting method has been drawn to the rightmost trough of the iron feature; in contrast the spline fitting method continues to be applied to the leftmost peak, as this reflects the original feature best.}
	\label{fig:comparefits2}
\end{figure*}

This visual comparison suggests that blending is handled by the template method, though it is hard to verify for the heavily blended spectra. Figure \ref{fig:1998bwtracingvels} also offers an explanation for the large errors in the early time results from template fitting; it is harder to determine where the minimum of a heavily blended line is, but it relatively simple once de-blending begins around 25 days after T0. By contrast the spline fitting method continues to follow the bluer sub-feature after de-blending, which explains why the velocity remains high. Figures \ref{fig:comparefits1} and \ref{fig:comparefits2} show the fitted features for both the spline fitting and template fitting methods at T0+8 and T0+23 days for GRB980425-SN1998bw. It is initially unclear what spectral feature the template fitting method is really fitting, though the fit appears to trace the spectrum well. At later times, the template fitting method clearly attempts to replicate the de-blending, while the spline fitting method continues to follow the same feature. In principle it would be possible to fit the emerging trough at 4900 \AA \, using the spline-fitting method, but there is no obvious route to match the template method at early times, as no feature is visible for spline fitting. Doing so would have resulted in a velocity  of $\sim$16000 km/s, in excellent agreement with the template fitting velocity.

\begin{figure}[h!]
	\centering
	\includegraphics[width=\linewidth]{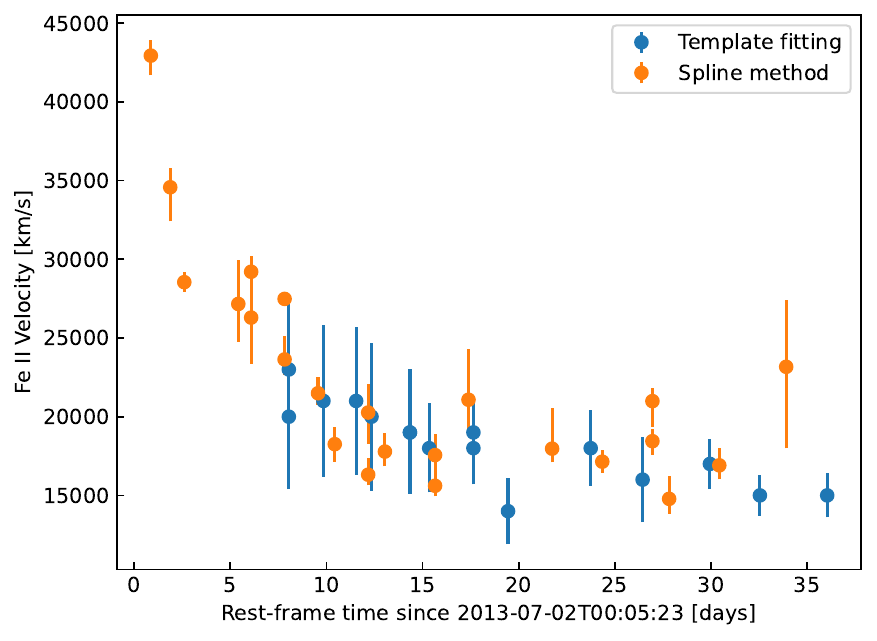}
	\caption[Template fitting vs. spline fitting velocities GRB130702A-SN2013dx]{Comparison of results produced by the spline-fitting and template-fitting procedures for GRB130702A-SN2013dx; template fitting velocities from \cite{Modjaz.2016}. The velocities produced by both methods are identical within their respective error bars.}
	\label{fig:modjazsplinecompare2013dx}
\end{figure}

Figure \ref{fig:modjazsplinecompare2013dx} shows the results of spline and template fitting for GRB130207A-SN2013dx. Unlike the case of GRB980425-SN1998bw, the two methods produce nearly identical velocities and velocity evolution; Figure \ref{fig:1998bwtracingvels}, confirms that both methods are tracing the same feature. Curiously  this feature appears to be just as heavily blended as the iron feature of GRB980425-SN1998bw. In fact Figs. \ref{fig:modjazsplinecompare1998bw} and \ref{fig:modjazsplinecompare2013dx} show that these two events have nearly identical velocities at early times\footnote{As measured by the template-fitting method.}, supporting the case for similar levels of blending. This prompts the question of what caused the disagreement between the two methods for GRB980425-SN1998bw.

One possibility is that the difference in velocity between these two methods is influenced by the local continuum, rather than the just the level of blending. The continuum in the spectra of these events may be made up of a thermal component, host-galaxy component and/or afterglow component. The iron feature is a triplet, composed of three lines at 4924 \AA, 5018 \AA\, and 5169 \AA. Assuming roughly equal line-strengths, if the iron feature falls in a region with a blue continuum, then the 5169 \AA\,line will likely be closest to the wavelength of the overall feature minimum. In contrast, if a red continuum lies beneath the iron lines, the 4924 \AA\,line may be closer to the minimum wavelength of the blended feature. This would not affect the template fitting method, as the spectra are flattened prior to fitting. However, the spline fitting method would measure a similar velocity to the template method in the blue continuum case, but a much higher velocity in the red continuum case. This would also serve to explain the diversity of shapes seen for iron features; some Ic-BL spectra show no clear minimum for the iron feature, typically in the case of a red continuum. 

There is some evidence for this paradigm in the spectra presented here. The spectra of SN1998bw have been corrected for host contamination \citep{Patat.2001q8t}, and the SN2013dx spectra have had the afterglow and host-galaxy contribution removed. Since SN1998bw exhibits well developed supernova features after $\sim$1 week, the continuum is likely due to the thermal SN component. This means the continua are determined by the effective temperatures of both supernovae. Figure \ref{fig:1998bwtracingvels} shows a key difference between the early spectra: SN2013dx has a bluer spectrum than 1998bw. This suggests that SN1998bw may be a cooler supernova than SN2013dx, which had a temperature of 16000 $k$ at 9.3 days \citep{Toy.2016}, giving it a peak wavelength of $\sim$1800 \AA. This would place the iron lines on a blue continuum, where the two methods are more likely to agree; this is what is observed for the velocities of this event. In contrast, the red continuum of SN1998bw would lead to a large discrepancy between the methods, as was seen in Fig. \ref{fig:modjazsplinecompare1998bw}. Theoretically, if a red continuum causes the minimum to be closer to the 4924 \AA\,or the 5018 \AA\, iron line, then a discrepancy between the methods of 10000-15000 km/s should be expected. This matches very well the 12000 km/s difference which is observed in Fig. \ref{fig:modjazsplinecompare1998bw}. In summary, blending seems to be well handled by the template fitting method, but it is not necessarily the case that blending will always affect the spline fitting method in a significant manner. Instead, the severity of this impact is due to the underlying continuum of the supernova. 

\section{Discussion}\label{sec:discussion}
There have been several attempts to quantify the expansion velocities of Ic-BL supernovae over the years. Many of these attempts rely on just two methods: the spline fitting \citep[e.g.][]{Finneran.2024B} or the template fitting method \citep{Modjaz.2016}. Whilst both methods had been widely applied, little attention had been devoted to which method is the most performant. This analysis has sought to answer that question by quantifying the additional sources of uncertainty present in both methods, and performing direct comparisons using test cases. A number of additional sources of error were identified by this analysis, and the validity of claims made about these methods have been tested. In this section an interpretation of the results is provided, and recommendations for best practice are offered.

\subsection{Lessons from direct comparison of the methods}

The analysis presented in Sect. \ref{sec:directcompare} compared the velocities produced by both methods for two GRB-SNe. This investigation confirmed that the template fitting method is capable of handling cases of severe blending, as was suggested by \cite{Modjaz.2016}. However, it was noted that significant errors are commonplace prior to de-blending. These errors are in the range of 2000-5000 km/s depending on the event and epoch. It was believed that spline fitting would fail for both test-events, as they both show similar levels of blending of the iron triplet. In the case of GRB980425-SN1998bw this belief was borne out. However, for GRB130702A-SN2013dx, the spline fitting method seems to be unaffected by blending. This suggests that different blending conditions exist which impact both methods differently. A proposal was put forward which explains the impact of blending with regard to the slope of the underlying continuum. Assuming continua which are dominated by thermal effects, it is possible to gauge how often blending may be an issue for spline fitting. \cite{Taddia.2019} present the blackbody temperature evolution for a sample of Ic-BL supernovae with and without gamma-ray bursts. Their results suggest that temperatures decline to a plateau of 5000-6000 K within the first 20 days after the explosion. This suggests that, at least during this period, many Ic-BLs will have iron features on blue continua. For this reason, it is likely reasonable to apply the spline fitting method for many Ic-BLs. This also means that the velocities measured in \cite{Finneran.2024B} are not affected by issues with blending in all cases. Furthermore, it was discovered that the shape of the velocity evolution for GRB980425-SN1998bw was identical for the two methods, despite issues arising from the blending and the underlying continuum. This is because the region which forms the iron feature forms each line in the iron triplet, and so they all undergo the same evolution. This means that velocity evolution studies for the iron feature are likely to produce very similar results, provided that one line is followed throughout the evolution. If different lines are followed once de-blending occurs, there will be discontinuities in the velocity evolution. 

In terms of absolute velocities, in cases where the continuum is biased against the spline fitting method, there may be differences in velocity of up to 15000 km/s between the two methods. Cumulatively these events would increase the average velocity for a population of Ic-BLs. Since the optical properties of Ic-BLs with and without GRBs are similar, this probably does not impact comparisons between these populations. Resolution of blending issues alone likely is not enough to separate the two populations. However, when comparing these populations to Ic supernovae, spline fitting might lead to a larger difference than exists in reality (assuming that the true values are similar to those seen with the template fitting method). The magnitude of this effect is dependent on the number of Ic-BLs where the continuum will cause this problem. Further work is needed to determine how often this occurs. This work should include modelling of the SED or spectra of a large population of Ic-BLs with and without GRBs.

Given the issues posed by blending, it is tempting to stick with the template fitting method. However, as pointed out by \cite{Modjaz.2016}, this method struggles to fit the silicon feature of Ic-BL supernovae. If performing an analysis of multiple features, it is advantageous to use the same method on all features, as the magnitude and sources of error will be consistent for each group of features. In this case, mitigation strategies are needed to reduce the impacts of blending and the continuum. One approach may be to remove the blackbody contribution from the spectrum prior to fitting. Since the continuum in the blue to UV region is likely not purely blackbody, so this approach may meet with challenges. Another tactic would be to fit the iron feature `in reverse', by starting with the latest non-nebular spectra, and identifying the 5169 \AA\,line, before following it backward in time until it disappears. This would likely produce the correct evolution and velocity for this line, since it tends to de-blend prior to the other two lines in the triplet. The disadvantage of course is that this will reduce the number of useable spectra for velocity measurements, because the line will likely be blended at early times. 

\subsection{Dealing with additional sources of error}

The remainder of this analysis was devoted to determining what the additional sources of error in each method might contribute to the overall uncertainty. In Sect. \ref{sec:smoothingerrors} the impacts of smoothing algorithms on the final result was determined for each method. In the case of the spline-fitting method, the fine-tuning of smoothing parameters was found to have a larger impact for low resolution or low S/N spectra; in these cases, errors may approach 3000-4000 km/s. In higher resolution/higher S/N cases, the typical error due to variation of the smoothing parameters is of order 1000 km/s. In most cases these errors are comparable to the one sigma errors produced by the method. For the template fitting method, smoothing may constitute an additional uncertainty of around 500-1000 km/s. In many cases the effect of this is marginal, because the errors produced by template fitting tend to be between 2000-5000 km/s at early times. However, at late times this source of error may become more relevant. In general however, smoothing algorithms do not significantly affect the result for a population, except in cases of extreme parameter choice which can be avoided through sensible visual examination of the spectra. This is probably because the presence of broad features in the Ic-BL spectrum provides more leeway with smoothing parameters, since these broad features are unlikely to be lost when smoothed.

The presence of a phase-shifts when using the the template fitting method, reviewed in Sect. \ref{sec:phase}, may shift the velocity by 500-2000 km/s (or more in some cases). These types of errors are likely to be very common in practice, since the template fitting method requires you to know the phase of each spectrum. The phase cannot always be determined with arbitrary accuracy, either because V-band data are not available or because a broad peak exists. Although it would be possible to use the explosion epoch for GRB-SNe, this tactic is probably not generally applicable for Ic-BLs, since finding the explosion epoch can often be harder than finding the peak. If the spectral templates used in this method were available every day rather than every 2 days, it would go some way to solving this issue. In theory this would be achievable with the existing Ic spectra used to make these templates. Phase shifts cannot be ignored at late times, since their contribution becomes similar to that of the typical error on the template fitting method. The impacts of phase shift will be present across the whole spectral evolution, as it arises when the peak time is not determined very precisely. This could mean that the shape of the evolution will be impacted, by either making the decay faster and then shallower or slower and then steeper. It is important to note that this effect is not detectable without prior knowledge of the velocity. This is one reason why verifying the results of any of these methods against simulations should be a priority in future. 

The spline fitting method tends to under-estimate the error on velocity. This typically contributes an extra 1000 km/s to the velocity, as determined from the residuals for the power-law and broken power-law fits. It is not entirely clear what the source of these errors is. They may be a result of slight differences in spectral resolution or alignment between different spectra. It cannot necessarily be avoided when using a heterogeneous dataset, so quantifying it is essential. However, even including this error, the spline fitting method is more precise than the template fitting method. Accuracy of the method was verified by comparing it to the template fitting method for SN2013dx, where the effects of blending are minimal. Given the benefits of the applicability of the spline fitting method to all the features in the Ic-BL spectrum, it remains a useful tool for measuring the velocities of Ic-BLs. 

\subsection{Unquantifiable errors}
The large errors produced by the template fitting method are due to a wide range of factors, which were not discussed here. The construction of Ic template spectra almost certainly has an impact on the velocity errors. Data presented in \cite{Modjaz.2016} shows that Ic supernovae reach a velocity plateau of around 8000 km/s, 20 days after V-band maximum light. Prior to this, their velocity declines from around 13000 km/s. Given that spectra are averaged in 5 day windows, there may be significant velocity differences in the pre 20 days window. Furthermore it was observed during the use of the template fitting method that poor fits are sometimes produced, as measured by the $\chi_r^{2}$ statistic or by visual examination. Both of these effects could make it difficult for the MCMC analysis to converge to a precise solution. However, the exact mechanism behind the large errors produced by this method are beyond the scope of this analysis. It is simply something worth bearing in mind for cases where spline fitting might be applicable. 

\section{Conclusion}\label{sec:conclusion}
This paper has reviewed the two major methods used to measure the velocity of Ic-BL supernovae. These are often applied without regard to their sources of error or true benefits. The focus of this work was to quantify and develop strategies to deal with these errors, and to identify whether either method performs better in all cases. This was accomplished through the use of test cases and exploration of the parameter spaces of each method. It was found that both methods are susceptible to uncertainty associated with smoothing, but that this source of error tends to be small relative to the overall error. Accounting for its impact may be more important for the spline fitting method because it has smaller errors than template fitting. In light of this, there is no motivation to improve smoothing algorithms or to fine tune smoothing parameters beyond a simple sanity check. 

 In the case of the template fitting method, the potential for phase discrepancies creates an additional source of error, which may contribute an additional $\sim$1000 km/s to the final error when added in quadrature. The size of this error seems to increase the further the given spectrum is from peak light; at peak, phase effects are probably negligible. Phase effects arise mainly because of the limited choice of template spectra, which forces the rounding of phase. However, in some cases the phase may be impacted by an inaccurate supernova peak time, as this method uses the V-band maximum for its epoch. One proposed solution for this is to use the GRB trigger time epoch where possible. However, this method cannot be applied to non GRB-associated Ic-BLs. In these cases the peak time is no worse than the explosion time, but it is easier to determine.
 
 For the spline fitting method, the largest additional source of uncertainty appears to be due to random scatter of the measured velocities. This could perhaps be caused by alignment issues between spectra, or instrumental differences. This typically contributes around 1000 km/s of additional error. This is significant in the context of this method, as it tends to produce smaller errors than template fitting. 
 
A direct comparison of both methods was used to verify the claim that the template fitting method can handle blended features. In instances where a red continuum is present in the region of the iron feature, the template fitting method performs reasonably well, but the spline fitting method over-estimates the velocity. In instances where a blue continuum is present both methods can be used. This means that, assuming that the template fitting velocities are correct, the spline fitting method over-estimates the velocity for a certain percentage of Ic-BLs. However, it may be possible to identify such cases by simple examination of the spectra. Two possible ways to reduce the impact of blending on the spline fitting method could be: flatting the spectra in a similar way to the template fitting method; or, in the case of a large sample, to apply the spline fitting method but trace features from the late-time spectra rather than beginning with early features. Overall the template fitting method seems to the most optimal method to use when dealing with cases of severe blending.

In summary, there does not seem to be a clear optimum method to measure the velocity of a Ic-BL supernova. Although the spline fitting method produces more precise estimates of velocity in some instances, it fails to handle blending in all cases, resulting in higher velocity measurements relative to template fitting. This may be due to the underlying blackbody temperature of the spectrum. The template fitting method can handle these cases, but with greater uncertainty in velocity, which will affect its ability to constrain the velocity evolution of SNe, especially in the case of large population studies. In any case, both methods seem to produce identical velocity evolution in all instances.

\begin{acknowledgements}
GF and AMC acknowledge support from the UCD Ad Astra programme. This work has made extensive use of WISeREP (available at: \url{https://www.wiserep.org}) as one source for our Ic-BL spectra. In order to access data from WISeREP we used the WISeREP API, created by Tomás E. Müller Bravo; we thank them for providing a modified code from which we built our data-collection pipeline. GRB-SN data was gathered from the GRBSN webtool, which can be found at \url{https://grbsn.watchertelescope.ie/}. This work made use of the data products generated by the NYU SN group, and 
released under DOI:10.5281/zenodo.58767, 
available at \url{https://github.com/nyusngroup/SESNspectraLib}.\end{acknowledgements}

\bibliography{paper3v2.bib}

\begin{appendix}
\onecolumn
\section{Additional violin and smoothing plots}\label{sec:appendix}

\begin{figure*}[h!]
	\centering
	\includegraphics[width=\linewidth]{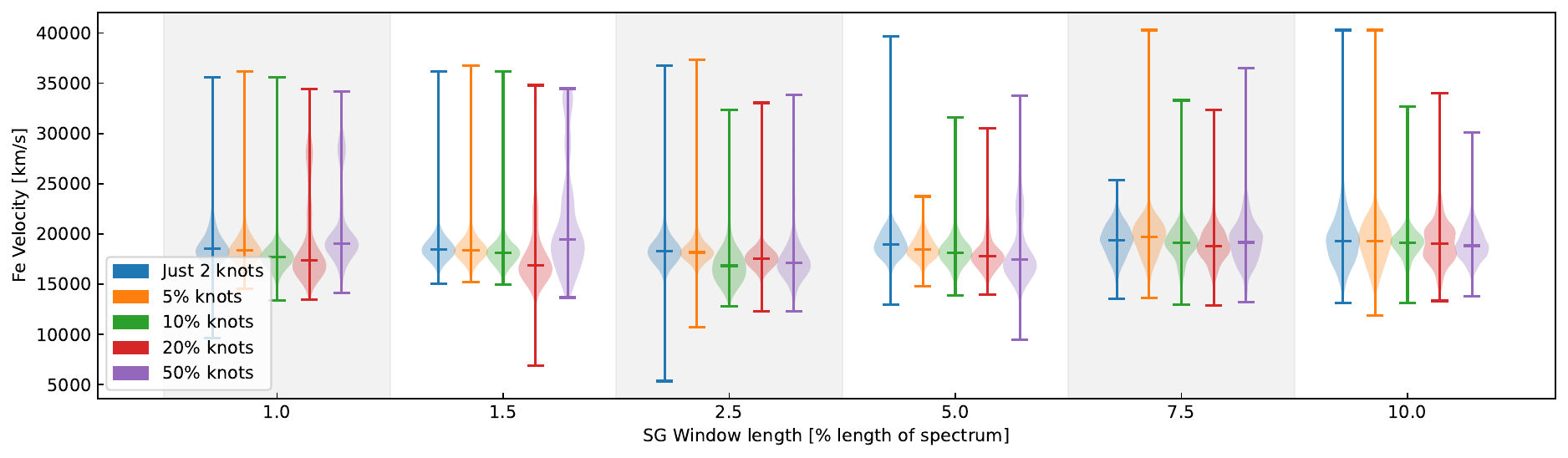}
	\caption[Variation of Fe II velocity with filter width and spline density for a sample spectrum of SN2016P]{Similar to Fig. \ref{fig:knotsmoothSi1998bw}, but instead showing the Fe II velocities of SN2016P; which is an example of a low-resolution, very low S/N spectrum. It appears that the low S/N may have washed out the impact of smoothing on the velocity, the range of all distributions is large, and the typical one-sigma error is 3000-5000 km/s. The effect of the spline density is also altered. Using the case of 2 knots makes little difference here, and there is no preferred spline density. The choice of smoothing parameters may shift the mean velocity by $\sim$2000-3000 km/s in such cases.}
	\label{fig:knotsmoothFe2016P}
\end{figure*}

\begin{figure}[h!]
	\includegraphics[width=\linewidth]{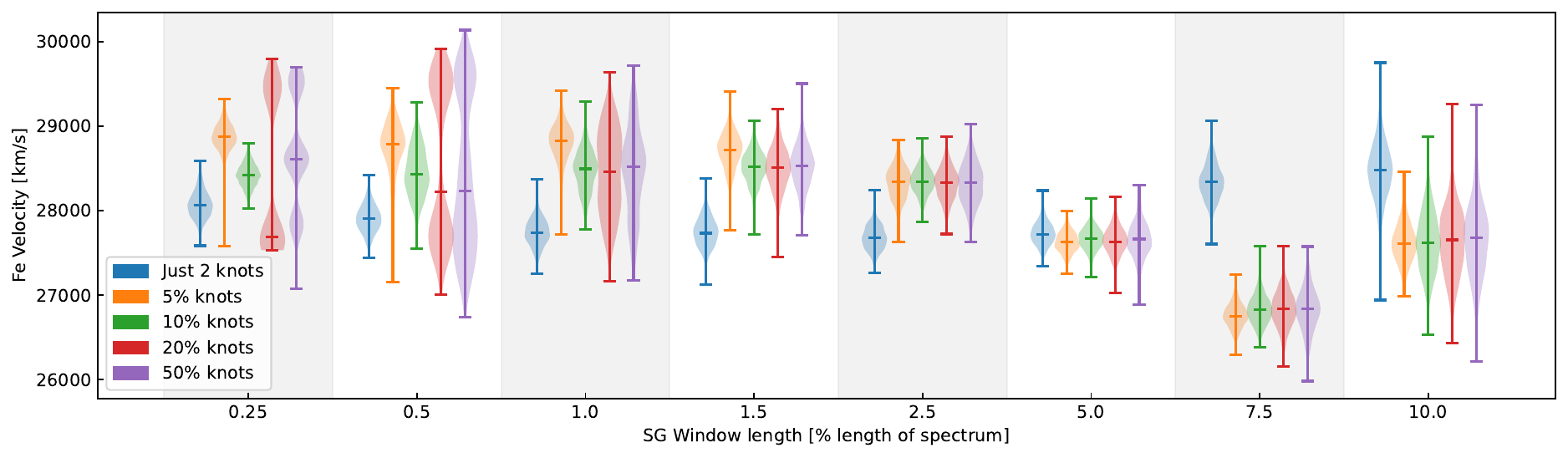}
	\caption[Variation of Fe II velocity with filter width and spline density for a sample spectrum of GRB980425-SN1998bw]{Similar to Fig. \ref{fig:knotsmoothSi1998bw}, but instead showing the change in Fe II velocities of GRB980425-SN1998bw with filter width and spline density; in the case of a spectrum with a high resolution and high S/N. The results are broadly similar to Fig. \ref{fig:knotsmoothSi1998bw}. Here the variation in mean velocity due to fine tuning of smoothing parameters is $\sim$1000 km/s.}
	\label{fig:knotsmooth1998bwFe}
\end{figure}

\begin{figure}[h!]
	\includegraphics[width=\linewidth]{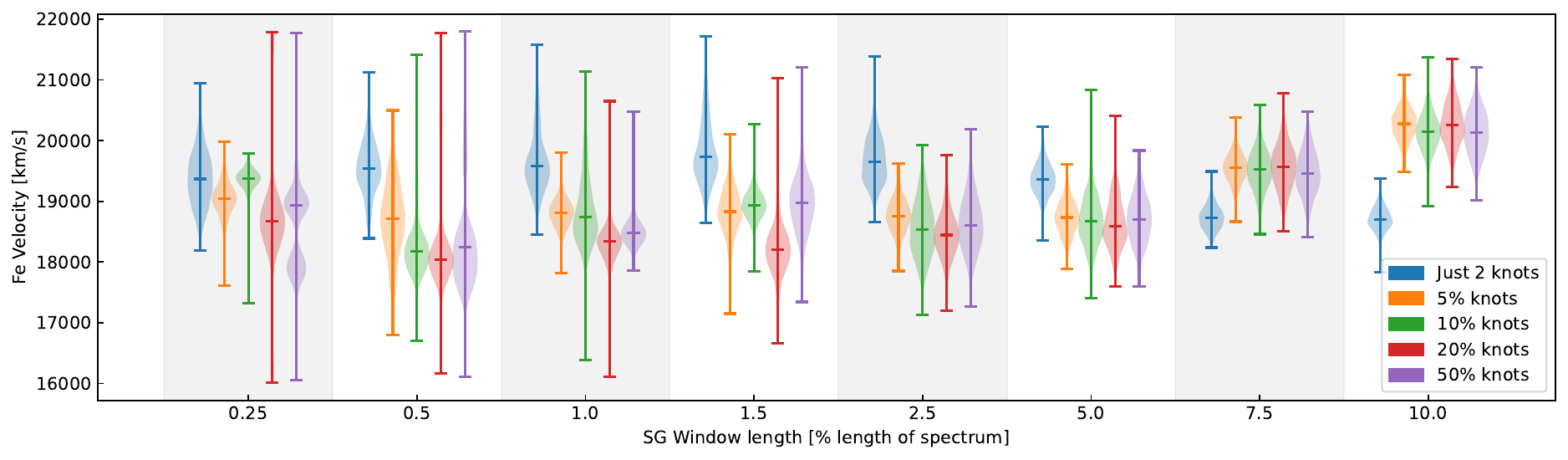}
	\caption[Variation of Fe II velocity with filter width and spline density for a sample spectrum of GRB030329-SN2003jd]{Similar to Fig. \ref{fig:knotsmoothSi1998bw}, but instead showing the change in Fe II velocities of GRB030329-SN2003jd with filter width and spline density; in the case of a spectrum that has a high-resolution and a reasonable S/N. The results are similar to those in Fig. \ref{fig:knotsmoothSi1998bw}. In this case the fine tuning of the smoothing parameters causes variation in the mean velocity of 1000-1500 km/s.}
	\label{fig:knotsmooth2003jdFe}
\end{figure}

\begin{figure}[h!]
	\includegraphics[width=\linewidth]{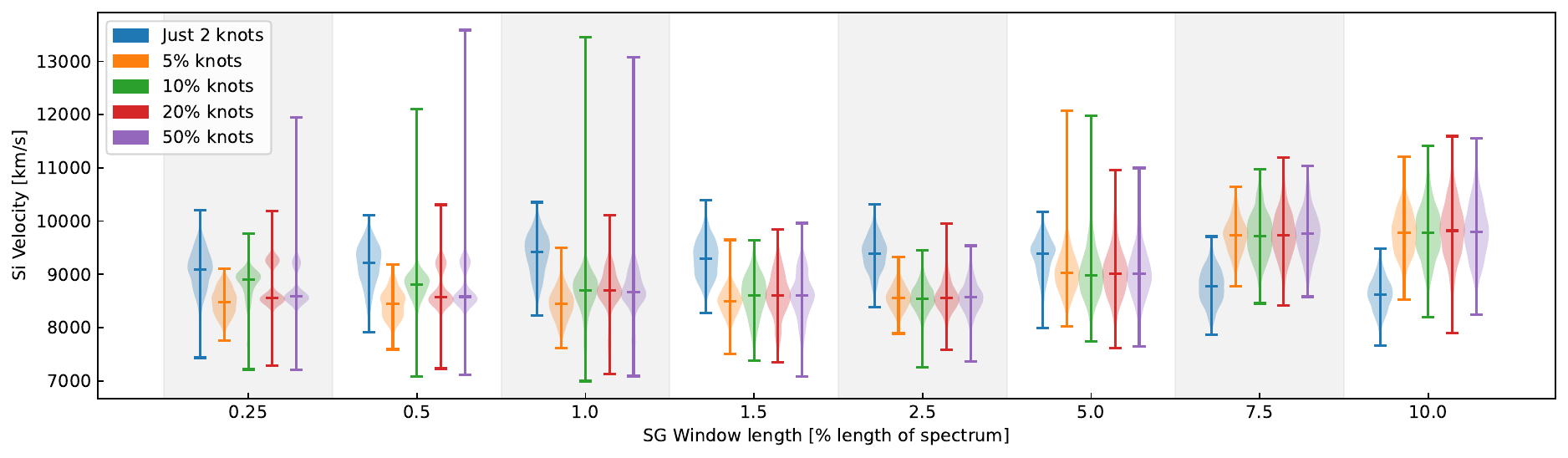}
	\caption[Variation of Si II velocity with filter width and spline density for a sample spectrum of GRB030329-SN2003jd]{Similar to Fig. \ref{fig:knotsmoothSi1998bw}, but instead showing the change in Si II velocities of GRB030329-SN2003jd with filter width and spline density; in the case of a spectrum that has a high-resolution and a reasonable S/N. Optimum filter width is 1.5-5\%, with no preference for knot density except that 2 knots should not be used. In this case the fine tuning of the smoothing parameters causes variation in the mean velocity of $<$1000 km/s.}
	\label{fig:knotsmooth2003jdSi}
\end{figure}

\begin{figure}[h!]
	\includegraphics[width=\linewidth]{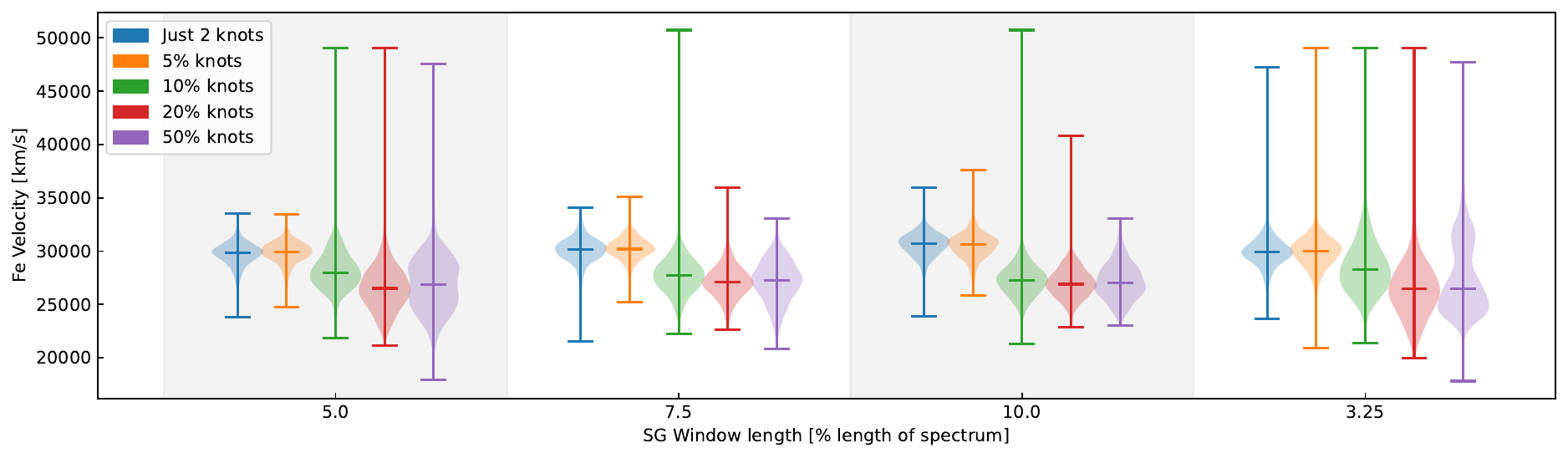}
	\caption[Variation of Fe II velocity with filter width and spline density for a sample spectrum of SN2020bvc]{Similar to Fig. \ref{fig:knotsmoothSi1998bw}, but instead showing the change in Fe II velocities of SN2020bvc with filter width and spline density; in the case of a spectrum that has a low-resolution and a high S/N. The lower resolution appears to have resulted in larger ranges for some of the violin plots, and typical one sigma errors are 3000-4000 km/s. In this instance, as for SN2016P it is difficult to choose optimum parameters, though it appears 5\% filter width and 5\% knots may be suitable. Varying the smoothing parameters in this instance may lead to mean velocity changes of order 3000 km/s.}
	\label{fig:knotsmooth2020bvcFe}
\end{figure}

\begin{figure}[h!]
	\includegraphics[width=\linewidth]{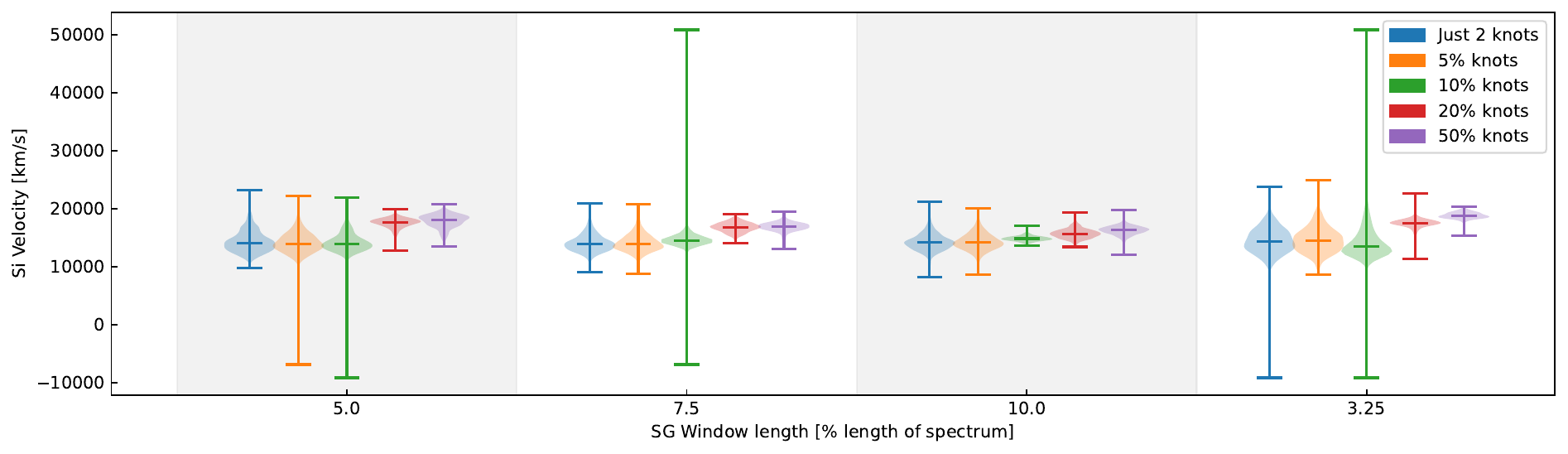}
	\caption[Similar to Fig. \ref{fig:knotsmoothSi1998bw} showing similar results for the Si II feature of SN2020bvc]{Similar to Fig. \ref{fig:knotsmoothSi1998bw}, showing the change in Si II velocities for SN2020bvc with filter width and spline density; in the case of a spectrum that has a low-resolution and a high S/N. Optimum parameters appear to be 20\% knot density with any filter width. It is particularly clear that 10\% knots performs poorly in this instance. Variation of these parameters may contribute an error of $\sim$4000 km/s, on top of typical 1 sigma errors of $\sim$2500 km/s. This large error is likely a result of the low resolution, which also explains why a higher knot percentage is required.}
	\label{fig:knotsmooth2020bvcSi}
\end{figure}

\begin{figure}[h!]

	\centering
	\begin{subfigure}{0.33\textwidth}
	\includegraphics[width=\linewidth, trim=1.2cm 0cm 1.2cm 0.4cm, clip]{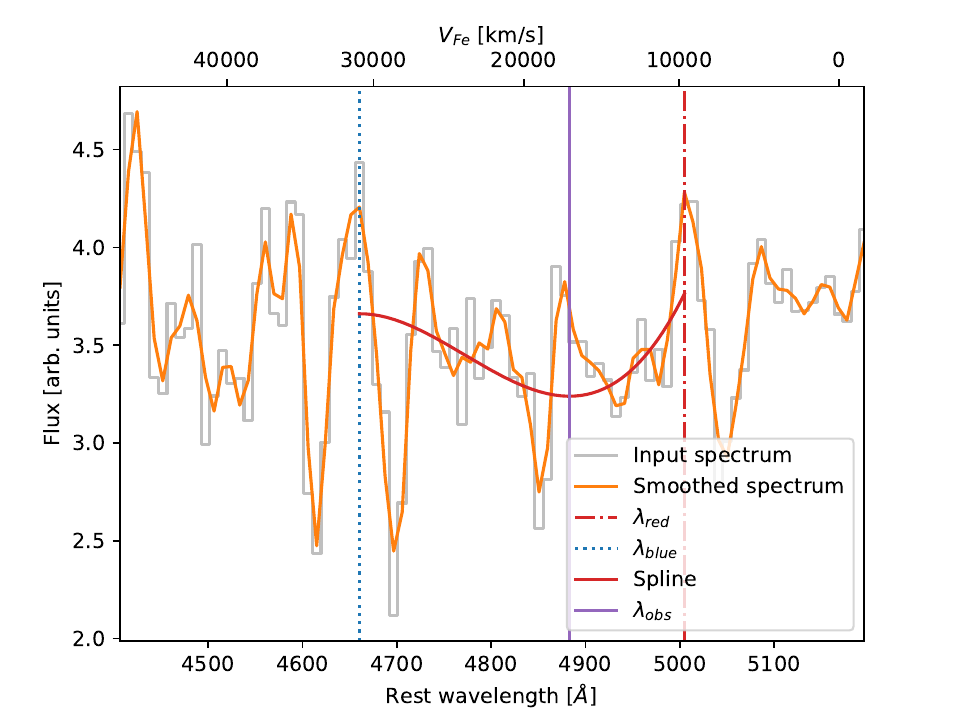}
	\caption{Knots: 2; SG: 1\%}
	\end{subfigure}
	\begin{subfigure}{0.33\textwidth}
	\includegraphics[width=\linewidth, trim=1.2cm 0cm 1.2cm 0.4cm, clip]{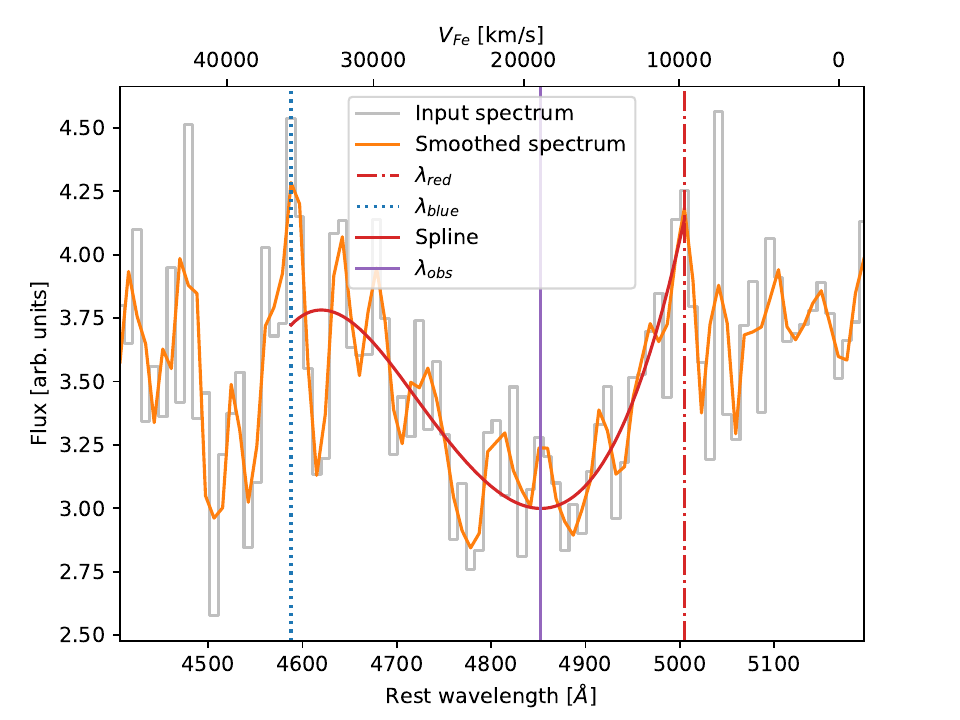}
	\caption{Knots: 10\%; SG: 1\%}
	\end{subfigure}
	\begin{subfigure}{0.33\textwidth}
	\includegraphics[width=\linewidth, trim=1.2cm 0cm 1.2cm 0.4cm, clip]{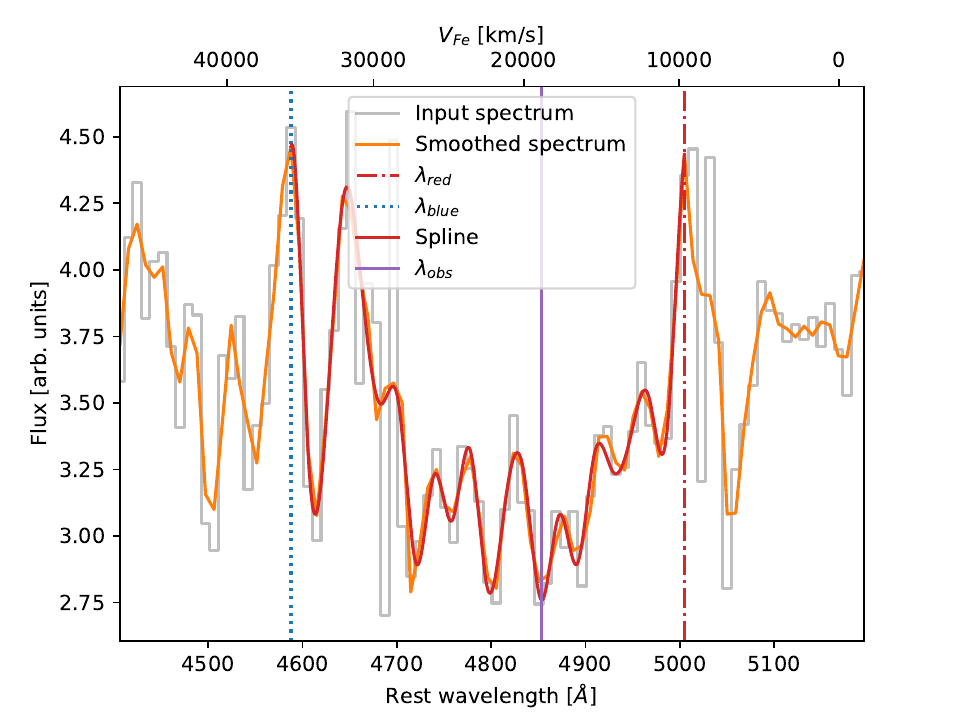}
	\caption{Knots: 50\%; SG: 1\%}
	\end{subfigure}
	
	\vspace{0.2cm}
	\begin{subfigure}{0.33\textwidth}
	\includegraphics[width=\linewidth, trim=1.2cm 0cm 1.2cm 0.4cm, clip]{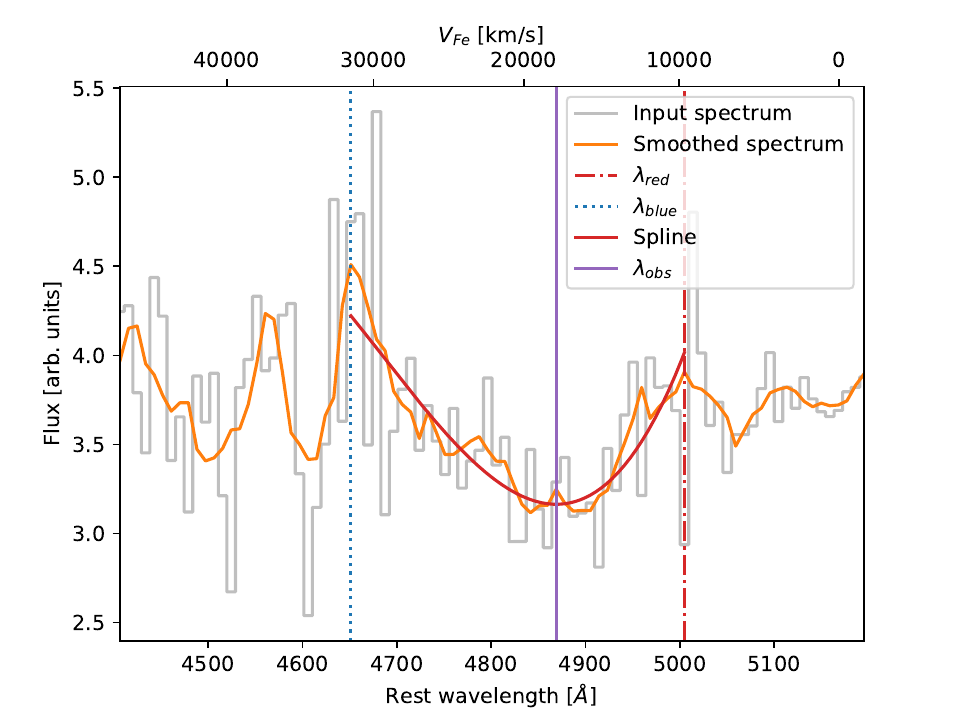}
	\caption{Knots: 2; SG: 2.5\%}
	\end{subfigure}
	\begin{subfigure}{0.33\textwidth}
	\includegraphics[width=\linewidth, trim=1.2cm 0cm 1.2cm 0.4cm, clip]{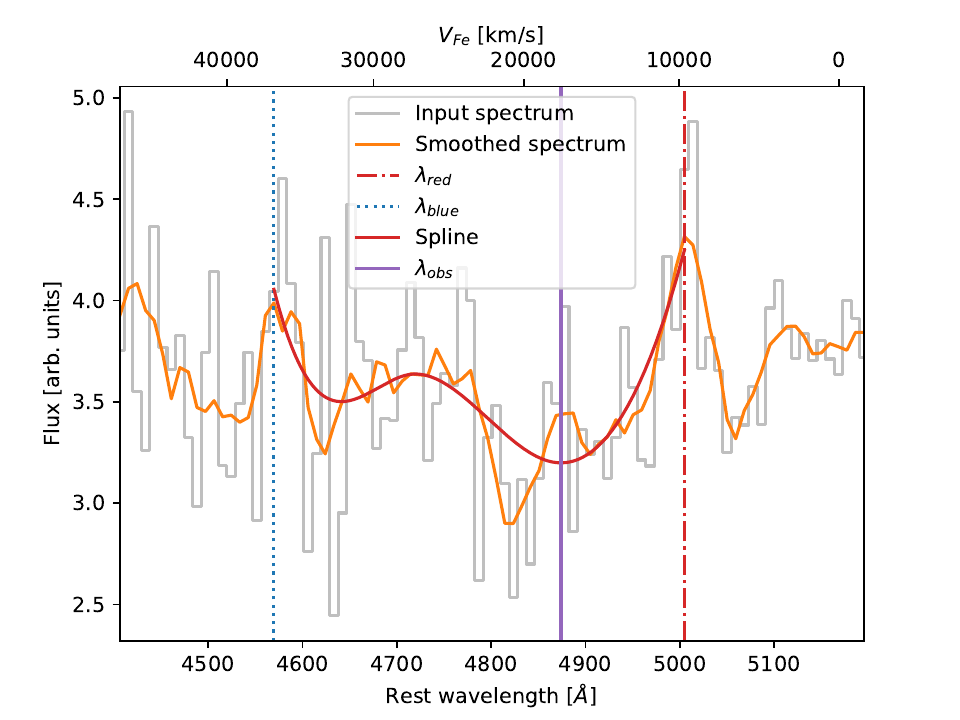}
	\caption{Knots: 10\%; SG: 2.5\%}
	\end{subfigure}
	\begin{subfigure}{0.33\textwidth}
	\includegraphics[width=\linewidth, trim=1.2cm 0cm 1.2cm 0.4cm, clip]{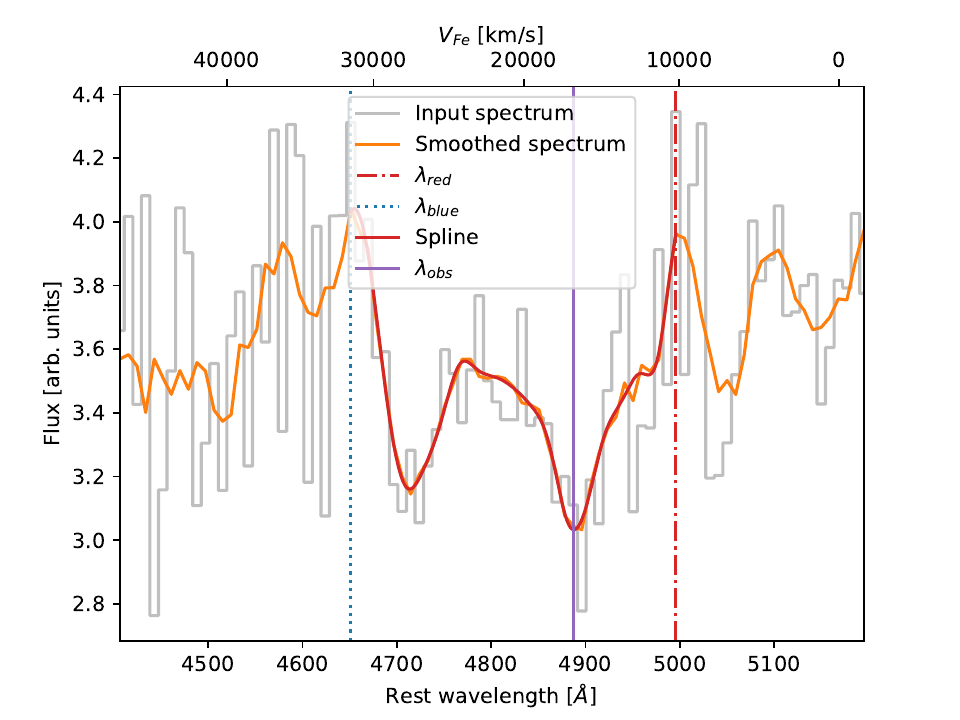}
	\caption{Knots: 50\%; SG: 2.5\%}
	\end{subfigure}
	
	\vspace{0.2cm}
	\begin{subfigure}{0.33\textwidth}
	\includegraphics[width=\linewidth, trim=1.2cm 0cm 1.2cm 0.4cm, clip]{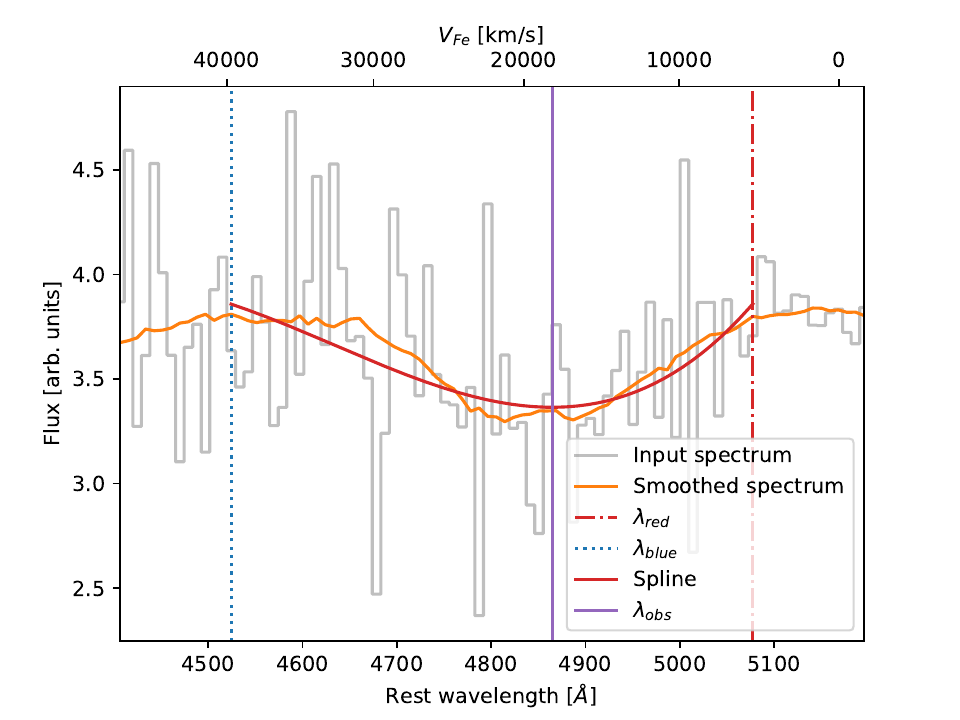}
	\caption{Knots:  0\%; SG: 10\%}
	\end{subfigure}
	\begin{subfigure}{0.33\textwidth}
	\includegraphics[width=\linewidth, trim=1.2cm 0cm 1.2cm 0.4cm, clip]{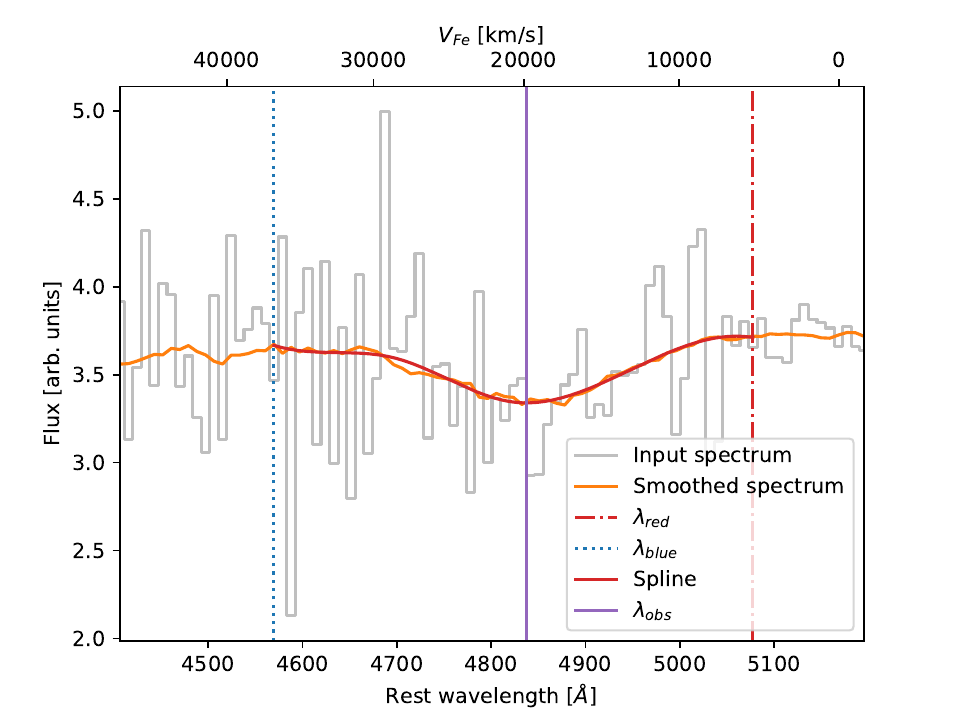}
	\caption{Knots: 10\%; SG: 10\%}
	\end{subfigure}
	\begin{subfigure}{0.33\textwidth}
	\includegraphics[width=\linewidth, trim=1.2cm 0cm 1.2cm 0.4cm, clip]{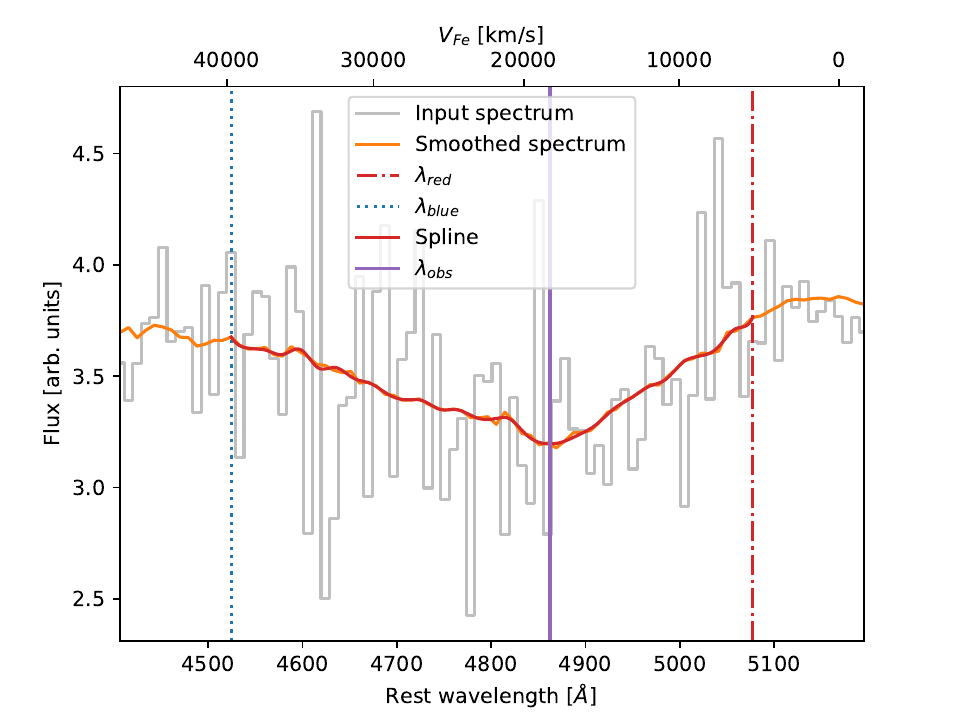}
	\caption{Knots: 50\%; SG: 10\%}
	\end{subfigure}

\caption[Impact of filter width and spline density on the shape and fidelity of the spline fitting method - SN2016P]{Similar to Fig. \ref{fig:1998bwknotswidth} showing the Fe II region of SN2016P; which is a typical example of a low-resolution spectrum with a low S/N. Increases to filter width create a smoother spectrum, which eventually leads to the smearing out of real features. The fidelity of the spline to the smoothed spectrum is improved by increasing the number of knots used for the spline, but eventually meets diminishing returns. In this instance the method clearly struggles with the low S/N, with the minimum wavelength shifting significantly between fits. Optimum parameters are likely a high level of smoothing coupled with a medium knot density of $\sim$10\%. A full description of the sub-figures is available in \cite{Finneran.2024B}}

\label{fig:2016Pknotswidth}
\end{figure}

\begin{figure}[h!]

	\centering
	\begin{subfigure}{0.33\textwidth}
		\includegraphics[width=\linewidth, trim=1.2cm 0cm 1.2cm 0.4cm, clip]{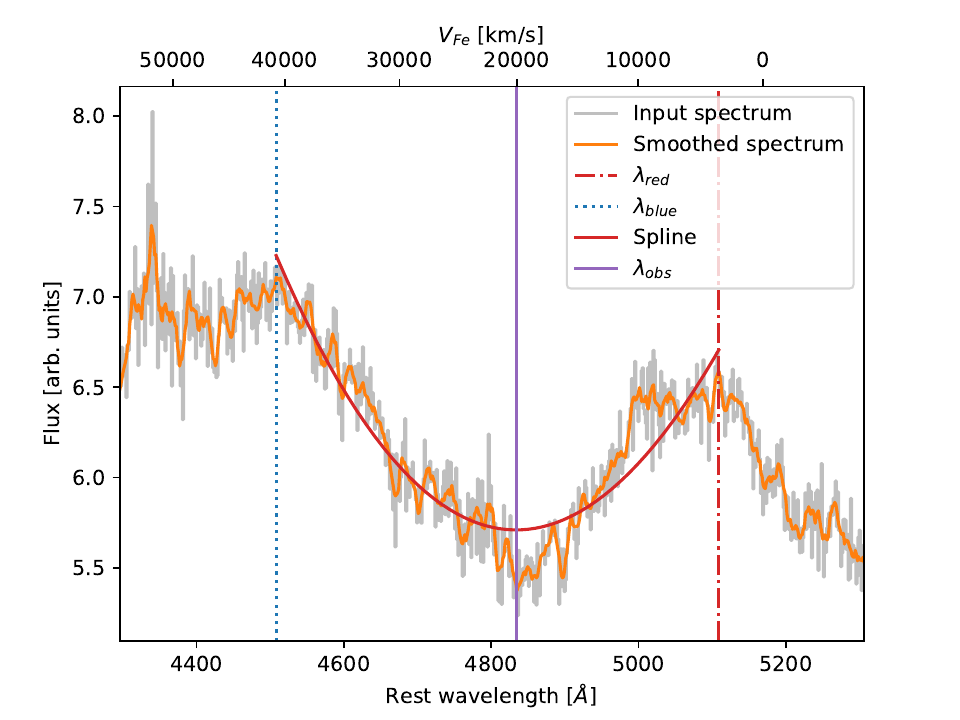}
		\caption{Knots: 2; SG: 0.5\%}
	\end{subfigure}
	\begin{subfigure}{0.33\textwidth}
		\includegraphics[width=\linewidth, trim=1.2cm 0cm 1.2cm 0.4cm, clip]{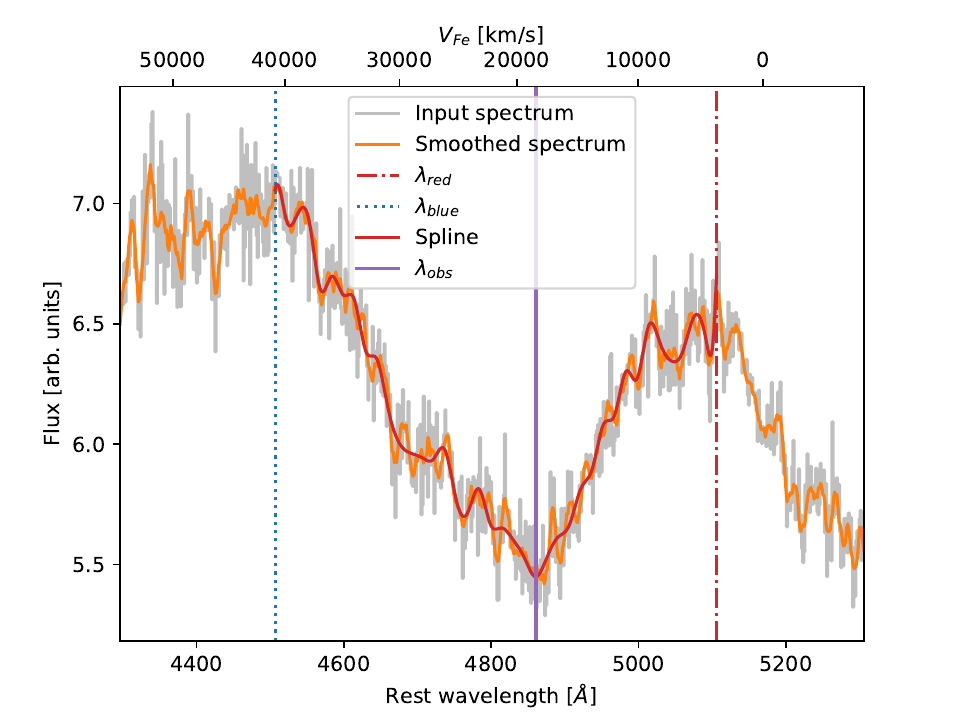}
		
		\caption{Knots: 10\%; SG: 0.5\%}
	\end{subfigure}
	\begin{subfigure}{0.33\textwidth}
		\includegraphics[width=\linewidth, trim=1.2cm 0cm 1.2cm 0.4cm, clip]{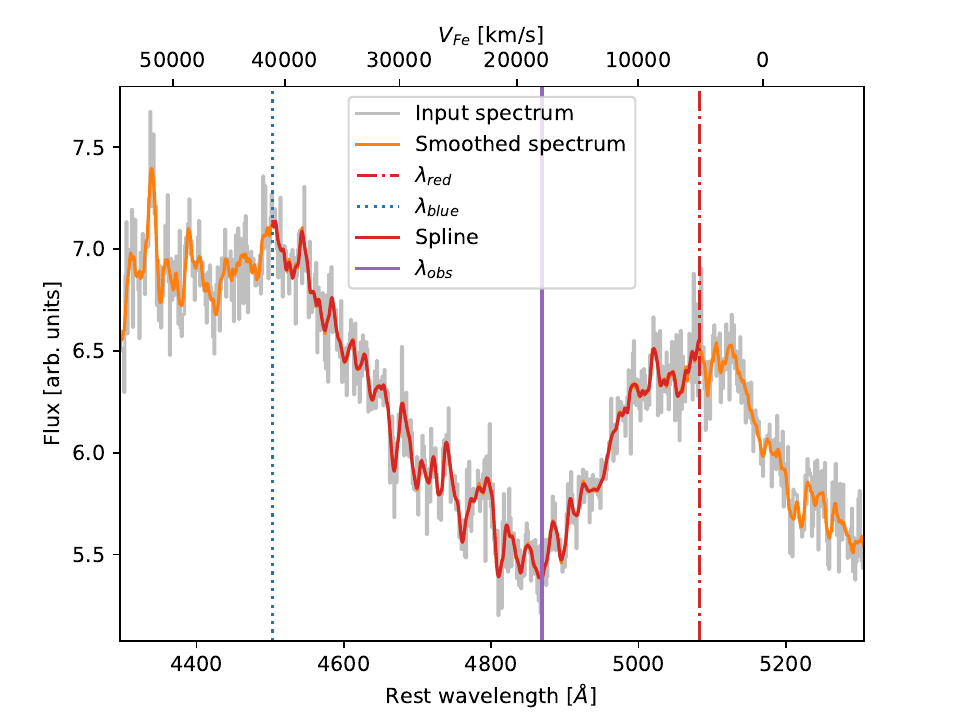}
		
		\caption{Knots: 50\%; SG: 0.5\%}
	\end{subfigure}

	\vspace{0.2cm}
	\begin{subfigure}{0.33\textwidth}
		\includegraphics[width=\linewidth, trim=1.2cm 0cm 1.2cm 0.4cm, clip]{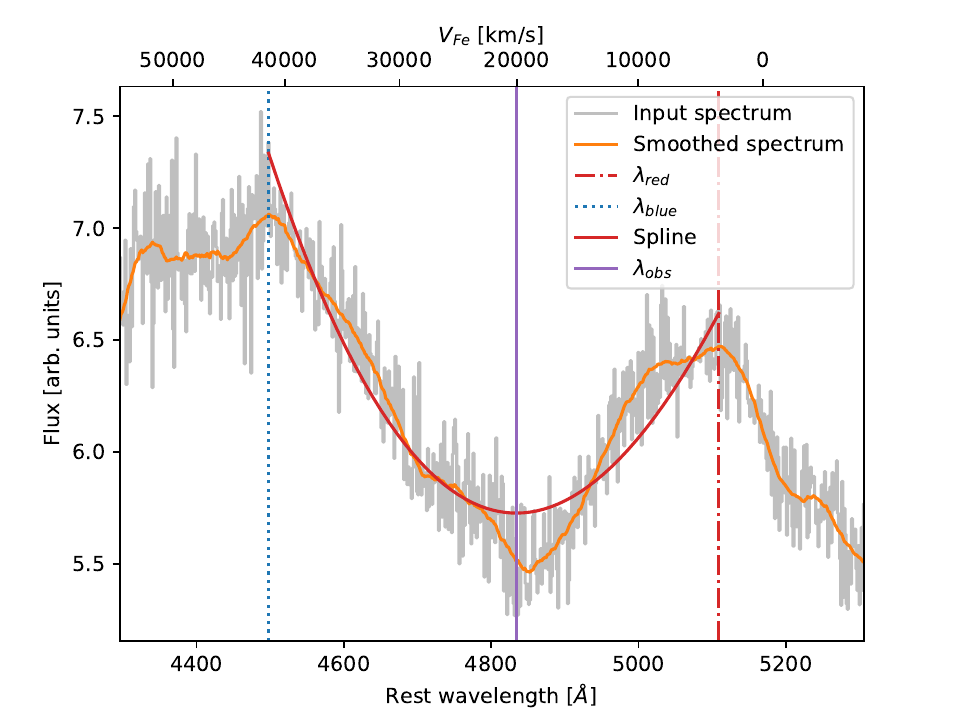}
		
		\caption{Knots: 2; SG: 2.5\%}
	\end{subfigure}
	\begin{subfigure}{0.33\textwidth}
		\includegraphics[width=\linewidth, trim=1.2cm 0cm 1.2cm 0.4cm, clip]{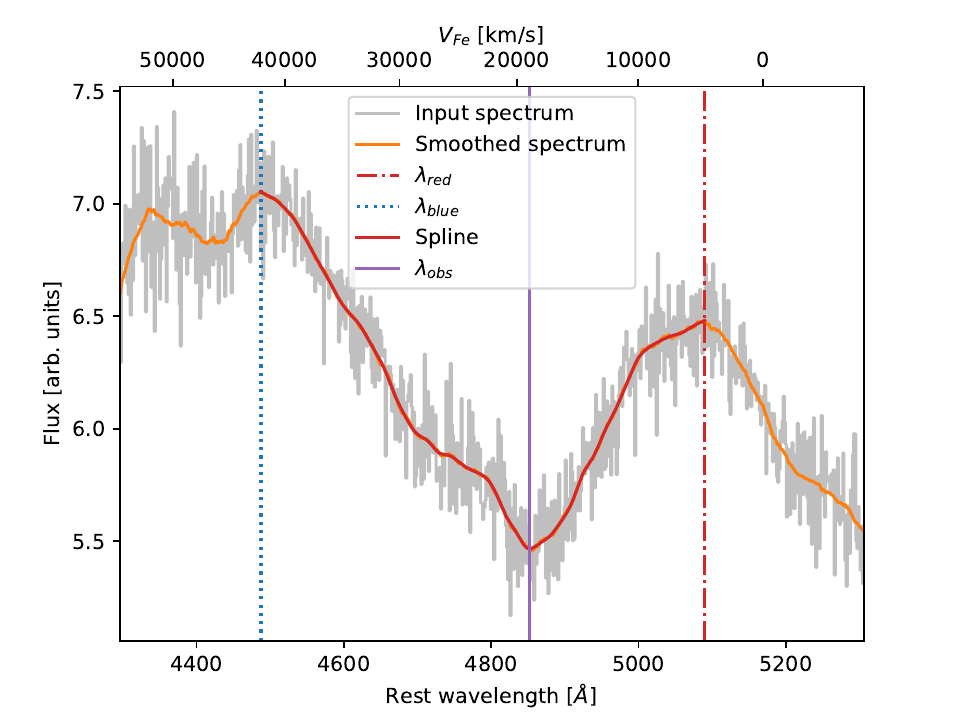}
		
		\caption{Knots: 10\%; SG: 2.5\%}
	\end{subfigure}
	\begin{subfigure}{0.33\textwidth}
		\includegraphics[width=\linewidth, trim=1.2cm 0cm 1.2cm 0.4cm, clip]{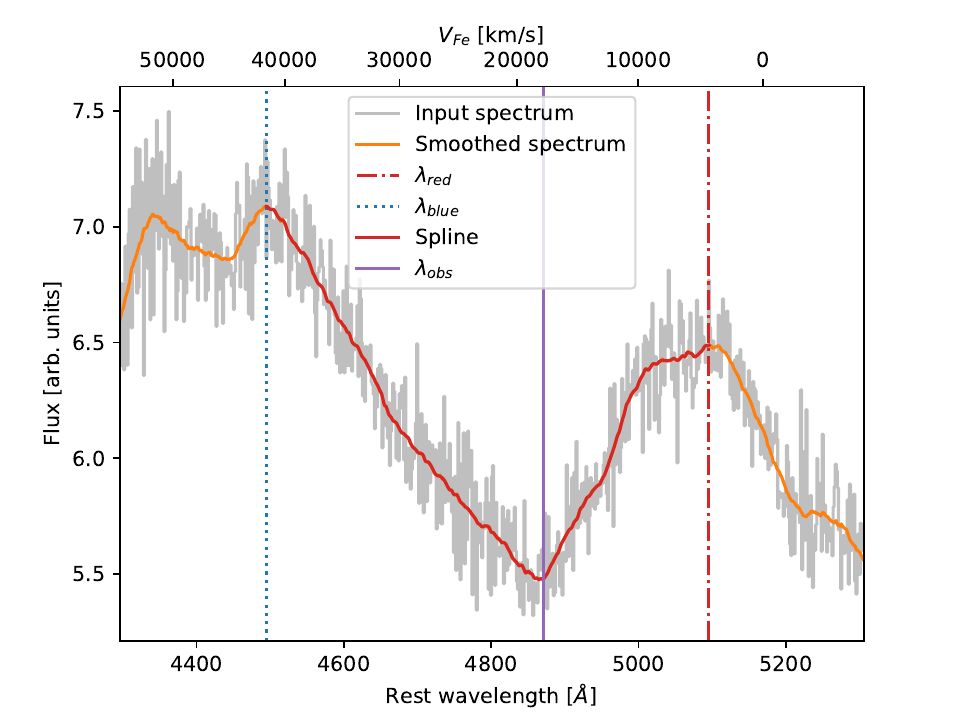}
		
		\caption{Knots: 50\%; SG: 2.5\%}
	\end{subfigure}
	
	\vspace{0.2cm}
	\begin{subfigure}{0.33\textwidth}
		\includegraphics[width=\linewidth, trim=1.2cm 0cm 1.2cm 0.4cm, clip]{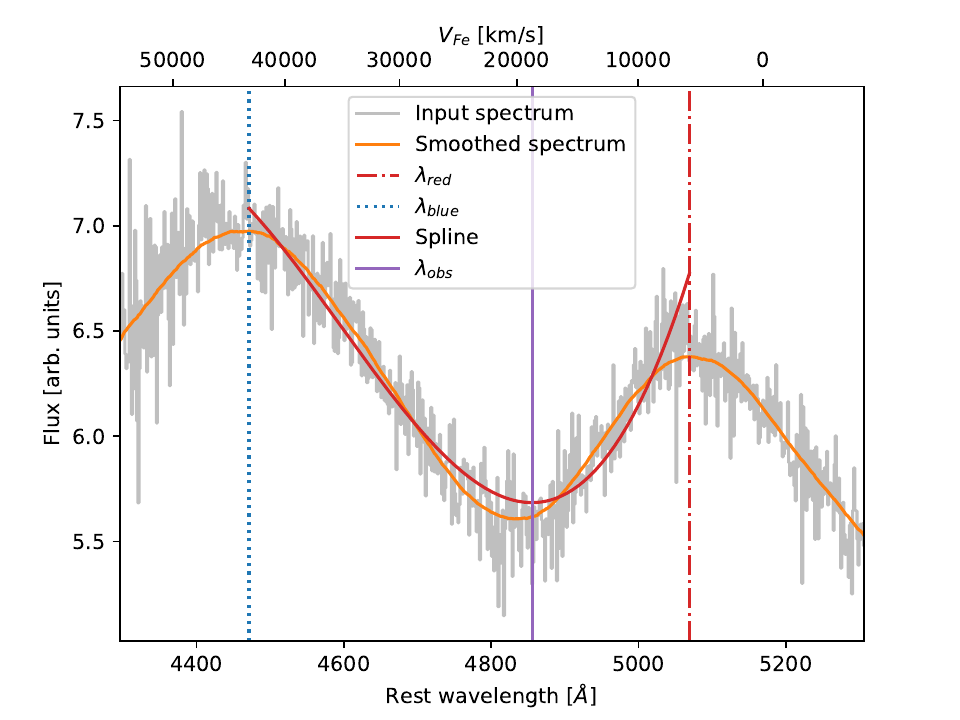}
		
		\caption{Knots:  0\%; SG: 10\%}
	\end{subfigure}
	\begin{subfigure}{0.33\textwidth}
		\includegraphics[width=\linewidth, trim=1.2cm 0cm 1.2cm 0.4cm, clip]{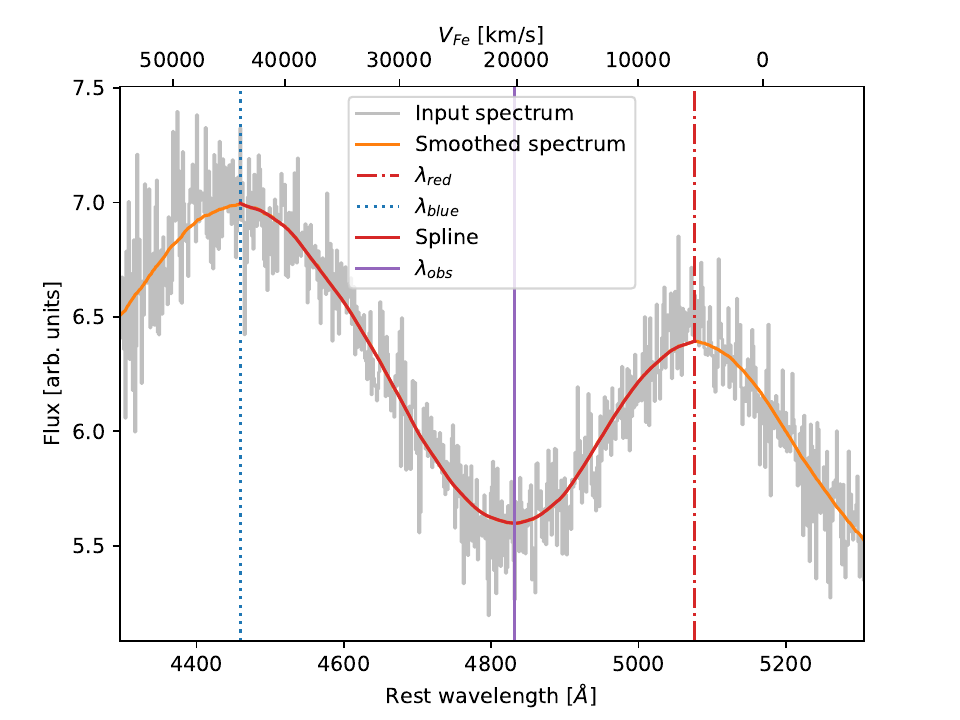}
		\caption{Knots: 10\%; SG: 10\%}
	\end{subfigure}
	\begin{subfigure}{0.33\textwidth}
		\includegraphics[width=\linewidth, trim=1.2cm 0cm 1.2cm 0.4cm, clip]{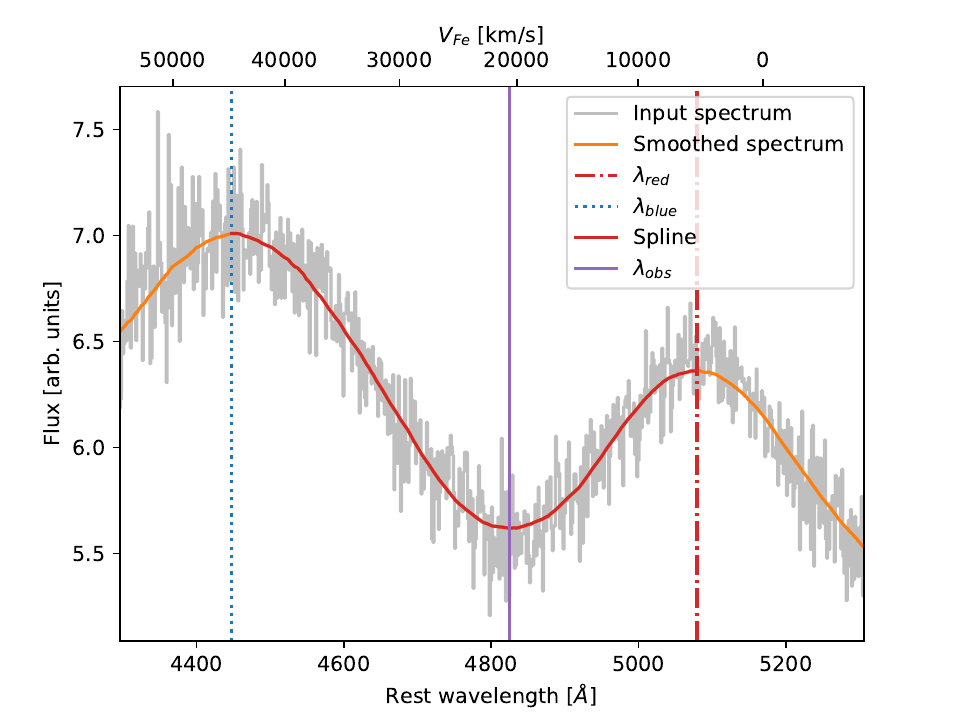}
		\caption{Knots: 50\%; SG: 10\%}

	\end{subfigure}

\caption[Impact of changing the filter width and number of spline knots on the shape and fidelity of the spline fitting method - GRB030329-SN2003jd]{Similar to Fig. \ref{fig:1998bwknotswidth} showing the Fe II region for GRB030329-SN2003jd; which is a typical example of a high resolution spectrum with good S/N. Increases to filter width create a smoother spectrum; eventually this leads to the smearing out of real features. The fidelity of the spline to the smoothed spectrum is improved by increasing the number of knots used for the spline, but eventually meets diminishing returns. This analysis suggests that an optimum spline density of 10\% should be paired with a filter width of $\sim$2.5\%.}
	\label{fig:2003jdknotswidth}
\end{figure}

\begin{figure}[h!]

	\centering
	\begin{subfigure}{0.33\textwidth}
		\includegraphics[width=\linewidth, trim=1.2cm 0cm 1.2cm 0.4cm, clip]{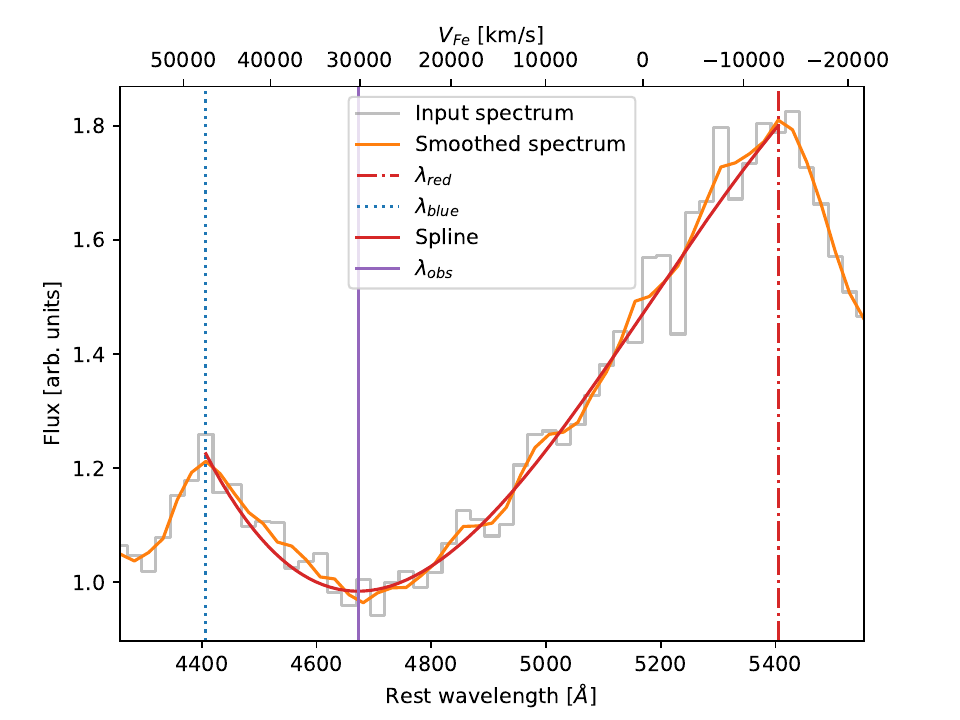}
		\caption{Knots: 2; SG: 3.25\%}
	\end{subfigure}
	\begin{subfigure}{0.33\textwidth}
		
		\includegraphics[width=\linewidth, trim=1.2cm 0cm 1.2cm 0.4cm, clip]{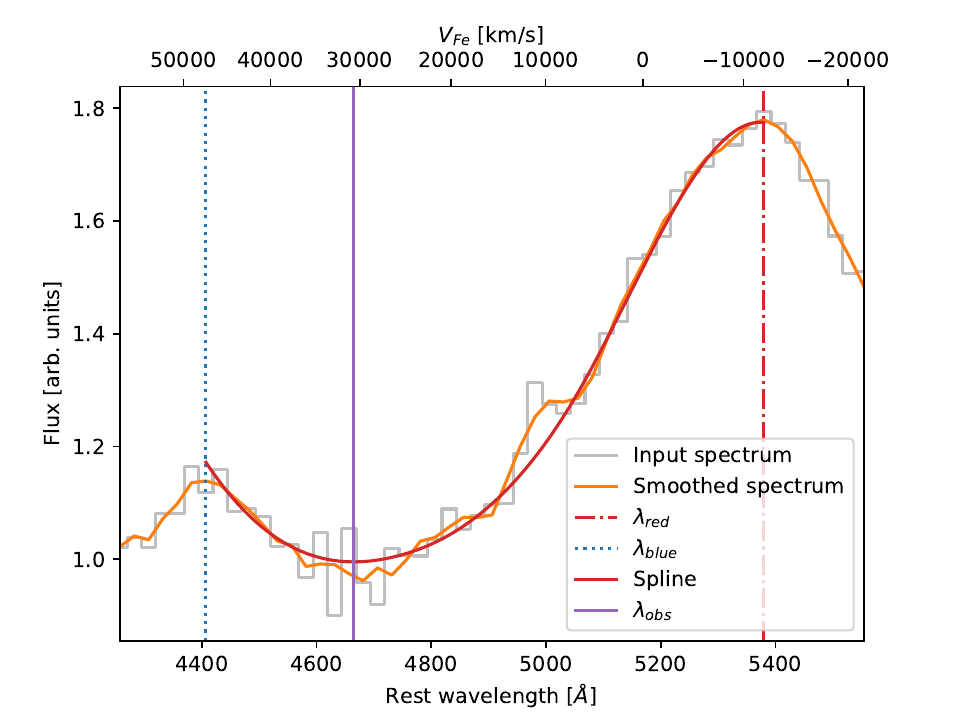}
		\caption{Knots: 10\%; SG: 3.25\%}
	\end{subfigure}
	\begin{subfigure}{0.33\textwidth}
	\includegraphics[width=\linewidth, trim=1.2cm 0cm 1.2cm 0.4cm, clip]{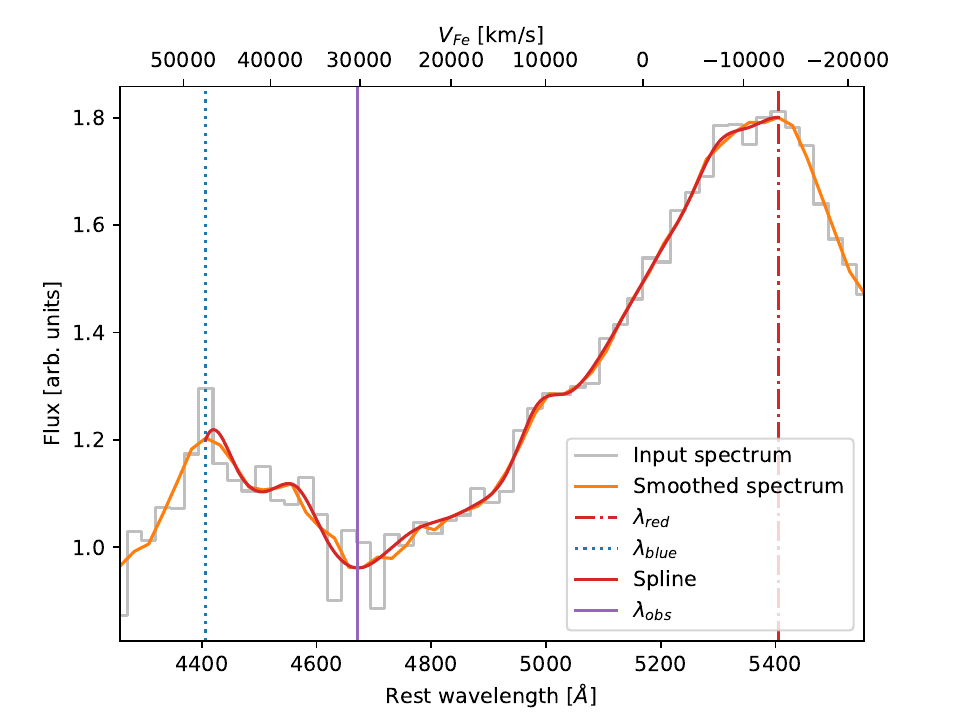}
		
		\caption{Knots: 50\%; SG: 3.25\%}
	\end{subfigure}
	
	\begin{subfigure}{0.33\textwidth}
		\includegraphics[width=\linewidth, trim=1.2cm 0cm 1.2cm 0.4cm, clip]{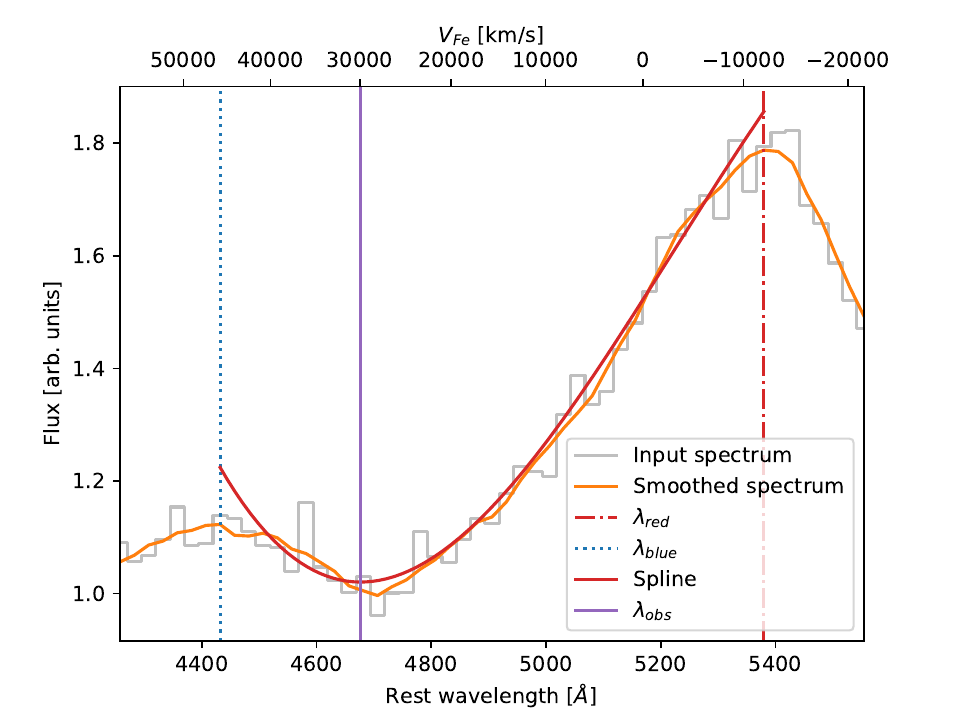}
		
		\caption{Knots: 2; SG: 5\%}
	\end{subfigure}
	\begin{subfigure}{0.33\textwidth}
		\includegraphics[width=\linewidth, trim=1.2cm 0cm 1.2cm 0.4cm, clip]{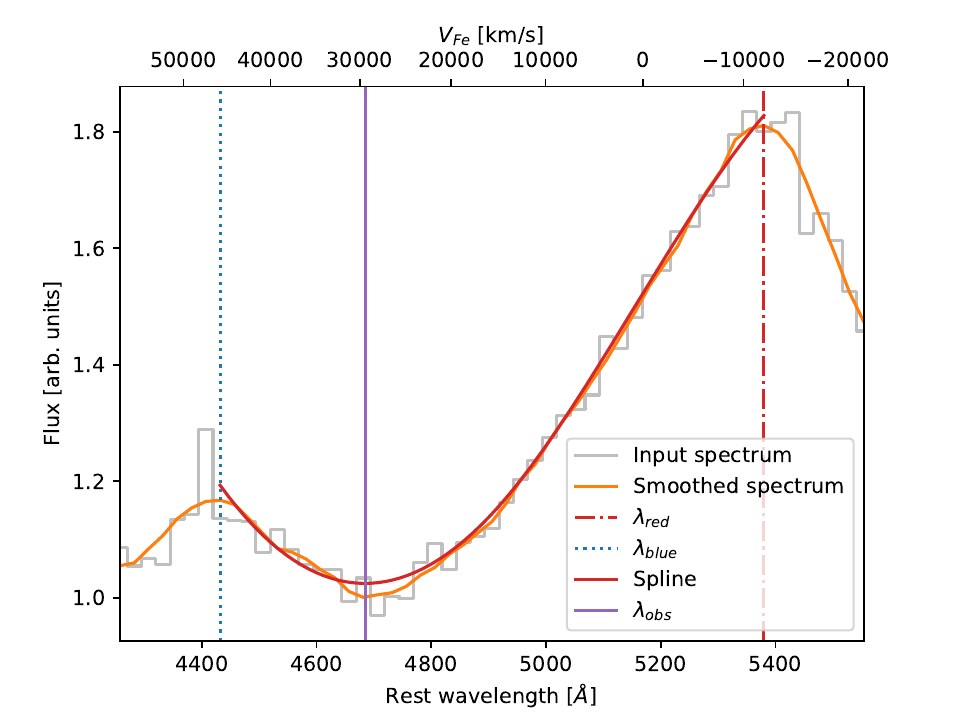}
		
		\caption{Knots: 10\%; SG: 5\%}
	\end{subfigure}
	\begin{subfigure}{0.33\textwidth}
		\includegraphics[width=\linewidth, trim=1.2cm 0cm 1.2cm 0.4cm, clip]{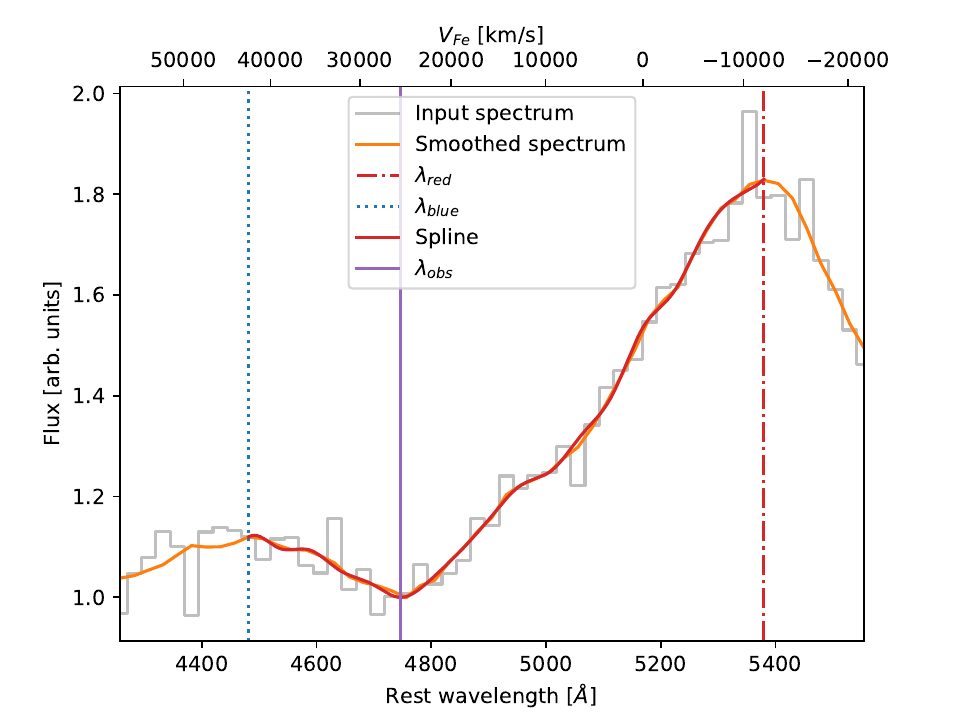}
		
		\caption{Knots: 50\%; SG: 5\%}
	\end{subfigure}
	
	\begin{subfigure}{0.33\textwidth}
		\includegraphics[width=\linewidth, trim=1.2cm 0cm 1.2cm 0.4cm, clip]{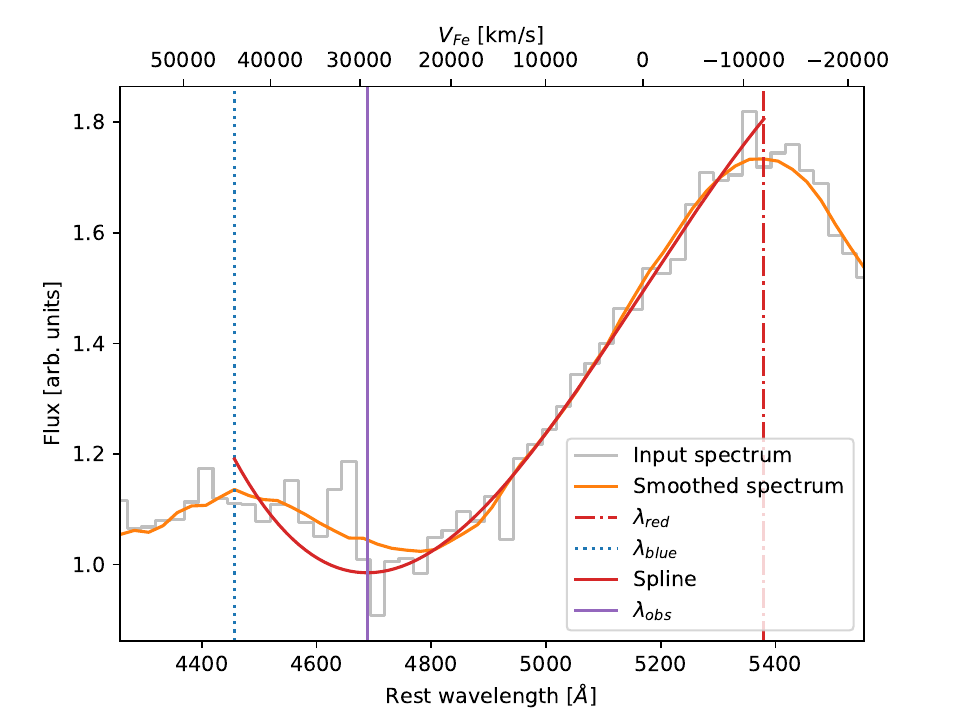}
		
		\caption{Knots: 2; SG: 10\%}
	\end{subfigure}
	\begin{subfigure}{0.33\textwidth}
		\includegraphics[width=\linewidth, trim=1.2cm 0cm 1.2cm 0.4cm, clip]{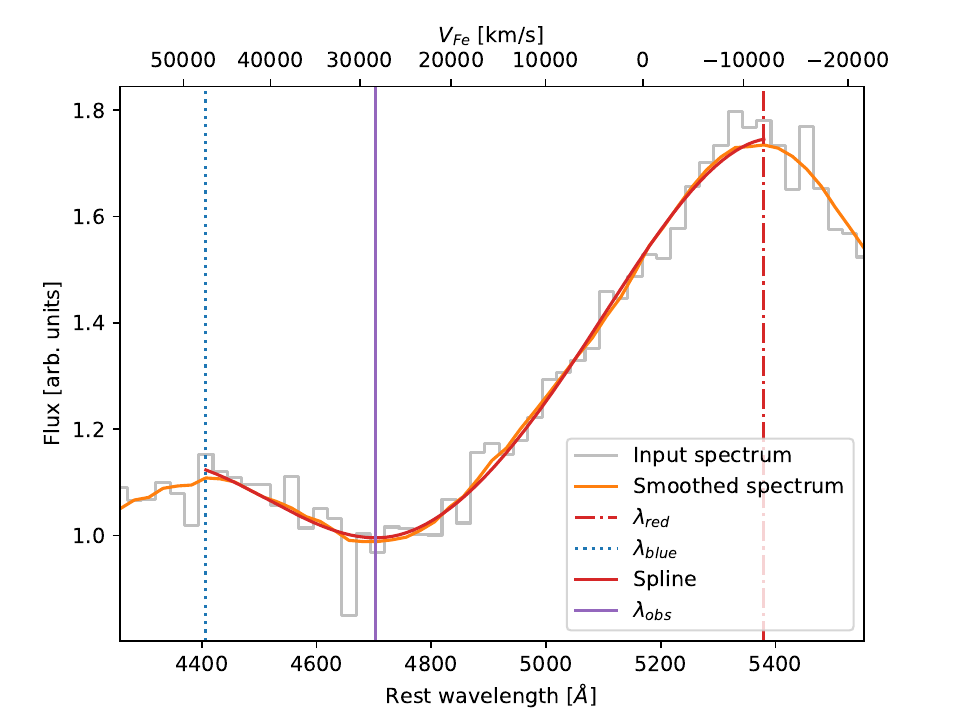}
		\caption{Knots: 10\%; SG: 10\%}

	\end{subfigure}
	\begin{subfigure}{0.33\textwidth}
		\includegraphics[width=\linewidth, trim=1.2cm 0cm 1.2cm 0.4cm, clip]{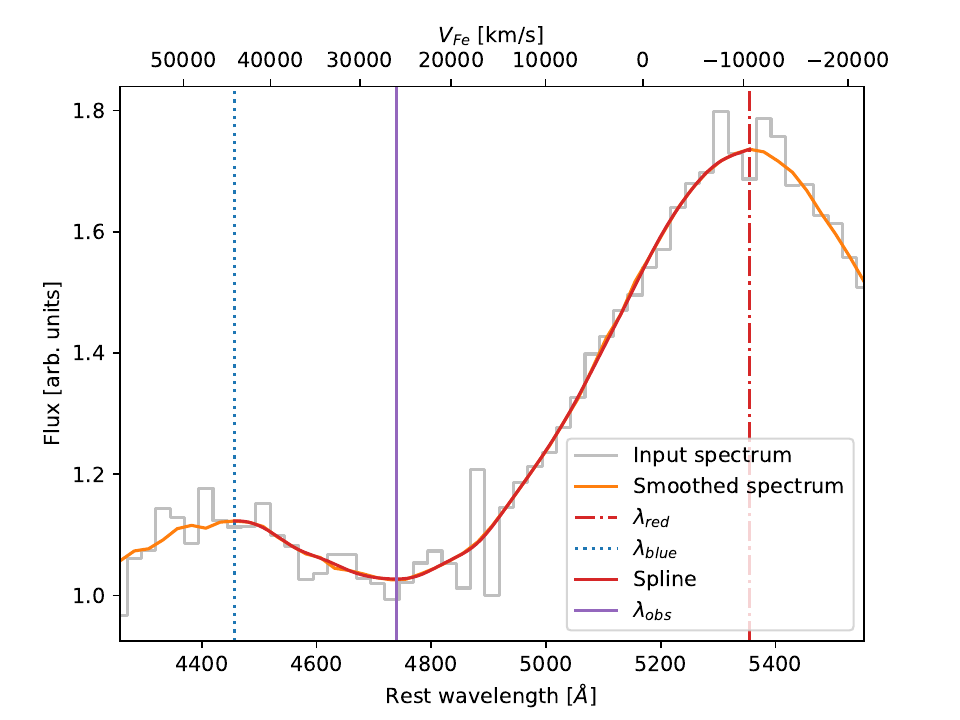}
		\caption{Knots: 50\%; SG: 10\%}

	\end{subfigure}

\caption[Impact of changing the filter width and number of spline knots on the shape and fidelity of the spline fitting method - SN2020bvc]{Same as Fig. \ref{fig:1998bwknotswidth} showing the Fe II region of SN2020bvc; which is a typical example of a low-resolution spectrum with good S/N. Increases to filter width create a smoother spectrum; eventually this leads to the smearing out of real features. The fidelity of the spline to the smoothed spectrum is improved by increasing the number of knots used for the spline, but eventually meets diminishing returns. This analysis suggests that an optimum spline density of 10\% should be paired with a filter width of $\sim$5\%.}
\label{fig:2020bvcknotswidth}
\end{figure}

\end{appendix}

\end{document}